\documentclass[aps,prd,nofootinbib,preprint,amsmath,amsfonts,showpacs,superscriptaddress]{revtex4-1}

\newcommand{\beq}{\begin{equation}}
\newcommand{\eeq}{\end{equation}}

\usepackage{epsfig}
\def\bs{\boldsymbol}

\usepackage[caption=false]{subfig}

\renewcommand\vec[1]{\ensuremath\boldsymbol{#1}} 	%vectors as bold letters
\allowdisplaybreaks 								%aligned equations can span through several pages

\usepackage{bbold}

\begin{document}

\title{Slepian models for Gaussian Random Landscapes}

\author{Jose J. Blanco-Pillado}
\affiliation{Department of Theoretical Physics, University of the Basque Country, 
UPV/EHU, 48080, Bilbao, Spain}
\affiliation{IKERBASQUE, Basque Foundation for Science, 48011, Bilbao, Spain}
\author{Kepa Sousa}
\affiliation{Institute of Theoretical Physics, Faculty of Mathematics and Physics, Charles University in Prague, V Holes\v{o}vi\v{c}k\'ach 2, Prague, Czech Republic}
\author{Mikel A. Urkiola}
\affiliation{Department of Theoretical Physics, University of the Basque Country, UPV/EHU, 48080, Bilbao, Spain}

\begin{abstract}
Phenomenologically interesting scalar potentials are highly atypical in generic  random landscapes.  We develop 
the mathematical techniques to generate constrained random potentials, i.e.  \emph{Slepian models},
which can globally  represent low-probability realizations of the landscape. We give analytical as well as 
numerical methods to construct these Slepian models for constrained realizations of a full Gaussian 
random field around critical as well as inflection points. We use these techniques to numerically  generate 
in an efficient way a large number of minima at arbitrary heights of the potential and calculate
their non-perturbative decay rate. Furthermore we also illustrate how to 
use these methods by obtaining statistical information about the distribution
of observables in an inflationary inflection point constructed within these models.
\end{abstract}

\maketitle

\section{Introduction}
The low energy description of many higher dimensional theories involve a large
number of fields (moduli fields) that need to be stabilized. This is normally achieved by the existence
of a potential that fixes the values of these fields to a local minimum of that potential
function. A typical example of this procedure can be found in String Theory compactification scenarios. In particular,
models of flux compactification have been shown to lead to an enormous set of possible $4d$ potentials
that can have many local minima.  The typical number of moduli fields in these
cases is quite large, reaching often
 the order of a few hundred. This makes  
prohibitively difficult to study these potentials in detail and one is forced to look
for simple models where the field space has been truncated to a small subset
of fields. Alternatively, one can try to study these models by taking a more
statistical approach, where the  scalar potential is regarded as a random field whose  sample space is the set of $4d$ low-energy effective potentials.
These ideas have been pursued in relation to the study
of the stability of critical points  in these potentials in \cite{Ashok:2003gk,Denef:2004ze,Denef:2004cf}, as well 
as the description of cosmological models for the early universe in \cite{Tegmark:2004qd,Aazami:2005jf,Easther:2005zr}.

In many of these studies one is interested in particular points of the landscape such
as, for example, a minimum with some value of its cosmological constant, or
an inflection point with a particular set of conditions in its derivatives necessary for it to sustain inflation. However,
depending on the restrictions imposed,  it may be very
difficult to obtain an example of the potential with these characteristics by producing
random realizations of the scalar potential. Indeed, metastable de Sitter vacua and inflationary points compatible with 
observations are very rare  in generic  landscapes, with probabilities scaling as $P \sim \text{exp}(-  N_f^p)$, where 
$N_f$ is the number of scalar fields in the theory, and $p>0$ is a number of order one 
\cite{Marsh:2011aa,Marsh:2013qca,Bachlechner:2014rqa,Sousa:2014qza,Pedro:2016sli,Freivogel:2016kxc}. To 
obtain  realizations with the desired properties, one can of course use a Taylor expansion around the
point in question and take into account the probability
distribution for its coefficients \cite{Masoumi:2016eag,Masoumi:2017gmh}. However  this becomes quite complicated as
one increases the number of fields and the field range that one is interested 
in\footnote{For another method of generating a specific class of constrained Gaussian random fields, 
see \cite{Bjorkmo:2018txh}.}. Moreover,   with this type of  procedures it is not possible to capture correctly the global 
properties of the scalar potential, which are essential to study  quantum decay processes in the landscape.  Here we 
present a different strategy to generate these potentials that
locally will be constrained to have  a particular form, but that globally will still
represent a faithful realization of the random landscape,  the so-called \emph{Slepian models} \cite{slepian1962one}.\\

Several different methods have been suggested as a way to represent these random
potentials in the landscape.  In this paper we will concentrate on potentials described by Gaussian Random
Fields (GRFs). This is based on the assumption that the $4d$ potential can be 
thought of as a sum of many different terms, of classical and quantum origin, coming from
the compactification mechanism rendering the final result a Gaussian random field. This type
of models have also been studied in connection to the distribution of vacua and its
stability \cite{Bachlechner:2014rqa,Bachlechner:2012at,Easther:2016ire} as well as
inflation \cite{Masoumi:2016eag,Masoumi:2017gmh,Masoumi:2017xbe,Blanco-Pillado:2017nin,Bjorkmo:2018txh}
in the landscape. As an illustration of   the mathematical techniques presented here for the construction of constrained
GRFs we develop Slepian models that are locally described by critical points (maxima,
minima and saddle points) as well as inflection points and
use these realizations to extract important statistical information about them. 

In particular, we will first study the quantum mechanical stability of local minima in these
landscapes. In order to do so, we will compute numerically the decay rate of these
minima using the quantum tunneling techniques first described in a series
of papers in \cite{Coleman:1977py,Callan:1977pt}. The result of this quantum instability 
is the creation of a bubble instanton that interpolates between the false vacuum and the 
true vacuum states. Using these Euclidean methods one can evaluate the probability 
of this decay channel and therefore estimate the lifetime of any specific vacuum. The 
calculation of these tunneling events in a multidimensional
potential is however notoriously difficult. Recently some work on this direction has been done
in relation to the stability of vacua in models with large number of dimensions in field space.
It has been argued that the probability of the decay depends exponentially on the number
of fields although the particular scaling is still uncertain 
\cite{Greene:2013ida,Aravind:2014aza,Aravind:2014pva,Dine:2015ioa}.

In this paper we will study these tunneling events in models of Gaussian random potentials.
In particular we are interested in studying the dependence of the tunnelling rate with the
height of the potential at the false vacuum. For large values of the cosmological constant
this calculation would be impossible without constraining methods, since
the number of these minima is negligible compared to the minima at lower
values of the field. Our techniques allowed us to efficiently generate the same
number of minima for different heights and have a good sample of cases from where 
we can extract statistical information.  The obtained distribution for the instanton actions $S_E$ (which 
determines the decay rate $\Gamma\sim  \mathrm{e}^{-S_E}$) is displayed in Figs. \ref{fig:actions_evol} 
and \ref{fig:actions_comp}, where we found that the average dependence of the decay rate  on the false vacuum height $V_{\text{fv}}$ is given by
\begin{equation}
\left\langle \log_{10} \left( \frac{S_E}{U_0^{-1} \Lambda^4} \right) \right\rangle \approx3.29 \exp \left( - 0.18 \frac{V_{\text{fv}}}{U_0}\right),
\nonumber 
\end{equation}
where $U_0$ and $\Lambda$ are the characteristic energy and length scale (in field space) of our potential respectively.   
The distribution for the Euclidean action becomes increasingly peaked around its mean, and thus  
more predictive, for larger values of $V_{\text{fv}}$. As we show in the main text, this enhancement 
of the predictability can be explained  using  Slepian models  for very atypical extrema of the potential, 
such as high  minima.

Our second application involves the generation of inflection points. These
are some of the most likely points in the landscape where cosmological inflation can happen. However this
does not mean that an arbitrary inflection point would lead to inflation. Obtaining a 
successful inflationary period consistent with the current cosmological observations
still requires some amount of  fine tuning of the potential around the inflection point.
Therefore,  to characterise the distribution of  observables for these inflationary models in the landscape    
one should again use some sort of constraining method, and look at a particular set of non-generic 
inflection points.  In the present paper we will explore the dependence of the observable 
parameters of inflation to its initial conditions in the landscape. In particular we 
will take the initial conditions for the fields to be the ones determined by the exit point of an instanton 
describing the transition from a nearby parent false vacuum. Note that in order to perform 
this analysis, one requires not only the knowledge of the potential around the inflection point 
but also its relation to nearby minima. Hence our method, which  accurately captures the global statistical properties 
of the potential,   is particularly suitable to carry out this investigation. It is worth noting that, to the 
best of our knowledge, this is the first time that an Slepian model for inflection points is presented 
in the literature.  The effect of the tunneling in the initial stages of inflation has also been discussed
in \cite{Freivogel:2005vv,BlancoPillado:2012cb,Blanco-Pillado:2015bha,Masoumi:2017gmh}.  

The remaining of the paper is organized as follows.  In section \ref{sec:preliminaries} we introduce
the notation that we will be using for describing our random potential function as a GRF.
In section \ref{sec:theory} we will outline the method for generating constrained random potentials as Slepian models.
In Section \ref{sec:tunneling}, we implement these ideas for a  $2d$ field space landscape  and generate
a large set of random potentials with a minimum at a specific point in  field space. This allows us to compute the tunneling paths from 
these minima and determine the statistics of the decay rate. In section \ref{sec:inflation}, we condition the  random  potential to 
have an inflection point suitable for inflation, and study the effect of the initial conditions
set by the tunneling process from a nearby minimum. We conclude in Section \ref{sec:conclusions} with 
some comments on the results and some further ideas that can be implemented with these numerical 
techniques. Some of the mathematical details and numerical proofs have been left for the Appendices.
In the present work, unless otherwise stated, we will use  reduced Planck  units $M_{\text{pl}}^{-2} = 8 \pi G/(\hbar c)=1$.

\section{Preliminaries for Gaussian Random Fields}
\label{sec:preliminaries}

In this paper we will take our random potential, $V(\bs{\phi})$, to be a Gaussian random
field defined over a $N$-dimensional field space, which we will parametrize with the vector 
$\bs{\phi}=\{\phi^i\}$, with  $i = 1, \ldots, N$. Furthermore, we will consider the probability 
distribution for the random potential  to be homogeneous and isotropic, so its covariance function 
will only depend on the distance between the points at which it is evaluated, in other words it is of the form
\beq
\left < V ({\bs{\phi_1}}) V (\bs{\phi_2})   \right> = C(|\bs{\phi_1}- \bs{\phi_2}|) ~.
\eeq
We will additionally require the potential to have a null mean:
\beq
\left < V (\bs{\phi}) \right> = 0~.
\label{eq:mean_grf}
\eeq

In the rest of the paper we will evaluate our expressions using the following simple 
covariance function:
\beq
C(\bs{\phi}) = U_0^2 \exp \left( - {{ {\bs{\phi}}^2}\over {2\Lambda^2}} \right)~,
\label{eq:cov_grf}
\eeq
for the case of $N=2$ field space dimensions. The parameter $U_0$ sets the energy scale of the potential while
$\Lambda$ represents the correlation length in field space. 
Generalizing this construction to other covariance functions, or a different number of dimensions in field space 
is straightforward \footnote{Note that this covariance function leads to a somewhat special form of the Hessian matrix for the
minima in this GRF (See for example the discussion of this point in \cite{Easther:2016ire}.) It would be interesting
to check whether this could have any quantitative effect on the conclusions of our paper.}.

In the following we will be interested in the value of the field and its derivatives at a particular point 
in field space,  which we can take to be $\bs{\phi}=0$ without loss of generality, and we will refer 
to it as the center of field space.   In order to simplify the notation we introduce the following definitions 
for the value of the potential and its derivatives: 
\begin{align*}
	u = V(\bs{\phi})|_{\vec{\phi} = \vec{0}} 
	\qquad \eta_i = \left. \frac{\partial V (\vec{\phi})}{\partial \phi^i} \right|_{\vec{\phi} = \vec{0}} 
	\qquad \zeta_{ij} = \left.  \frac{\partial^2V (\vec{\phi})}{\partial \phi^i \partial \phi^j } \right|_{\vec{\phi} = \vec{0}}
	\qquad \rho_{ijk} = \left.  \frac{\partial^3 V (\vec{\phi})}{\partial \phi^i \partial \phi^j \partial \phi^k} \right|_{\vec{\phi} = \vec{0}} ~.
\end{align*}

Furthermore, we will denote the eigenvalues of the Hessian matrix by $\lambda_i$ with $i=1,2$ which will
single out the directions $1,2$ in our field space. Note that   the  derivatives of the scalar potential  are also  
Gaussian random variables, and therefore any collection of the previous quantities  forms a Gaussian random vector.    
 In Appendix \ref{sec:correlations}  we will give the expressions for
the correlators between these different derivatives of the potential as a function of the derivatives
of the covariance function $C(\bs{\phi})$. These correlations will play an important role
in some parts of our discussions.

\section{Slepian Models for constrained Gaussian random fields}
\label{sec:theory}

A key point in our construction of the GRF rests on the fact that a
conditioned GRF maintains its Gaussian nature. More specifically, homogeneous and isotropic 
processes (such as the GRFs we are dealing with) can be conditioned using 
the Kac-Rice formula \cite{lindgren2012stationary} in order to obtain new 
mean and covariance functions which generate GRFs with the required 
constraints \footnote{See a brief description of the Kac-Rice formula in the current context
in Appendix \ref{sec:KR}.}.  The models for stochastic processes dealing with conditional events 
and crossings where pioneered by David Slepian \cite{slepian1962one}, and 
have thus been coined in the mathematical literature as \emph{Slepian models}.

We can describe these constrained processes in a generic form in the following way. For simplicity, let us 
consider first  a Gaussian random $p$-dimensional vector, composed of jointly Gaussian 
variables, $\bs{x}^T = (x_1, \ldots, x_p)$, whose probability distribution function (PDF) is given by,
\begin{equation}
f(\bs{x}) = \frac{1}{(2\pi)^{p/2} \sqrt{\det \Sigma}} \exp \left[ -\frac{1}{2} (\bs{x}-\bs{\mu})^T ~\Sigma^{-1} ~(\bs{x}-\bs{\mu}) \right]
\label{vector_pdf}
\end{equation}
where $\mu = \left\langle \bs{x} \right\rangle$ is the mean \emph{vector} and $\Sigma$ is the \emph{covariance matrix}, whose elements are given by 
\begin{equation}
\Sigma_{ab}= \left\langle (x_a - \mu_a)(x_b-\mu_b) \right\rangle.
\end{equation}
with $a,b =1,\ldots, p$.

Let us now consider the following decomposition of the random vector $\bs{x} = (\bs{x_1},\bs{x_2})$, where 
$\bs{x_2}$ are $p_c$ components of the vector $\bs{x}$ that will be constrained by a condition 
$\bs{x_2} = \bs{\tilde x}$, and $\bs{x_1}$ are the remaining $p-p_c$ unconstrained  elements.  Then one 
can show \cite{lindgren2012stationary,adler2009random} that the distribution probability for  $\bs{x_1}$ 
holding $\bs{x_2}$ fixed to the desired values  is given by,

\small
\begin{align}
{\tilde f} (\bs{x}_1|\bs{x}_2=\tilde{\bs{x}}) = \frac{1}{(2\pi)^{\frac{p-p_c}{2}} \sqrt{\det \tilde{\Sigma} }} \exp \left[ - \frac{1}{2} \left( \bs{x}_1 - \tilde{\vec\mu} \right)^T \tilde{\Sigma}^{-1}  \left( \bs{x}_1 - \tilde{\vec\mu} \right) \right],
\end{align}
\normalsize
which shows that the distribution for the variables $\bs{x_1}$ is indeed a Gaussian distribution but now 
with a mean and covariance functions given in terms of the original ones as
\begin{align}
	\vec{\tilde{\mu}} = \vec{\mu}_1 + \Sigma_{12} \Sigma_{22}^{-1} (\vec{\tilde{x}}-\vec{\mu}_2) \qquad , \qquad   \tilde{\Sigma} = \Sigma_{11} - \Sigma_{12} \Sigma_{22}^{-1} \Sigma_{21} ~,
	\label{eq:cond}
\end{align}
where  $\bs{\mu_1}$ and $\bs{\mu_2}$ are the means of the vectors $\bs{x_1}$ and $\bs{x_2}$ respectively, and
\begin{eqnarray}
\Sigma_{11} &=& \left\langle (\bs{x_1} - \bs{\mu_1})(\bs{x_1}-\bs{\mu_1}) \right\rangle, \nonumber \\
\Sigma_{12} &=& \Sigma_{21} = \left\langle (\bs{x_1} - \bs{\mu_1})(\bs{x_2}-\bs{\mu_2}) \right\rangle, \nonumber \\
\Sigma_{22} &=&  \left\langle (\bs{x_2} - \bs{\mu_2})(\bs{x_2}-\bs{\mu_2}) \right\rangle.
 \end{eqnarray}
 
 This is possible because one can always find a new Gaussian random vector $\bs{x'}=(\bs{x_1'},\bs{x_2'})$, connected 
to the original one with a non-singular  linear transformation $ \bs{x'} = A \cdot \bs{x}$, such that $\bs{x_2'=x_2}$ is 
uncorrelated to $\bs{x'_1}$. We show in Appendix \ref{sec:conditioning} a proof of this statement. In the rest of the paper
we will use this fact in several different ways, applying this technique for Gaussian random vectors made of different quantities 
of our potential.

\subsection{Slepian models for critical points}

In this  section we will use the methods described earlier to generate a Gaussian random field 
with a critical point with a specific height at the center, $\bs{\phi}= \bs{0}$. 
In other words, we will find a description 
of the new GRF conditioned so that the point at its center satisfies the following properties: $V(\bs{0})=u$ and $V'_i (\bs{0}) = \eta_i =0$ for $i=1,2$.  In order
to do this we will follow the prescription used in the mathematical literature for 
maxima in GRF  \cite{Lindgren} and adapt it to our case.
Let us start by introducing the following Gaussian random vector:
\beq
\vec{x} = \lbrace  V(\bs{\phi}_1), \ldots , V(\bs{\phi}_q),V(\bs{0}) ,\eta_1,\eta_2,\zeta_{11},\zeta_{22},\zeta_{12}\rbrace
\label{eq:gaussian_vec_ders}
\eeq
where we denote by $\vec{\phi}_a$, with $a= 1, \ldots, q$, the position in field space of a discrete set of $q$ points.
 One can show that the Gaussian random vector $\bs{x}$ has zero mean, and a
probability distribution that can be readily computed using the form of the covariance function 
and its derivatives. This is a somewhat lengthy calculation and we have given the general
expression in Appendix \ref{sec:conditioned_grf_min}. According to the description for constrained Gaussian random vectors  given above this is all we 
need to obtain the new mean and covariance function for the new conditioned vector (and thus, also for the constrained GRF).

Using the results in Appendix \ref{sec:conditioned_grf_min}, one can show that the new mean function for the GRF 
with the constrained conditions is given by,
\begin{align}
\tilde{\mu}(\bs{\phi}) = e^{-\frac{\bs{\phi}^2}{2\Lambda^2}} \left[ u \left( 1 + \frac{\bs{\phi}^2}{2\Lambda^2} \right) + \frac{1}{2} \sum_{i=1}^2 \phi_i^2 \lambda_i \right].
\label{mean-critical}
\end{align}
This result corresponds to the particular choice of covariance function in Eq. (\ref{eq:cov_grf}),  and is written in terms  of the
the value of the field $V(\bs{0})=u$ and the eigenvalues of the Hessian matrix at the center, $\lambda_i$, which are  to be drawn from the distribution in Eq. \eqref{eq:P_u_lambda} below. The new covariance function is 
\begin{align}
\tilde{C}(\bs{\phi_1},\bs{\phi_2}) = U_0^2 \exp \left[ - \frac{|\bs{\phi_1}|^2 + |\bs{\phi_2}|^2}{2\Lambda^2} \right] \left( \exp \left[  \frac{\bs{\phi_1}\cdot \bs{\phi_2}}{\Lambda^2}  \right] - 1 - \frac{\bs{\phi_1}\cdot\bs{\phi_2}}{\Lambda^2}  - \frac{(\bs{\phi_1}\cdot\bs{\phi_2})^2}{2\Lambda^4}  \right)~,
\label{newcov}
\end{align}
which is no longer homogeneous, but it is still isotropic.

It is important to note that the eigenvalues of the Hessian are not statistically independent of the height
of the potential. This is intuitively clear since, for example, one would expect the typical
minimum at a large height to be quite shallow compared to the minima situated well bellow
the mean value of the potential. This expectation can be translated to the existence of important correlations
between the field and its second derivatives at a point, and in particular  at critical points. In order to 
take this effect into account one can calculate the joint probability distributions for  the Hessian 
eigenvalues ($\lambda_i$) and heights ($u$) at critical points to obtain\footnote{See the calculation
in Appendix \ref{sec:conditioned_grf_min}.}
\begin{align}
P_{u,\lambda} \ du \prod_i d\lambda_i= \mathcal{N} \exp \left[ -\frac{u^2}{2 U_0^2} \right]  |\lambda_1 - \lambda_2| \prod_{i=1}^2 |\lambda_i|  \exp \left[ - \left( \frac{\Lambda^2 \lambda_i + u}{2U_0} \right)^2 \right] d\lambda_i \ du~,
\label{eq:P_u_lambda}
\end{align}
where $\mathcal{N}$ is a normalizing constant. This distribution includes all types of critical points, 
namely maxima, minima and saddle points. Depending on the kind we are interested in, we simply 
need to impose positivity or negativity conditions on the values of each $\lambda_i$.

Using these results we can generate a Gaussian random field with a critical point with the desired 
properties by the following procedure. Let us consider for example a minimum with fixed height $u$. Our 
first step will be to generate a set of eigenvalues drawn from the distribution (\ref{eq:P_u_lambda}) 
taking into account the value of $u$, imposing the non-negativity condition $\lambda_i\ge0$, and fixing the normalization factor accordingly.
 
\begin{figure}
\includegraphics[scale=0.8]{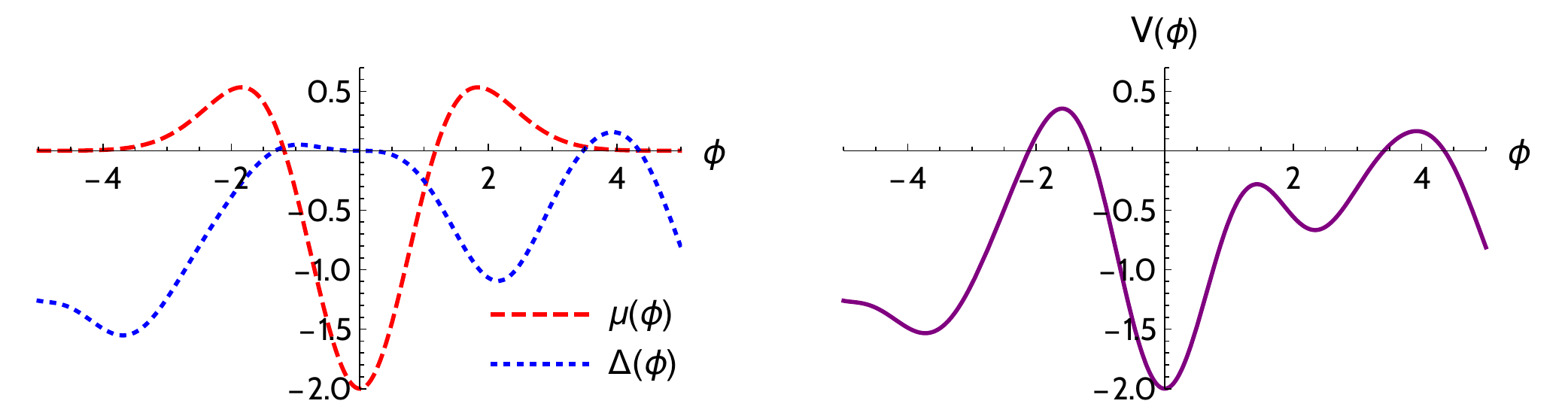}
\caption{A 1d example of a Slepian model of a constrained minimum in a GRF. We 
show, for a particular realization, the two separate components of the construction on the left, namely,  the constrained mean 
field form $\mu(\phi)$ in Eq. (\ref{mean-critical}) and the inhomogenous new GRF $\Delta(\phi)$ 
with covariance function given by Eq. (\ref{newcov}). The total GRF is shown on the right.} 
\label{1d-example}
\end{figure}

Using these values for $\lambda_i$ we can then generate realizations of  the potential using the expression
\beq
V(\bs{\phi}) = e^{-\frac{\bs{\phi}^2}{2\Lambda^2}} \left[ u \left( 1 + \frac{\bs{\phi}^2}{2\Lambda^2} \right) + \frac{1}{2} \sum_{i=1}^2 \phi_i^2 \lambda_i \right] + \Delta(\bs{\phi})
\label{eq:V_decomp}
\eeq
where we have denoted by $\Delta(\bs{\phi})$ an inhomogeneous, zero-mean Gaussian random field 
whose covariance function is given by $\tilde{C}(\bs{\phi_1},\bs{\phi_2})$ in Eq. (\ref{newcov}). We show in 
Fig. (\ref{1d-example}) an example of the different ingredients that make up a Slepian model for a local minimum in a $1d$ GRF. We can use a similar procedure to generate other critical points, such as saddle points with different
number of negative eigenvalues, by generating  the appropriate samples of $\lambda_i$'s. \\

An important conclusion that can be derived from the Slepian model \eqref{eq:V_decomp},  first noticed in \cite{Lindgren}, is 
that for highly non-generic extrema  $|u|\gg U_0$ (such as very low maxima or high minima), the shape of this GRF becomes 
very deterministic around the critical point, and it is described very accurately by the first two terms in Eq. \eqref{eq:V_decomp}. Indeed, 
one can see from Eq. \eqref{newcov} that the standard deviation of the random component $\Delta(\bs{\phi})$ is always 
smaller than $U_0$, and that it approaches zero  near the extremum located at $\bs{\phi} = \bs{0}$ (see also Fig. \ref{1d-example}). 
Therefore the last  contribution in \eqref{eq:V_decomp}  can be neglected in a neighbourhood of the  extremum where 
$|\Delta(\bs{\phi})|\lesssim U_0  \ll |V(\bs{0})|$ holds. On the other hand, in the limit $|u|\gg U_0$ the eigenvalue distribution 
of the Hessian \eqref{eq:P_u_lambda} is approximately given by\footnote{Note that for very high minima $u>0$ and 
$\lambda_i>0$, while for very low maxima all signs are reversed.} 
\begin{align}
P_{\lambda} \ d\lambda_1 \ d\lambda_2   \sim |\lambda_1 - \lambda_2|  |\lambda_1| |\lambda_2|  \exp \left[ - \frac{\Lambda^2 |(\lambda_1+ \lambda_2) u|}{2 U_0^2 } \right] d\lambda_1 \ d\lambda_2~,
\label{eq:P_u_lambda_approx}
\end{align}
which indicates that in this limit the magnitude of the eigenvalues is very suppressed $|\lambda_i| \ll U_0/\Lambda^2$. Then, as we 
mentioned above, for highly non-generic extrema the decomposition \eqref{eq:V_decomp} is dominated by its deterministic 
part (the first term), what makes these Slepian models very predictive in those situations. As we shall see bellow, this result is 
particularly important when we consider the distribution of non-perturbative decay rates from minima with a large vacuum energy. For 
an example of a realization with a high minimum see figure \ref{fig:grf_examples}(a). 

This deterministic character of large fluctuations of  Gaussian Random Fields  plays an important role in various areas of 
Cosmology, such as the analysis of the CMB data, and the study of  Large Scale Structure formation in the 
universe (see e.g. \cite{Bucher_2012,Marcos_Caballero_2017,Bardeen:1985tr, Bertschinger:1987qp,1993ApJ415L5G}).

\subsection{Slepian models for inflection points}

As we discussed in the Introduction, we are also interested in inflection points in the landscape.
The reason is that in a cosmological context these points could be one of the regions of the
potential that give rise to a cosmological inflationary period. However, in order
to be compatible with the latest cosmological observations, one needs to restrict the
 form of these inflection points. This leads us to consider an inflection 
point at $\bs{\phi}=0$ as a realization of the GRF with a small gradient of the potential in 
the $\phi_1$ direction, denoted by $\eta_1$, and the rest of the coefficients of the Taylor expansion
of the field around that point of the form
\beq
\label{inflection-cond}
	\eta_2 = 0 \quad ; \quad \lambda_1 = 0 \quad ; \quad \lambda_2 >0 \quad ; \quad \eta_1 \cdot \rho_{111} > 0~.
\eeq

The intuitive picture of these choices is clear, we are looking for a one dimensional inflection point
that allows the slow-roll conditions to be satisfied along the direction $\phi^1$ while the perpendicular directions
have positive curvature. In other words, we are looking for a potential where inflation is effectively
one dimensional locally. This also explains the last condition, which is imposed in order to allow for enough 
slow-roll inflation in the vicinity of this inflection point. 

This is admittedly a very particular form of the potential around the inflection point and, even though
it could be interesting to identify this type of points in a GRF in other contexts, we have not
seen any studies of this class of constrained points on GRFs in the mathematical literature. However, it is not 
difficult to follow a similar procedure to the one  for critical points in order to obtain Slepian
models in this case. The first thing we should do is to enlarge the form of our initial
Gaussian random vector \eqref{eq:gaussian_vec_ders}, since we now want to constrain not only first derivatives but
second derivatives as well. This suggests that we should take the vector of the form,
\beq
\bs{x}= \lbrace  V(\bs{\phi}_1), \ldots , V(\bs{\phi}_q),V(\bs{0}) ,\eta_1,\eta_2,\zeta_{11},\zeta_{22},\zeta_{12}, \rho_{111} , \rho_{122} , \rho_{222} , \rho_{112}
\rbrace
\eeq
which, similarly to the critical point case, can now be conditioned 
to have the desired properties given in Eq. (\ref{inflection-cond})

Following the computations given in the Appendix \ref{sec:conditioned_grf_ip} one arrives to the result that a GRF 
with an inflection point at $\bs{\phi}=\bs{0}$ is described by the expression
\begin{equation}
V(\bs{\phi})   
= \exp \left[ - \frac{\bs{\phi}^2}{2\Lambda^2} \right] \left( (u + \bs{\phi}\cdot\bs{\eta}) \left( 1 + \frac{\bs{\phi}^2}{2\Lambda^2} \right) + \frac{1}{2} \sum_{i=1}^2 \lambda_i {\phi_i}^2  + \frac{1}{6} \sum_{i,j,k=1}^{2} {\phi}_i {\phi}_j {\phi}_k \rho_{ijk} \right) + \Gamma(\bs{\phi}) \ , 
\end{equation}
where $\Gamma (\vec{\phi})$ is an inhomogeneous zero-mean GRF with covariance function
\begin{equation}
\tilde{C}(\bs{\phi_1},\bs{\phi_2}) = U_0^2 \exp \left[- \frac{|\bs{\phi_1}|^2 + |\bs{\phi_2}|^2}{2\Lambda^2} \right] \left( \exp \left[ \frac{\bs{\phi_1}\cdot \bs{\phi_2}}{\Lambda^2} \right] - 1 - \frac{\bs{\phi_1}\cdot\bs{\phi_2}}{\Lambda^2} - \frac{(\bs{\phi_1}\cdot\bs{\phi_2})^2}{2\Lambda^4} - \frac{(\bs{\phi_1}\cdot\bs{\phi_2})^3}{6\Lambda^6} \right).
\end{equation}
In these expressions  $u$, $\lambda_i$ and $\rho_{ijk}$ should be   drawn from the joint probability distribution for heights, 
 first, second and third derivatives of the potential at inflection points\footnote{See the computation of
these distributions in Appendix \ref{sec:conditioned_grf_ip}.}
\begin{align}
P_{\text{inf}} ~ du~ d\lambda_2 ~d\eta_1 ~d{\bs \rho}= \mathcal{N} |\lambda_2|^2 |\rho_{111}| \ P ( u,  \lambda_2 \ \left|  \ \lambda_1 = 0 \right. ) \ P \left( \eta_1, \rho_{ijk} \ \left| \ \eta_2 = 0 \right. \right) ~ du~ d\lambda_2 ~d\eta_1 ~d{\bs \rho}
\label{eq:P_ip}
\end{align}
where
\begin{align}
& P ( u, \lambda_2 \ \left|  \ \lambda_1 = 0 \right. ) \ du \ d\lambda_2 = \mathcal{N} |\lambda_2| \exp \left[ - \frac{4u^2 - 2\Lambda^2 u \lambda_2 - \Lambda^4 \lambda_2^2}{2U_0}  \right] \ du \ d\lambda_2 , \nonumber\\[10pt]
& P \left( \eta_1, \rho_{ijk} \ \left| \ \eta_2 = 0 \right. \right) \ d\eta_1 \ d\rho_{ijk} = \nonumber \\[5pt]
& \hspace{30pt} \mathcal{N} \exp \left[ - \frac{\Lambda^2}{12 U_0^2} \left( 18 \eta_1^2 + 6 \Lambda^2 \eta_1 (\rho_{111} + \rho_{122}) + \Lambda^4 \sum_{i,j,k=1}^2 \rho_{ijk}^2 \right) \right] \ d\eta_1 \ d\rho_{ijk}~.
\label{inflection-correlations}
\end{align}
In the last distribution, the condition $\eta_1 \cdot \rho_{111} > 0$ should also be imposed if one is interested in `inflationary' inflection points. 

We have checked the accuracy of  these distributions by numerically computing them from a large set of generic (unconstrained)
GRF examples. We have identified all the inflection points of our sample, and used this information to compute the
distributions of the parameters of the inflection points we are interested in. See Appendix \ref{sec:numerical} for
the details of these numerical checks, which are summarised in figure \ref{fig:ip_dist}.

\subsection{2D numerical implementation}

All GRFs generated for this work were constructed following the Karhunen-Love
expansion (see e.g. \cite{adler2009random}), which is briefly described in Appendix \ref{sec:numerical}. This algorithm generates values for a GRF discretized over a lattice which is to be interpolated afterwards. 

Based in the criteria developed in \cite{Masoumi:2017gmh}, we used 5 lattice points 
per correlation length (25 per length squared). The resulting grid was then interpolated 
with fourth-order splines in order to analyse up to third-order derivatives of the field as 
faithfully as possible. The generated GRFs were found to follow successfully the initial 
mean and covariance function, as well as other properties such as the distribution of 
critical points and eigenvalues thereof.

\begin{figure}
	\centering
	\subfloat[]{
		\includegraphics[width=0.7\textwidth]{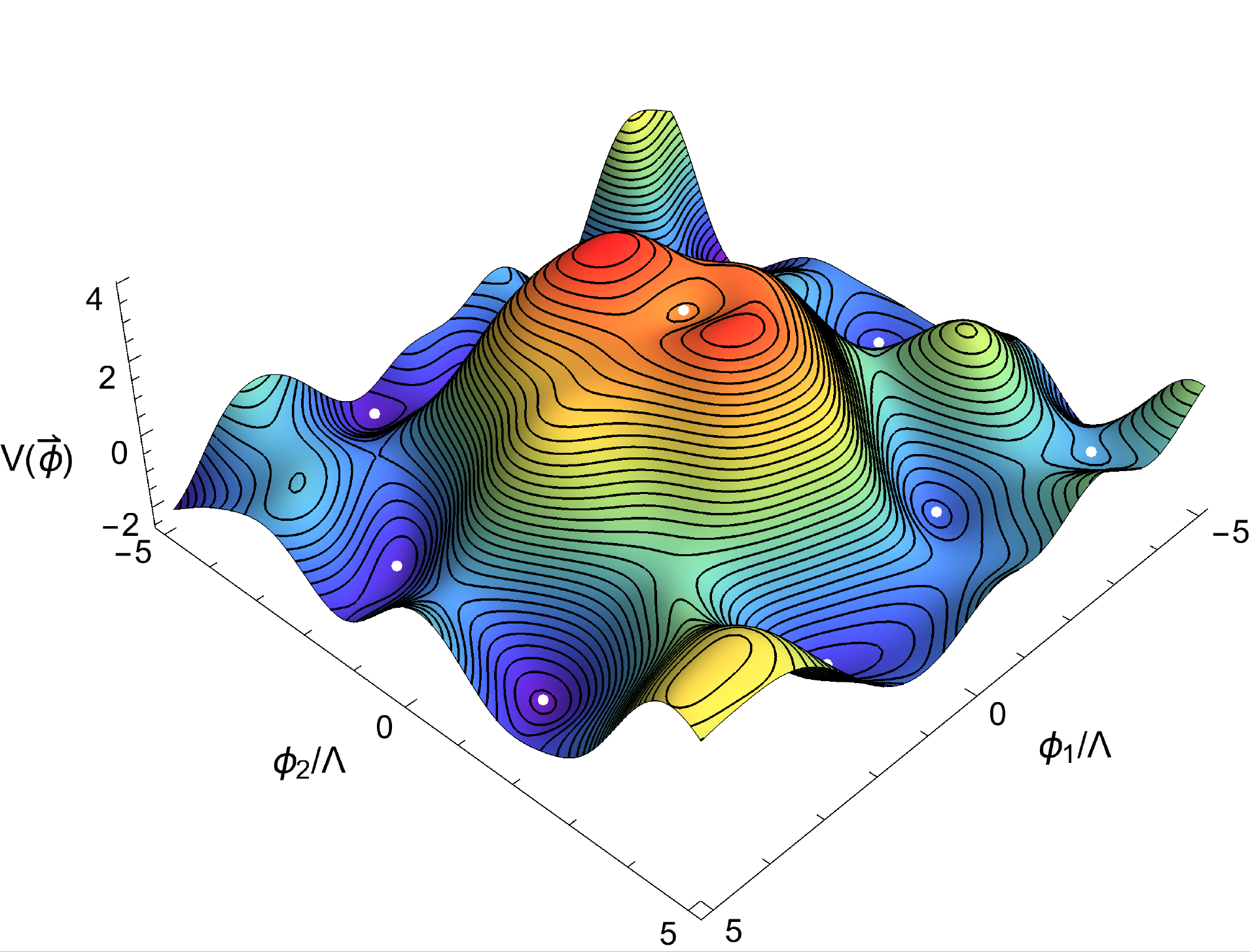}
	}
	\\	
	\subfloat[]{
		\includegraphics[width=0.7\textwidth]{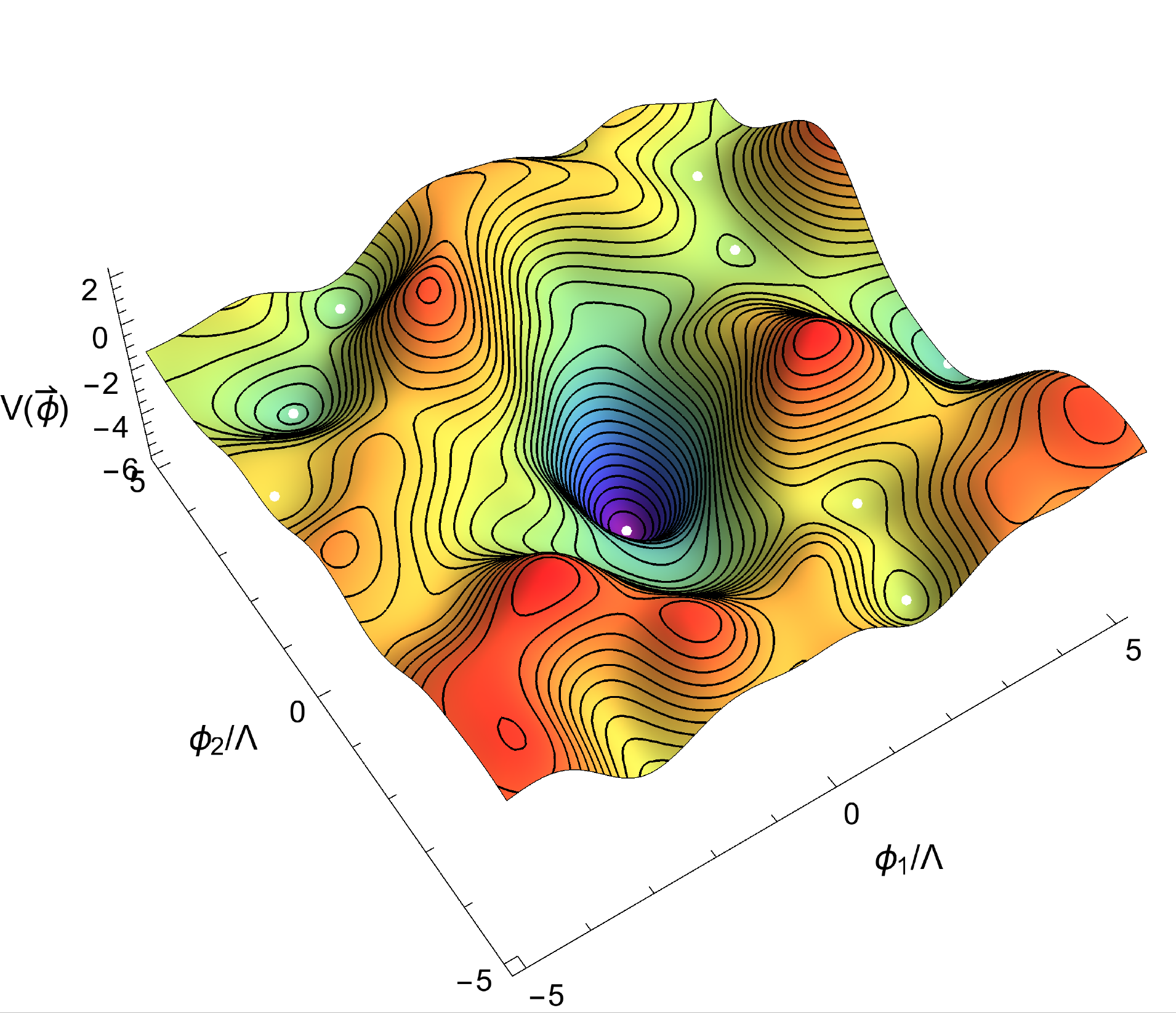}
	}
	\caption{A pair of realizations of a $2d$ Gaussian random fields with zero mean, covariance 
		function (\ref{eq:cov_grf}), and conditioned to have a minimum at center of height (in units of $U_0$) 
		4 (a) and  -4 (b). The higher the minimum is, the lower its eigenvalues will typically be and 
		vice versa (see text). The location of the minima of each realization has been marked with a white 
		dot. }
	\label{fig:grf_examples}
\end{figure}

Two examples of (rather extreme) GRFs generated following the steps in this section have been plotted in figure \ref{fig:grf_examples}. 

\section{Tunneling in a Gaussian random landscape}
\label{sec:tunneling}

A Gaussian random landscape possesses a large number of perturbatively
stable minima. However, we know that quantum mechanically these 
vacua are not completely stable and can decay by the nucleation of a
bubble of the new state. This means that a typical vacuum in our landscape
will have many channels to decay into, each of them with a different
probability. Here we would like to study the statistics of these decay
channels in a controlled way by generating a large number of GRF realizations,
and analyse their dependence on the parameters of the central minimum.

In order to do that we will use the instanton techniques first described by
Coleman and collaborators  \cite{Coleman:1977py} where 
 it was shown that for a given minimum of the potential the decay probability  per unit time and per unit 
volume is given by
\begin{equation}
	\Gamma / V \sim A e^{-S_E} 
	\label{eq:decay_rate}
\end{equation}
where $S_E$ is the Euclidean action for the bounce solution that interpolates between
the new state and the original one\footnote{Here we will not be concerned with the pre factor
$A$. See \cite{Callan:1977pt} for a detailed description of its computation.}.

In the absence of gravity, one can show that the most likely decay channel is given by the
$O(4)$-symmetric instanton solution in a 4-dimensional Euclidean spacetime; therefore 
we will be interested in solving the following set of Euclidean equations of motion
\begin{align}
	\phi_i'' + \frac{3}{r} \phi_i' = \frac{\partial V(\bs{\phi})}{\partial \phi_i} \  , 
	\label{eq:eom}
\end{align} 
where the prime denotes a derivative with respect to the radial coordinate in 4-dimensional Euclidean
spacetime, $r$, and we have assumed that the 
fields $\vec{\phi}(r) = \lbrace \phi^1 (r),\ldots,\phi^N (r) \rbrace$ 
are canonically normalized. Finally the boundary conditions are
\begin{align}
	\vec{\phi} (\infty) = \vec{\phi}_{FV} \qquad , \qquad \phi_i'(0) = 0.
\end{align}
where $\bs{\phi}_{FV}$ is the location of the false vacuum in field space, the minimum of the 
potential from which the decay happens.
Once the field equations have been solved, the action in the exponent of (\ref{eq:decay_rate}) reads
\begin{align}
	S_E=2\pi^2 \int_0^{\infty} dr \ r^3 \left[ \frac{1}{2} |\vec\phi '|^2 + V(\vec{\phi}) - V(\vec{\phi}_{FV}) \right]~.
\end{align}

Computing the coupled system of the instanton equations (\ref{eq:eom}) is no easy task; 
particularly, as the dimensionality of the field space grows, the solutions tend to be increasingly 
 unstable. There are, however, several publicly available algorithms in the literature 
to tackle the problem (see, e.g., \cite{Wainwright:2011kj,Athron:2019nbd}); additionally,
some alternative methods have been recently proposed to find the action and escape point for the
instanton, as in \cite{Espinosa:2018hue,Espinosa:2018szu}. 

\begin{figure}
\includegraphics[scale=0.8]{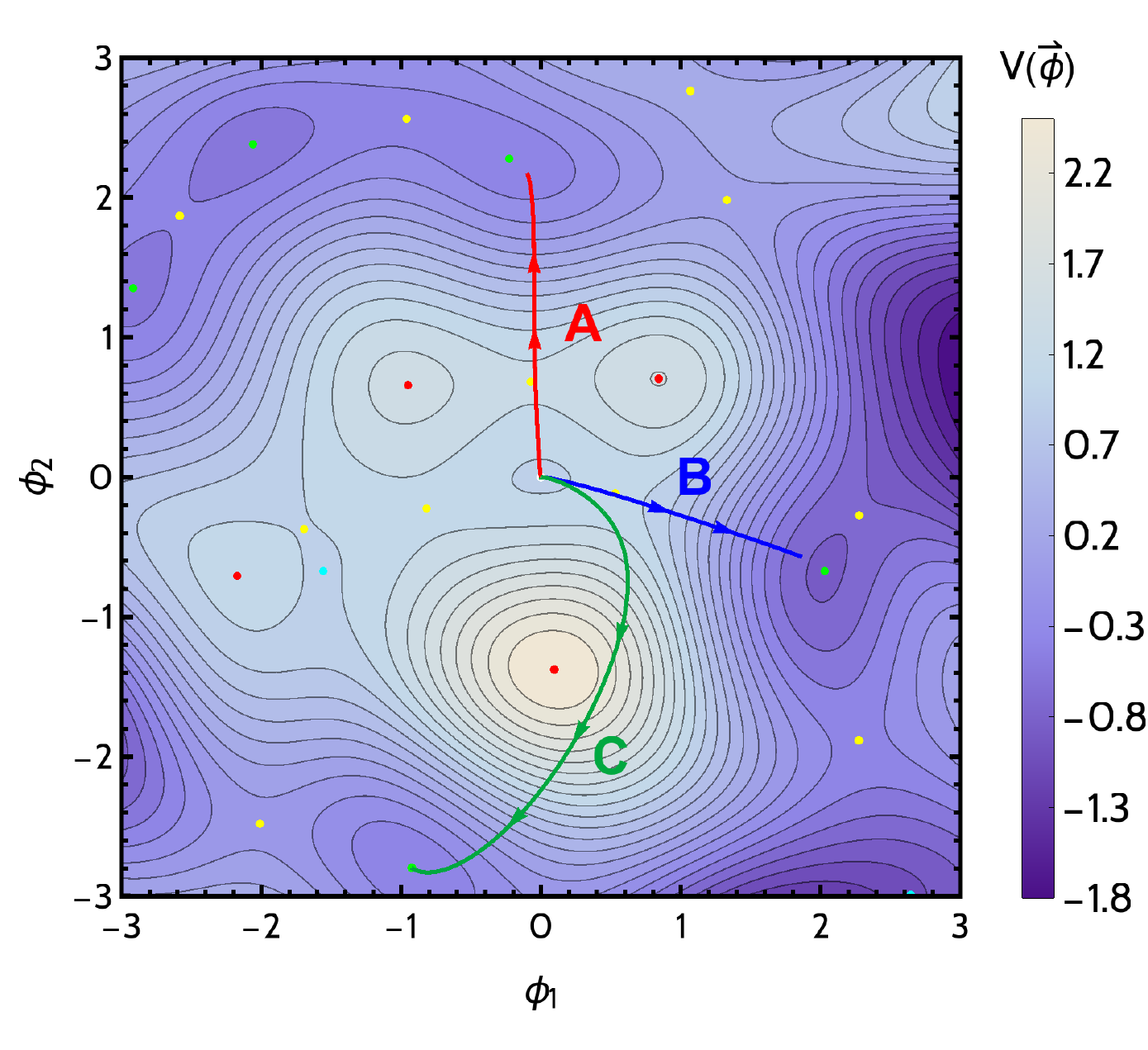}
\caption{A typical example of the considered tunneling events. After generating a GRF with a minimum 
at the center with height 1 (in terms of $U_0$), we compute possible tunnelings with AnyBubble. The plot 
shows the GRF along with its minima (green), saddles (yellow), maxima (red) and inflection points (blue) as
well as 3 of the instanton trajectories in field space for 3 decay channels. } 
\label{fig:tunnel_examples}
\end{figure}

\begin{figure}
\includegraphics[scale=1.0]{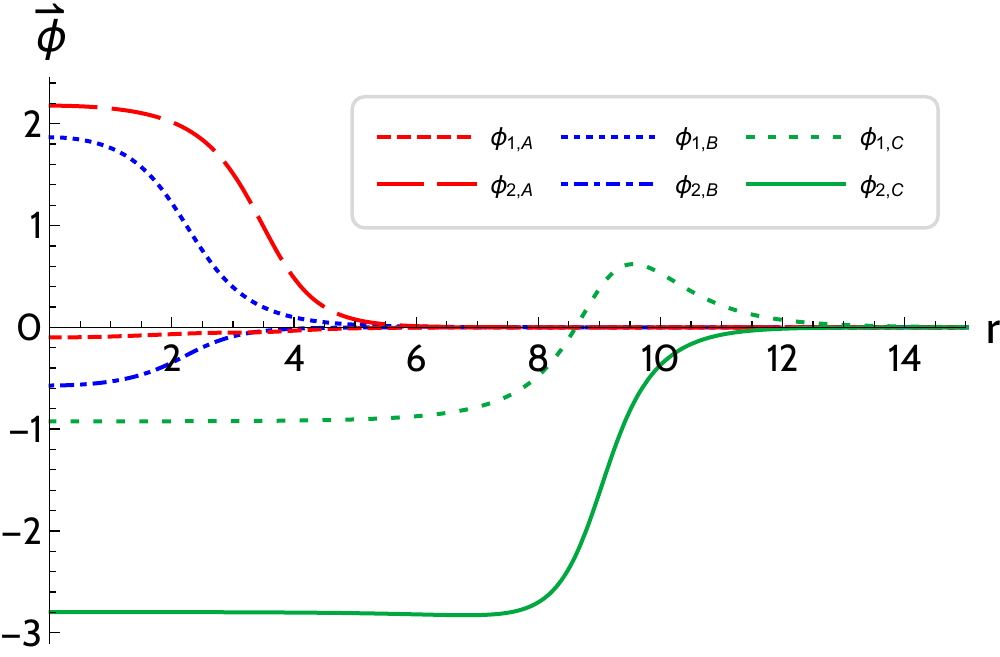}
\caption{Field trajectories for the decays channels shown in Fig. (\ref{fig:tunnel_examples}) in terms
of the distance $r$ in Euclidean space.} 
\label{fig:tunnel_examples_trajectories}
\end{figure}

In this work, we use AnyBubble \cite{Masoumi:2016wot} to compute the instanton actions for our 
realizations. AnyBubble is a Mathematica Package based on efficient numerical methods for the 
solution and optimization of the tunneling equations, see \cite{Masoumi:2016wot} for details.

In order to obtain statistics of the tunneling action in terms of the properties of 
the central minimum, we sampled false vacua with heights between -2 and 5 (in units of 
$U_0$, see Eq.~(\ref{eq:cov_grf})) in uniform intervals.  
As explained in \cite{Masoumi:2017gmh}, we can write the Euclidean action as 
\begin{align}
S_E = \frac{\Lambda^4}{U_0} \bar{S}
\end{align}
so that $\bar{S}$ corresponds to the Euclidean action of a potential with covariance function (\ref{eq:cov_grf}) with $U_0=\Lambda=1$. Unless otherwise specified, all histograms corresponding to the action are given in terms of $\bar{S}$ due to numerical simplicity.

Following the procedure of the Slepian models described the previous sections, for each value of the false vacuum height, we generated $2\cdot 10^4$ 
Gaussian random field realizations centered around the minimum. All of these 
minima have the correct distribution of the Hessian eigenvalues, and the potentials are quite different from one another as
one moves away from the minimum by one correlation length. This means that each realization has different vacua
situated in different directions and lengths from the false vacuum, although the typical number of minima below
$V(\bs{\phi}_{FV})$ is quite similar in all cases.

We can readily see the power of the machinery described in the previous section when constraining the field 
to have a minimum with a vacuum energy  higher than $1.5 U_0$. If we tried  to find a minimum higher than that drawing samples from an unconstrained GRF, we would need to 
generate tens (if not hundreds) of random fields before finding a single minimum satisfying that 
condition, see figure \ref{cp_histograms}(a) in Appendix \ref{sec:numerical}. For example, from equations (\ref{u_sp}), we can easily check that the probability of any minimum being higher than $5 U_0$ is $\mathcal{O}(10^{-16})$, so finding one by chance happens to be quite remarkable. With the aid of conditioning 
methods, we are able to construct very efficiently large samples of random fields subject to a condition as difficult to meet as this one.

In order to study tunneling processes on each generated example, we identified all the minima 
near the center of field space  and computed the tunneling rate between the central minimum (which 
always acts as a false vacuum, in our analysis) to all lower minima.  An example of this procedure 
is plotted in figures \ref{fig:tunnel_examples} and \ref{fig:tunnel_examples_trajectories}, where we show the paths followed in field space by the different instanton
decay channels\footnote{We have also identified the rest of critical points as well as inflection points with different
colours in all of our GRF realizations. }. We have only considered tunneling to minima around the center
to avoid problematic issues with minima close to the boundaries of our realizations. 

\subsection{Statistics of the instanton action}

\begin{figure}
	\centering
	\includegraphics[width=0.7\textwidth]{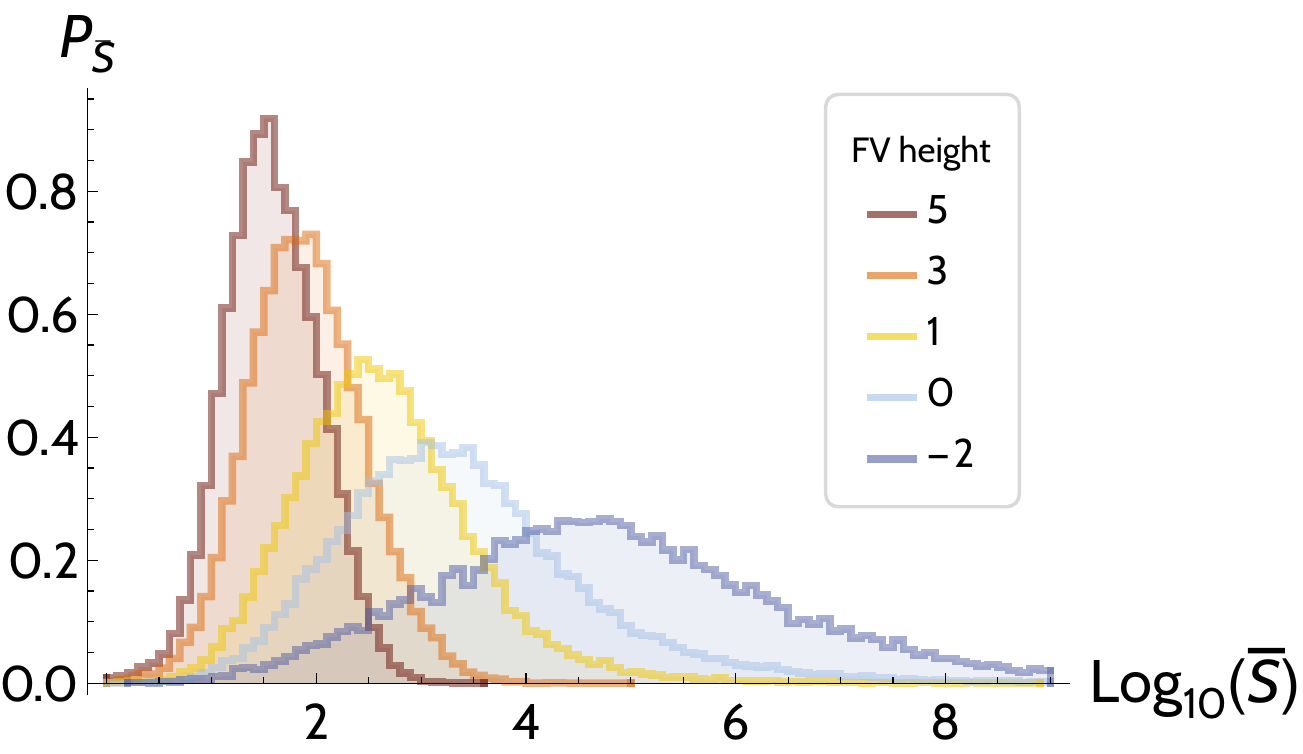}
	\caption{Obtained distribution of tunneling action ($\bar{S}$) in terms of false vacuum height.}
	\label{fig:actions_evol}
\end{figure}
\begin{figure}
	\includegraphics[width=0.7\textwidth]{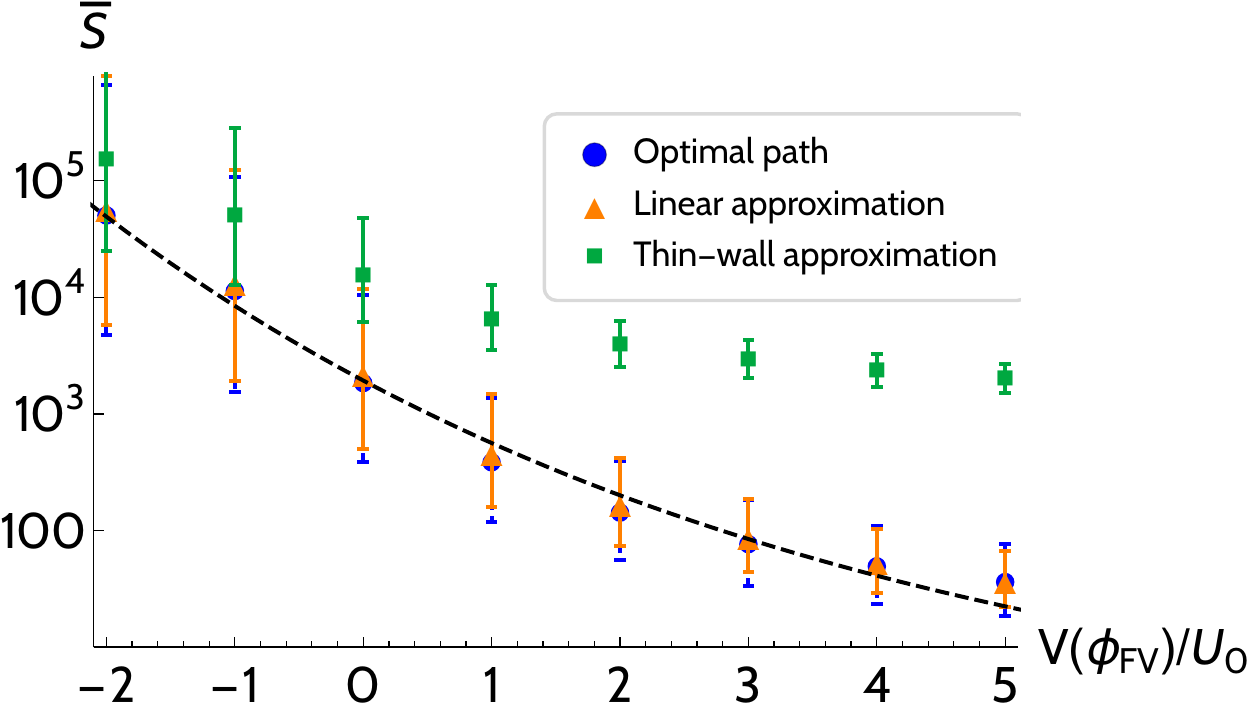}
	\caption{Evolution of the median of the action, with error bars representing data between the first and third quartiles of 
		each distribution, for the optimal path, the linear (straight-path) and the thin wall approximations, along with 
		a fitting curve (see (\ref{eq:action_fit})).}
	\label{fig:actions_comp}
\end{figure}

\subsubsection{Dependence with the height}

Figure \ref{fig:actions_evol} shows the resulting distributions\footnote{Unless otherwise specified, all histograms represent the normalized probability distribution function of the obtained results.} for the tunneling action, for 
different values of the false vacuum height. There is an interesting correlation between the mean and width 
of this distribution and the height of the false vacuum. Namely, we find that the higher the false 
vacuum is the lower the action and thus, the higher the probability of tunneling is. This behaviour 
is quite intuitive; as we can see from the examples in Fig.~\ref{fig:grf_examples}, tunneling from a 
minimum high up in field space requires crossing a lower barrier to the true vacuum, which in turn results in 
a lower action for those transitions. Figure \ref{fig:actions_comp} (blue dots) shows the median of each distribution  along with the range of 
actions between the first and third quartiles. We see, once again, that higher false vacua lead to accumulation 
over lower values of the action.  

The obtained data for each potential height was found to be easily fitted to a log-normal distribution. More specifically,  the logarithm of the median of each 
distribution $\bar{S}_{\text{med}}$ (which, in this case, is very similar to the mean of $\log_{10} \bar S$) can be fitted by the following
expression
\begin{align}
	\left\langle \log_{10} \bar{S}_{\text{med}} \right\rangle \approx 3.29 \exp\left(- 0.18 \frac{V_{\text{fv}}}{U_0}\right)
	\label{eq:action_fit}
\end{align}
where $V_{\text{fv}}$ stands for the height of the false vacuum. As we see from figure~\ref{fig:actions_comp}, 
increasing $V_{\text{fv}}$ reduces the width of the distribution significantly, thus increasing the predictive 
power of (\ref{eq:action_fit}) for the expected value of the action. This enhancement of the predictability of the Slepian model for large values of $V_{\text{fv}}$ corresponds  precisely to what we anticipated in the previous section. Indeed,  there we showed that near high minima  the random potential becomes dominated by the first term in the decomposition \eqref{eq:V_decomp}, and therefore  the landscape is very deterministic in a neighbourhood of false vacua with large $V_\text{fv}$. Consistent with this result, when studying the non-perturbative stability  from these vacua we observe a reduction of the variance of  tunneling actions for large  heights of the false vacuum. This agreement also suggests that in the case of minima with a large  $V_{\text{fv}}$ the value of the instanton action is dominated by the local structure of the minimum. We will provide further evidence for this claim below.

\subsection{Approximations for the calculation of the action}

Due to the inherent instability of the equations to be solved to compute tunneling profiles, it is clear that as 
we increase the domain and dimensionality of the potential under study, the required computational time to 
solve the system will grow accordingly. Evidently, this makes the study of higher-dimensional GRFs and their tunneling 
properties almost prohibitive in this sense. Motivated by these limitations, we turn to computing several different 
approximations of tunneling actions suggested in the literature, and compare them with our exact results.

\subsubsection{Thin wall approximation}

The thin-wall prescription was already discussed in the original papers by Coleman in \cite{Coleman:1977py}.
In this approximation the instanton action is given in terms of the difference between potential at the false 
vacuum ($V_\text{fv}$) and true vacuum ($V_\text{tv}$) and $\sigma$, the tension of the wall interpolating between them, namely,
\begin{align}
	\bar{S}_{\text{tw}} = \frac{27 \pi^2 \sigma^4}{2 (V_{\text{fv}}-V_{\text{tv}})^3}, \qquad \sigma = \int_{\bs{\phi}_{TV}}^{\bs{\phi}_{FV}} d\phi \ \sqrt{2 (V(\bs{\phi}) - V(\bs{\phi}_{TV}))}~.
	\label{eq:thin_wall}
\end{align}
This approximation  is accurate as long as the difference between $V_{\text{fv}}$ and $V_{\text{tv}}$ is small.

We evaluated (\ref{eq:thin_wall}) for each bounce we previously found with AnyBubble in order to check this expression 
and its predictive power for GRFs. In the computation we   restricted the field to a straight line in field space connecting the true and false 
vacua. Figure \ref{fig:actions_comp} shows the evolution of the median of $\bar{S}_{\text{tw}}$ as a function of the 
false-vacuum height. While the width and median of the distribution in this case follow the same pattern as the 
optimal action, the values diverge rapidly from the optimal ones as the false vacuum height increases.  This is
not too surprising since, as one increases the height of the false vacuum minimum, the field can tunnel to a minimum
with quite different values of the potential, what violates one of the premises of the thin wall approximation.

\subsubsection{Straight-path approximation}

While the thin-wall prescription provides a solid upper bound on the bounce action \cite{Brown:2017cca}, it does 
not provide any useful estimation on the actual value on the bounce in our case. This fact calls for an alternative way to estimate 
the action, mostly for higher-dimensional landscapes. 

A straightforward simplification to this problem was introduced in \cite{Dasgupta:1996qu}, which we will denote by 
\emph{straight-path approximation}. This prescription is based on reducing the field space to a single straight line connecting the false and true vacua, thus making 
the problem of tunneling effectively one-dimensional. As can be seen from figure~\ref{fig:tunnel_examples}, this approximation 
may not be too unreasonable. Even though there are some paths which do curve over the field space, many 
(if not most) of them follow a straight trajectory in field space. Note, however, that this restriction in field space may yield 
effective potentials where the bounce does not exist or might even correspond to a different bounce in the full theory. For 
more details on the properties of this approximation, see \cite{Masoumi:2017trx}.

For each optimal path, we considered a straight line in the two-dimensional GRF connecting the true and false vacua, and 
computed the corresponding estimate of the action, $\bar{S}_{\text{sp}}$, in each case. In principle, $\bar{S}_{\text{sp}}$ represents an upper bound 
on the optimal action $\bar{S}$, as the former only considers variations of the action in the direction of the straight 
path \cite{Dasgupta:1996qu}. It is thus expected (and explicitly shown in \cite{Masoumi:2017trx}) that this approximation
will diverge from the full solution as the dimensionality of the potential is increased.

We found that in this case the distribution of actions in terms of false vacuum height  is identical to the optimal one shown 
in Fig.~\ref{fig:actions_evol}, though slightly shifted to higher values. As we can see from Fig.~\ref{fig:actions_comp}, the 
change in the median is minimal when the straight-path approximation is considered.
Although, as we just mentioned,  
the straight-path approximation is not expected to give precise results for potentials in a higher field space dimension, this result suggests 
that  it would  be interesting to explore the validity of this method with GRFs in higher dimensions. Indeed, due to the computational 
complexity of such an analysis, a rough statistical estimate of the decay rate obtained with this approximation would still be very valuable.

\subsection{The lowest action}

In many circumstances one will be interested in the lowest action for a particular
kind of minima. This will of course correspond to the path that would dominate the decay
for those minima. In this subsection we will investigate the characteristics of such
trajectories in field space.

\subsubsection{Exit angle}

\begin{figure}
	\centering
	\includegraphics[width=0.8\textwidth]{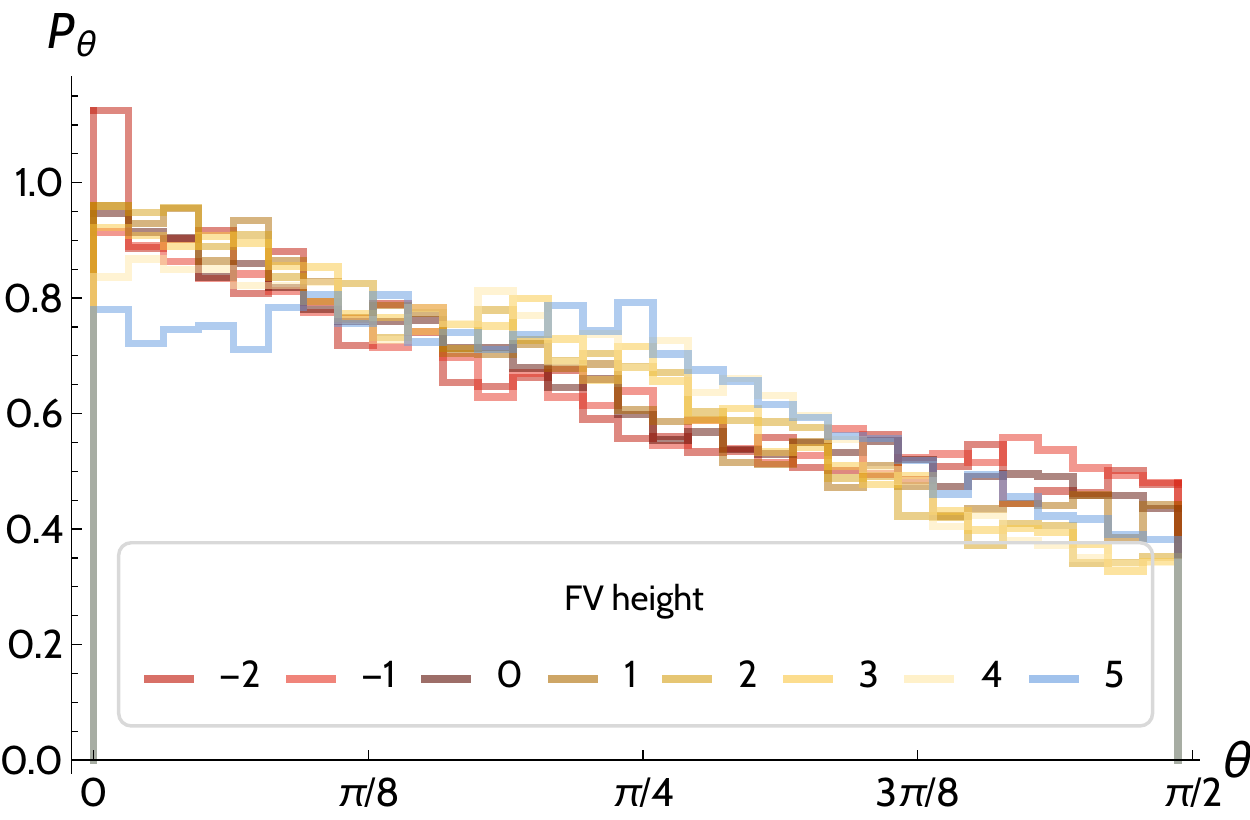}
	\caption{Exit angle distribution with respect to direction of the lowest eigenvalue 
	for the instanton path of the most probable decay channel in each generated potential.}	\label{fig:angle}
\end{figure}

An intuitive way to think about the most likely decay process would be to imagine that
the tunneling occurs along the trajectory with the lowest barrier. One can check this idea
in our case by first identifying the angle (in our $2d$ field space), $\theta$, that the instanton trajectory 
makes with respect to the direction of the lowest eigenvalue of the Hessian at the minimum. 
A distribution of such angles obtained for different values of the height is plotted in 
figure~\ref{fig:angle}. We see that there is a clear tendency of the tunnelings to occur 
around $\theta \approx 0$ but the correlation is not very strong.

\subsubsection{Estimating the lowest action}

 The correlation of the instanton path with the 
lowest eigenvalue direction at the false vacuum  suggests that one
can try to estimate the lowest action by analyzing the potential along the 
lowest eigenvalue direction alone. This has been recently proposed
in the context of the landscape in \cite{Dine:2015ioa}. In the following we will
use our large sample of realizations to test this idea in detail in our
$2d$ GRF model of the landscape.

 In order to evaluate the instanton action along the lowest eigenvalue direction we 
 first take a slice of the potential along that direction and fit it to be of the form,
\begin{equation}
V_{le}(\phi_1) = V_0 + \frac{1}{2} \lambda_1\phi_1^2 + \frac{1}{3!} \rho_{111} \phi_1^3 + \frac{1}{4!} \delta \phi_1^4~.
\end{equation}

Note that this procedure does not guarantee that the resulting
one-dimensional potential is suitable for a tunneling process.
In fact, in many cases the potential constructed this way does not
have a lower minimum along this direction and therefore it cannot
be used to estimate the decay rate. In the following we will only
compute the instanton action in the successful cases where this
$1d$ truncation gives an acceptable form, what in particular requires $\rho_{111}<0$.

Considering this simple form of the potential as the most likely exit path
for the decay transition we can estimate the instanton action. In order to do that we will 
use the parametrization of the Euclidean action for the bounce that
was obtained by Sarid in \cite{Sarid:1998sn}. In our notation this becomes,
\begin{align}
		\bar{S}_{S}=
		\left\lbrace
		\begin{array}{ll}
		\frac{18 \lambda_1}{\rho{111}^2} \left( 45.4 - 46.1 + \frac{2 \pi^2}{12 (1-4 \kappa)^3} + \frac{16.5}{(1-4\kappa)^2} + \frac{28}{1-4\kappa} \right), & \kappa >0 \\
		\frac{18 \lambda_1}{\rho{111}^2}  45.4 \left(1+ (\frac{136.2}{2\pi^2})^{1.1} |\kappa|^{1.1} \right)^{-1/1.1}, & \kappa \leq 0 \\
		\end{array}
		\right.
\label{eq:Sarid_app}
\end{align}
where
\begin{align}
		\kappa = \frac{3}{4} \delta \frac{\lambda_1}{\rho_{111}^2} ~.
\end{align}

\begin{figure}
	\centering
	\includegraphics[width=0.8\textwidth]{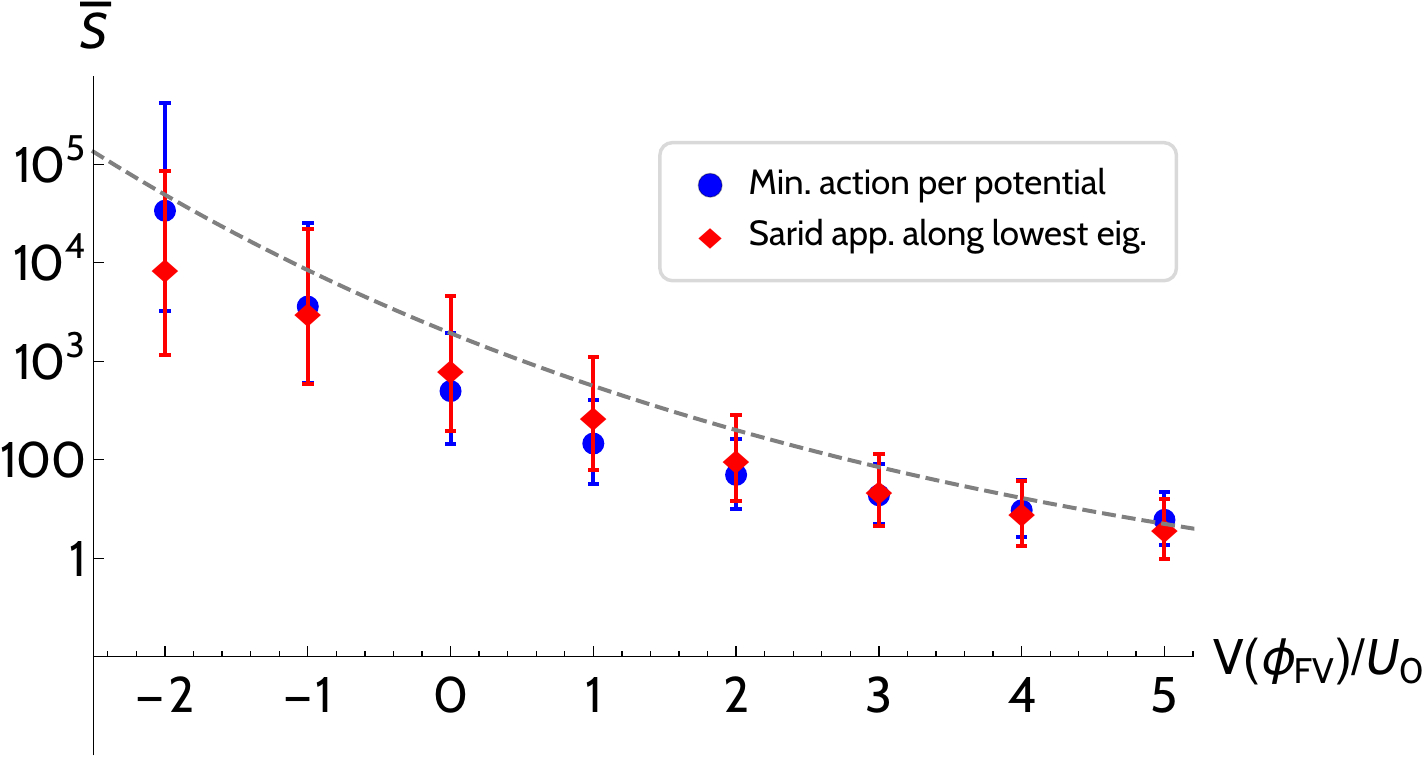}
	\caption{Distribution of lowest action per potential and Sarid approximation  \cite{Sarid:1998sn}  along the lowest barrier direction, in terms of false vacuum height. The fit in Eq. (\ref{eq:action_fit}) is shown for comparison with previous results. }	\label{fig:Saridapprox}
\end{figure}

We show in figure \ref{fig:Saridapprox} the distributions of the lowest action from the exact computation 
and compare it to this estimate along the lowest barrier direction. We notice that the 
agreement between these two results is pretty good, what suggest that one can use this approximation to estimate 
the decay rate of vacua in a Gaussian random landscape. Moreover, it is worth noting that this approximation depends 
only on the local structure of the minimum, precisely where the Slepian model has a large predictive power for large values 
of $V_\text{fv}$. The expression \eqref{eq:Sarid_app} becomes increasingly accurate for large values of the false vacuum energy $V_{\text{fv}}$, what 
indicates that in this regime instanton action is mostly  determined by the local form of the minimum. On the other hand,   
according to the Slepian model, the scalar potential around all high minima should look very similar in all realizations, 
with its shape dominated by the first  term in \eqref{eq:V_decomp}. This explains why the distribution of instanton actions 
becomes more deterministic (Fig. \ref{fig:actions_evol}) for larger values of $V_{\text{fv}}$, and therefore also the agreement 
between the Sarid approximation \eqref{eq:Sarid_app} for the \emph{lowest action} and our  fit  in Eq. \eqref{eq:action_fit} 
 \emph{for the median} of the distribution.

 It would be interesting to check if this good agreement persists on a much larger landscape with hundreds of 
directions in field space\footnote{Note that in our calculation we kept the quartic term of the
potential while in reference \cite{Dine:2015ioa} the authors drop this term arguing that for large
number of fields (N) this coefficient becomes irrelevant. We have checked that in our case this
is not the case and in order to obtain a good agreement it is necessary to take this term into account. This is
due to the fact that we have limited our investigation to the N=2 case.}, and whether the approximation  \eqref{eq:Sarid_app} can be used in combination with our Slepian model make robust predictions regarding the tunneling rates of high vacua.

\section{Inflation in a Slepian Random Landscape}
\label{sec:inflation}

Up to now we have been using all the software and mathematical tools described above for the computation 
of bounce profiles and actions with Gaussian random fields conditioned to have a minimum at $\bs{\phi}= \bs{0}$. In this section, 
we turn to studying constrained GRFs with inflection points at the origin of field space focusing on their application to 
cosmological inflation. 

Inflation in random potentials has already been extensively studied 
\cite{Masoumi:2016eag,Masoumi:2017gmh,Masoumi:2017xbe,Blanco-Pillado:2017nin,Bjorkmo:2017nzd}. More specifically, inflation 
around inflection points has received special attention for being capable of sustaining enough e-folds to make 
contact with observations, while taking place in a small region of field space with an effectively one-dimensional potential. 

While most of the obtained results and distributions seem promising, they have only been tested within Taylor 
expansions around these points, instead of using full GRFs. As we mentioned before, such methods do not capture correctly 
the global features of the potential, what is  essential for characterising the non-perturbative stability of vacua.  Therefore, this 
procedure  is unsuitable for studying models of inflation where the initial conditions are determined  by the decay of a parent false vacuum.  

In this section we will apply Slepian models  to constrain Gaussian random fields to have inflection point with the desired 
properties to sustain inflation, and then we will study the dependence of its cosmological observables on the initial 
conditions, set by different realizations of the parent vacuum.

\subsection{1D Inflection point inflation}

Let us briefly review the main results for one-dimensional inflection-point inflation 
(see \cite{Baumann:2007ah,Blanco-Pillado:2017nin} and references therein for more details). Let us consider a potential of the form,
\begin{equation}
V(\phi) = u + \eta \phi + \frac{1}{6} \rho  \phi^3 ~,
\end{equation}
where, in order to satisfy the slow-roll conditions around the inflection point, we will
assume that $\eta \ll u$. Note that  we do not need to assume that the third
derivative is too small. In fact, following typical conditions for a GRF we will consider
the case where $ u  \ll \rho$. Taking this into account one can show that slow-roll 
inflation conditions will be satisfied in the interval
\begin{align}
	-\frac{u}{\rho} < \phi < \frac{u}{\rho}~,
\end{align}
which together with the condition  $ u  \ll \rho$ implies that we are describing 
small field inflation. Using the slow-roll conditions, it is easy to check
that the expected number of e-folds, $N_{\text{exp}}$, that can be achieved 
within that region is
\begin{align}
	N_{\text{exp}} = \int_{-u/\rho}^{u/\rho} \frac{d\phi}{\sqrt{2\epsilon}} \approx \pi \sqrt{2} \frac{u}{\sqrt{\eta \rho}}-4 \equiv N_{\text{max}}-4.
	\label{eq:N_max}
\end{align}
where $\epsilon = (V''(\phi)/\sqrt{2}V(\phi))^2$ and $N_{\text{max}}$ is the maximal number of e-folds achievable in the whole potential. Moreover, defining 
\begin{align}
	x \equiv \pi \frac{N_{\text{CMB}}}{N_{\text{max}}} \qquad ; \qquad y \equiv \frac{N_{\text{max}}}{2\pi},
\end{align}
where $N_{\text{CMB}} $ is the e-fold number at which the CMB scales leave the horizon, 
the spectral index of scalar perturbations can be shown to be given by 
\begin{align}
	n_s = 1 + \frac{2}{y} \left(\frac{\tan x - y}{1+y \tan x}\right)~.
	\label{eq:n_s}
\end{align}

Finally, the amplitude of scalar perturbations can be expressed as
\begin{equation}
\Delta^2_{\cal R} = \frac{1}{12 \pi^2} \frac{V^3(\phi)}{V'(\phi)^2} \approx \frac{N_{\text{CMB}}^4 \rho^2}{48 \pi^2 u} f^2(x,y)
\end{equation}
where
\begin{equation}
f(x,y)= \frac{\cos^2(x) (y \tan(x) +1)^2}{x^2(y^2+1)}.
\end{equation}
satisfies $f(x,y)\sim 1$ for $y\gg1$ and $x\sim 1$.

With these expressions at hand, we can easily obtain a set of parameters for the inflection
point ($u, \eta$ and $\rho$) that are in agreement with the current cosmological
observations, namely, $N_{\text{exp}} > N_{\text{CMB}} \approx 50$, $n_s \approx 0.965$ 
and $\Delta^2_{\cal R} \approx 2\times 10^{-9}$ (see Eq. \eqref{eq:infl_params} below).

\subsection{Numerical inflection points in a 2D Landscape}

We now want to embed  $1d$ inflection-point inflation in our $2d$ GRF landscape. In order to do that
we can follow the procedure explained in Section \ref{sec:theory}B for Slepian models in the case of  inflection
points. In the notation introduced earlier, the $1d$ parameters $\eta=\eta_1$ and $\rho = \rho_{111}$, correspond 
to the derivatives along the flat direction of the multidimensional inflection point.  Note that, 
in principle, $u$ and $\rho_{111}$ (when evaluated at the same point) are uncorrelated, but the same
is not true for $u$ and the second derivative along the inflaton direction $\lambda_1$; similarly 
$\eta_1$ and $\rho_{111}$ are also correlated, see Eq. (\ref{inflection-correlations}). Here we are interested in 
studying the global properties of the landscape on the cosmological
observables so we will focus on a particular type of inflection point where we have fixed its
$1d$ parameters\footnote{It is also interesting to study the effects of varying these parameters
together with the global properties of the GRF. We leave the details of this calculation for
a future publication.}.

Following the steps from the previous section, we built two-dimensional GRFs with an inflection point 
whose inflating direction has fixed features. In the forthcoming sections we set 
\begin{equation}
u=0.5 \ U_0 \quad, \quad \eta_1 = 6.8 \cdot 10^{-6} \ \frac{U_0}{\Lambda} \quad , \quad \rho_{111} = 2.5 \ \frac{U_0}{\Lambda^3}
\label{eq:infl_params}
\end{equation} 
where $U_0 = 6.0 \cdot 10^{-16} \ M_{\text{Pl}}^4$ and $\Lambda = 0.5 \ M_{\text{Pl}}$ define the energy scale and correlation length respectively, with the    Planck masses written explicitly for clarity.

Once $u$, $\eta$ and $\rho$ have been fixed, using the probability distributions listed in  
\eqref{eq:P_ip} and (\ref{inflection-correlations}), we can obtain the remaining parameters of the two-dimensional 
inflection point set at the origin of field space $\bs{\phi}=\bs{0}$, and generate in a efficient way a large sample of GRFs with the listed properties.\footnote{ Note that following our earlier definition of the inflection point in our $2d$ landscape, we have set $\eta_2 =0$ and $\lambda_2 >0$.}

\begin{figure}
	\centering
	\includegraphics[width=0.7\textwidth]{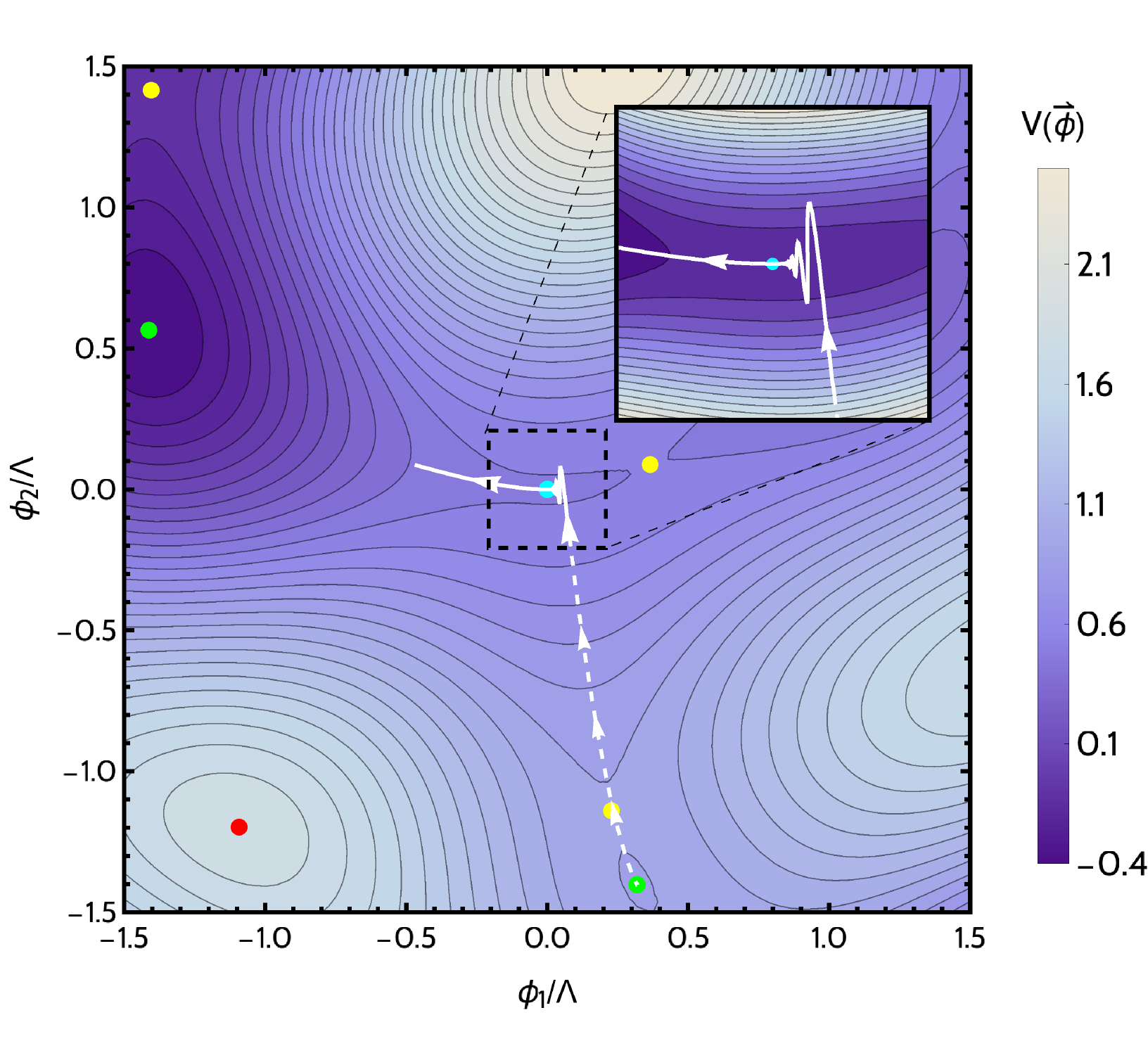}
	\caption{A Gaussian random  field conditioned to have an inflection point in the middle. The 
	dashed line represents the tunneling from a minimum to a lower inflection point. The inflationary 
	slow-roll phase starts at the exit point, inflates for around 124 e-folds following the solid line, and evolves towards the closest minimum. We only show the 
        inflationary part of the trajectory. Green, yellow and red dots represent minima, saddle points and maxima of the potential. The inflection point is marked with a blue dot. }
	\label{fig:grf_inflation}
\end{figure}

As an example, we show in figure \ref{fig:grf_inflation} a field constructed with the above constraints. 
We then used AnyBubble to tunnel from a higher false vacuum to the central inflection point. We note that
even though in every realization the inflection point has the same properties along the $\phi^1$ direction 
up to third order, the potentials are different away from that point. This means that the false vacuum, which decays
to the region around the inflection point, is located in a different place and it also has different features in each realization, e.g. vacuum energy and barrier height. Using
AnyBubble we  computed the exit points of a large set of realizations. After that we used these exit
points of the instanton decay as the starting points of a Lorentzian evolution of a FRW universe with this
potential.

In order to study the inflationary trajectory we used mTransport \cite{Dias:2015rca}, 
a Mathematica code developed to compute inflationary observables using the transport method. 
The cosmological evolution inside of a bubble universe created from tunneling is described
by an open FRW universe \cite{Coleman:1980aw}. Here, for simplicity, we used the flat-space approximation
for the evolution of the cosmological interior of the bubble\footnote{Note that in reality the initial cosmological
evolution is dominated by the spatial curvature of the open FRW slices that describe the bubble interior. This  
will have some effect on the initial stages of the evolution of the scalar field in a multidimensional potential. See \cite{BlancoPillado:2012cb,Masoumi:2017xbe} 
for a discussion of these effects.}.

In the example from figure \ref{fig:grf_inflation}, the dashed line represents the tunneling trajectory, 
while the solid one marks the inflationary one. We found this path to sustain a total of 124.1 e-folds 
and a spectral index of $n_s = 0.964$ at the observable scale. 

\subsection{Statistics of inflationary parameters}

\begin{figure}
	\centering
	\subfloat[]{
		\includegraphics[width=0.45\textwidth]{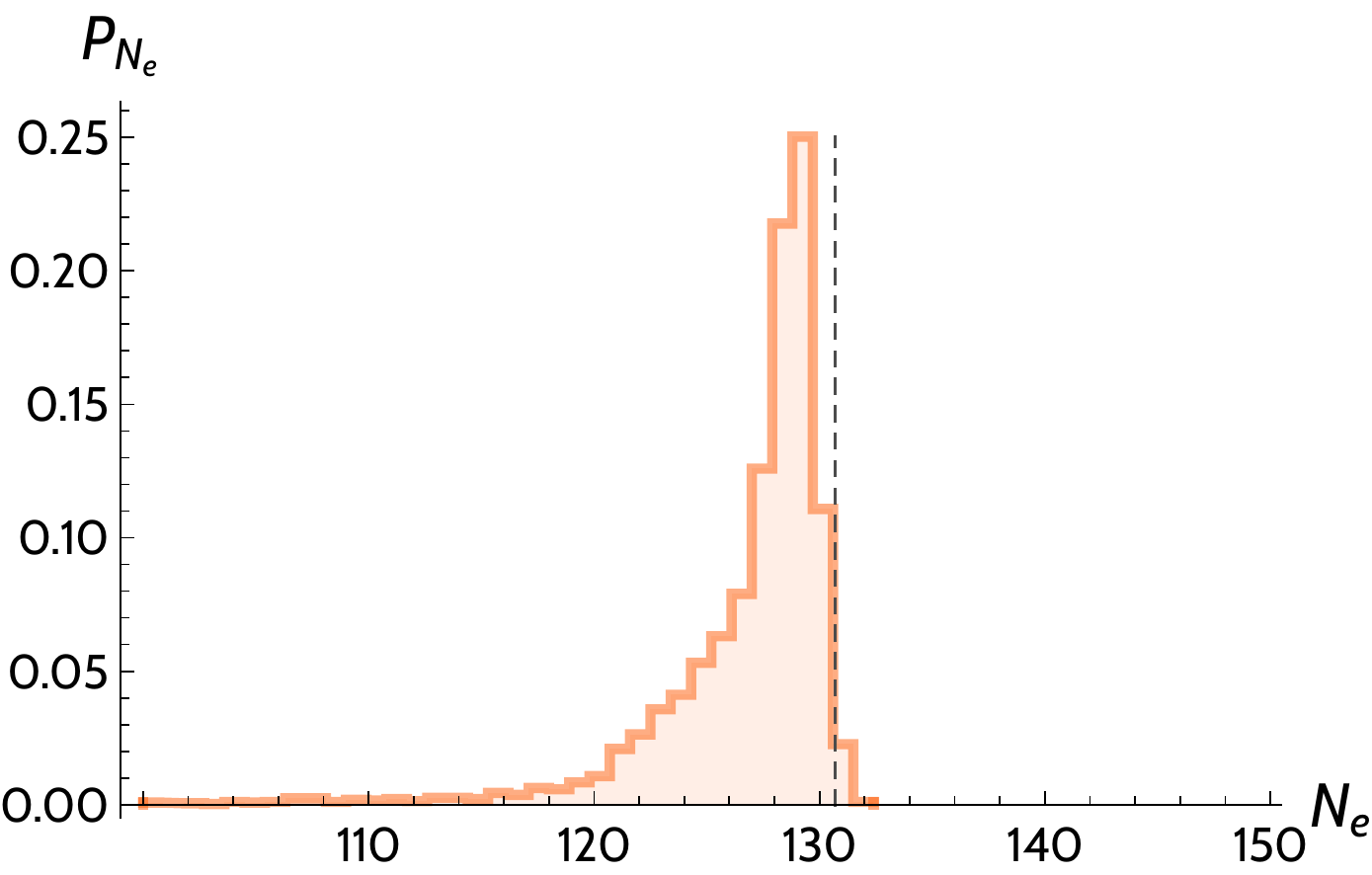}
	}
	\hfill	
	\subfloat[]{
		\includegraphics[width=0.45\textwidth]{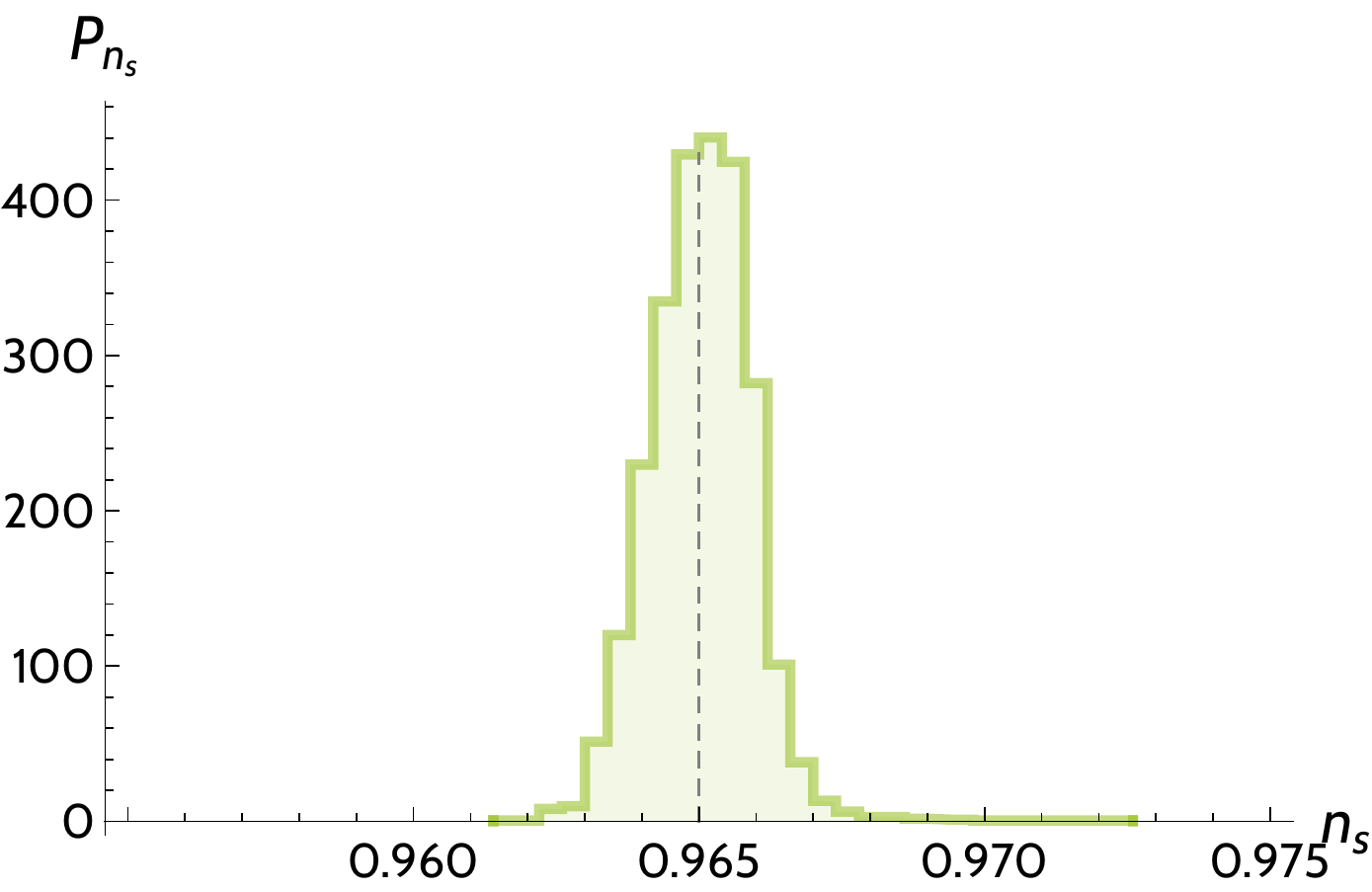}
	}
	\caption{(a) Distribution of number of e-folds, with $N_{\text{exp}}$ shown with a dashed line (b) Histogram of the obtained spectral index, with the analytic prediction marked with a dashed line. Both figures represent 4000 inflationary trajectories (see text).}
	\label{fig:inflation_params}
\end{figure}

\begin{figure}
	\centering
	\includegraphics[width=0.75\textwidth]{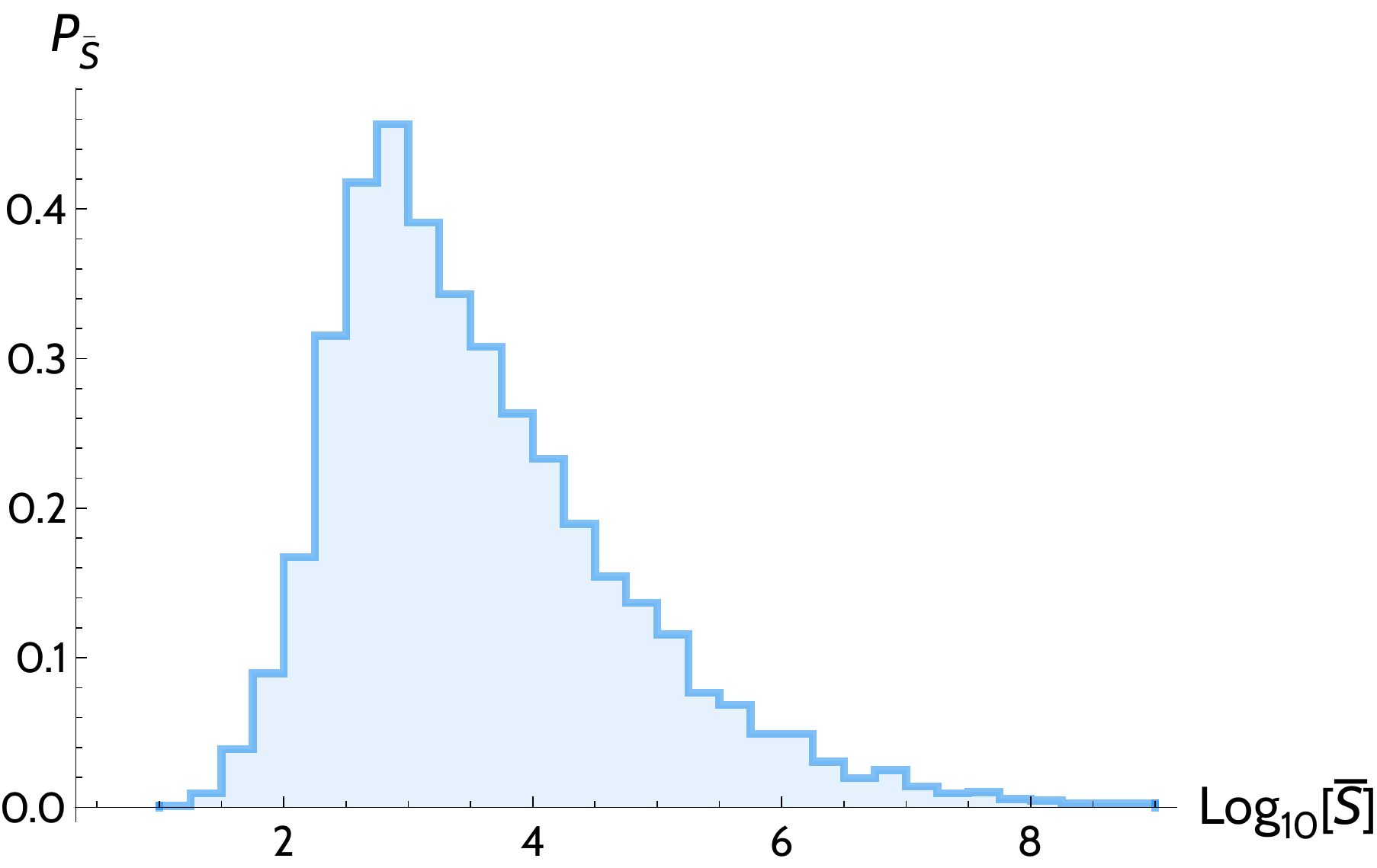}
	\caption{Distribution of the tunneling action from a minimum to the central inflection point, right before inflation begins.}
	\label{fig:inflation_action}
\end{figure}

In order to test the method described above to generate inflationary random fields, we generated 
5000 GRFs constrained to have an inflection point with the same properties as the one in the 
example of figure \ref{fig:grf_inflation} (see Eq. \eqref{eq:infl_params}). Next, in each of these realizations, we found all minima lying above the central inflection 
point and used anyBubble to compute the tunneling trajectory from the former to the latter 
in each case. Considering the exit point as the starting point of an inflationary phase, we used 
mTransport to find the number of e-folds, power spectrum, tensor-to-scalar ratio, spectral index 
and its running. The distributions of the e-fold number and the spectral index
are shown in figure \ref{fig:inflation_params}, for a pivot scale of 50 e-folds, whereas the action associated
to the tunneling to the inflection point is shown in Fig. \ref{fig:inflation_action}. This is a different distribution 
than the ones we found earlier, since the common factor in these decays is the final point and we do not impose anything about the
initial (false vacuum) state. It is interesting to see that this distribution is quite peaked around an action
of the order of $10^3$. 

We have also obtained the distributions for the amplitude of scalar perturbations, tensor-to-scalar ratio 
and running of spectral index which turned out the be centered around the values
\begin{equation}
\Delta^2_{\cal R} = (2.02 \pm 0.04)\cdot 10^{-9},\quad r=(8.0\pm 0.1)\cdot 10^{-9}\quad \text{and}\quad  \alpha = (-2.49 \pm 0.02)\cdot 10^{-3},
\label{eq:inflation_obs}
\end{equation}
respectively \footnote{The cosmological evolution of these Lorentzian trajectories continue
after inflation until they reach a lower minimum. We have not fine-tuned this minimum to be
in Minkowski space, so in general the evolution leads to eternal de Sitter or to an Anti-deSitter crunch. 
We are only interested in the statistics of the inflationary period so we have stopped this evolution
after the field leaves the slow-roll regime. }. Our results in this section are fully compatible with the
$1d$ studies in \cite{Masoumi:2017gmh}.

\begin{figure}
	\centering
	\includegraphics[width=0.75\textwidth]{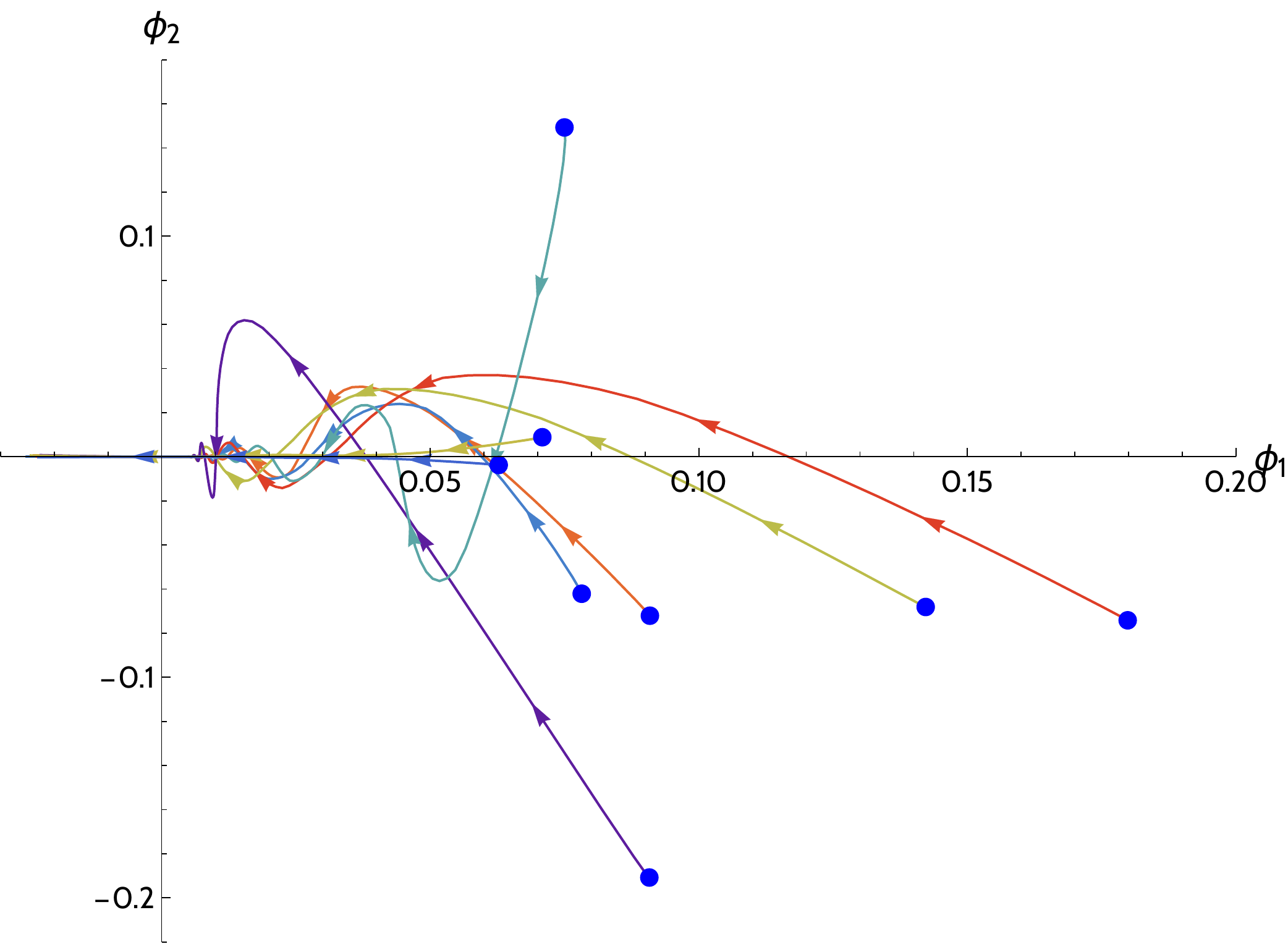}
	\caption{Showcase of several inflationary trajectories from different tunnelings to the central inflection point. Each exit point is marked
	by a blue dot.}
	\label{fig:inflation_paths}
\end{figure}

Finally, in Fig.~\ref{fig:inflation_paths}, we show several inflationary trajectories corresponding to 
tunnelings in different GRFs with an inflection point in the middle with the same features. Note that all trajectories,
no matter how far they start from, have a similar behavior. After oscillating in the vertical $\phi_2$ direction, 
they all stabilize around the inflection point and inflate along it. Most of the e-folds happen in the vicinity of the 
inflection point, as predicted by the analytic estimation.

We have obtained successful results from this analysis around $80\%$ of the times. The rest of the times the
procedure did not yield a cosmological solution in agreement with our universe either because 
inflation ended too soon or because the exit point was too far from the central inflection point and the inflaton
trajectory went astray. The successful paths show very good agreement with the $1d$ results presented
in the previous section. We see that even though some of the trajectories have some substantial
deviation from the $1d$ inflationary direction, the cosmological observables are still in pretty good
agreement with the single field inflection point inflation. The distributions of the results are
quite peaked around their central values, so we can conclude that the dependence of the observables on the initial
conditions seems to be quite mild. 

It is important to remember that all these realizations
have the same $1d$ inflection point parameters. In order to extract  the complete statistical information
about the predictions of a particular GRF we should combine these results with the
ones obtained from inflection points with other parameters with their correct statistical weight.
This is a much more numerically intensive problem and we leave it for a future 
publication.

\section{Summary and conclusions}
\label{sec:conclusions}

Slepian models are a powerful mathematical technique for modelling the statistics of random landscapes conditioned to satisfy a certain set of constraints. For this reason they are particularly useful to characterise phenomenologically interesting corners of the landscape, e.g. de Sitter vacua or inflationary regions consistent with the cosmological data, which are known to have highly suppressed probability to occur in generic random potentials. On the one hand, Slepian models provide a way to generate  numerically  large samples of a random landscape  
containing the region of phenomenological interest to be studied, regardless of the low probability of the realizations. 
On the other hand, this technique can also be used as an analytical description of conditioned random potentials, and thus to obtain valuable insight about  properties of the landscape around these regions of interest. A particularly attractive feature of Slepian models, as opposed for example to the use of  Taylor expansions, is that they can capture the global features of the random potential, and therefore they are specially useful for studying quantum mechanical instabilities in the landscape. 
In this paper we have presented the mathematical techniques for studying conditioned Gaussian random landscapes. We have applied these method to condition a $2d$ random potential to have a de Sitter minimum with a specific vacuum energy and also to study $2d$ landscapes containing an inflection point capable of sustaining a period of inflation compatible with the data. 

More specifically, regarding our discussion of de Sitter minima, we have considered the non-perturbative decay of these vacua to lower minima, and characterised the statistical distribution of their decay rate as a function of the height of the  false vacuum. For this purpose we have used our Slepian model to generate numerically large samples of vacua  with varying values of the vacuum energy, and then computed the corresponding decay rates both solving the full instanton equations, and using various approximate methods present in the literature: the thin-wall approximation \cite{Coleman:1977py}, the straight-path approximation \cite{Dasgupta:1996qu}, and the estimate proposed by  Sarid  \cite{Sarid:1998sn} for the lowest instanton action (see Eq. \eqref{eq:Sarid_app}).

Our analysis shows that the thin-wall approximation is in good qualitative agreement with the numerical results, but only provides an accurate estimate of the instanton action action  for minima with a relatively small vacuum energy. Indeed, consistently with  the  thin-wall prediction of the instaton action,  we observe that the decay rate increases (on average) for increasing values false vacuum height.  This can be understood noticing that, in a Gaussian random landscape, the barrier height that needs to be crossed  to escape from the vacuum  decreases when the vacuum energy of the minimum increases. However, for minima with a large vacuum energy  the tunneling typically occurs to much lower vacua, what violates the assumptions of the thin-wall approximation, and thus it cannot  provide a good quantitative estimate of the decay rate.

In the straight-path approximation one assumes the decay is effectively one-dimensional, so that it occurs along the line connecting the false and true vacua. We have shown this simplification agrees remarkably  well with the results of our full numerical analysis in all cases we studied in a $2d$ Gaussian landscape. It is interesting to check if this simplification still provides a rough estimate (see \cite{Masoumi:2017trx} for a discussion) for the instanton action in higher dimensional  landscapes where the numerical resolution of the full instanton equations becomes prohibitively difficult. In the particular case of a  Gaussian random landscape this approach is specially attractive, since the statistics of the random field along the straight line connecting the false and true vacua can be fully described by simply restricting its covariance function to that line. Therefore, if this method would prove useful to estimate  the non-perturbative stability of vacua in large-dimensional Gaussian landscapes, it would not be necessary to produce a sample the full higher dimensional GRF, it would suffice to generate one dimensional realizations of the random field with the same covariance. 

Regarding the estimate of Sarid \cite{Sarid:1998sn} for the lowest action (the most likely decay channel), our numerical analysis shows that this approximation provides an accurate quantitative estimate of the instanton action in the case of minima with a large vacuum energy. Interestingly, this estimate depends only on the form of the potential in a neighbourhood of the false vacuum which, according to the predictions of the Slepian model, does not experience large variations between different realizations. In plain words, all high minima look locally very similar to each other.   Indeed,  Gaussian random potentials  conditioned to have high minima exhibit a very deterministic shape in a large region around it,  which is dominated  by the first term in equation \eqref{eq:V_decomp}. As we argued in the main text, combining the estimate of \cite{Sarid:1998sn} for the lowest action, with the Slepian analysis one concludes that the distribution for the instanton actions should become increasingly peaked and deterministic for higher minima. 
Our numerical results, displayed in figures \ref{fig:actions_evol} and \ref{fig:Saridapprox},  match  perfectly this expectation. This  suggests that the estimate for the instanton action in Eq. \eqref{eq:Sarid_app}, in combination with the Slepian techniques, might also provide a very good prediction for the decay rates of high false vacua in higher dimensional landscapes. For this purpose, the alternative methods proposed in \cite{Bjorkmo:2018txh} to generate constrained multidimensional Gaussian random landscapes might also proof very useful.

With respect to our second application of Slepian models, the analysis of inflection point inflation in a Gaussian random landscape,  we have considered the dependence of the cosmological observables on the initial conditions for inflation. This initial conditions in our model are determined by the exit point of a quantum tunnelling process from a parent false vacuum.   This study would have been very difficult without the aid of our conditioning techniques, since  generating numerically a large sample of potentials with an inflection point with the right properties is exceedingly costly in terms of computation time. With our methods, however, we were able to generate easily a large number of realizations of the landscape with an inflection point capable of sustaining more that 60 e-folds of inflation, and with observables consistent with the current cosmological data.  Note also that the ability of Slepian models to reproduce faithfully the global features of the potential was also essential in this analysis,  in particular for modelling the preinflationary phase of  false vacuum decay. Our results are summarised by  figure \ref{fig:inflation_params} and  equation \eqref{eq:inflation_obs}, which display the computed values of the  cosmological observables. We see that the dependence of the inflationary parameters on the initial conditions is quite mild.  The obtained distributions for the observables are very peaked around their expected value in the $1d$ slow roll model where inflation happens around the inflection point. The typical realizations in our landscape have some variation on the observable parameters ranging between  $1\%$  and $10\%$ depending on the quantity under consideration. It is important to emphasise that in this study we kept fixed the local properties of the inflection point. In order to perform a complete characterisation  of inflection point inflation in a Gaussian landscape we would also need to study the effect of changing the inflection point parameters on the observables. We will leave this analysis for a later publication\footnote{A realistic study of the observable parameters of inflation in this model should also include a prescription to calculate their probability distribution in the multiverse. This will require the introduction of a measure. Here we have not discussed this issue any further. (See \cite{Freivogel:2011eg} for a detailed description of the proposed prescriptions.)}. 

Finally, it is worth mentioning that  the present work has potentially very interesting applications to characterise the landscape of $4d$ effective field theories in String Theory flux compactifications at tree-level. Actually, as was discussed in \cite{Ashok:2003gk},  the superpotential defining the  effective supergravity  description of  flux compactifications  can be modelled as a (complex)  Gaussian random field with a specific  covariance function  determined by the geometry of the compact dimensions. The superpotential encodes a large amount of information about the low energy theory: the critical points of the superpotential represent minima of the tree-level moduli potential;  the supersymmetry breaking scale is given by its absolute value;  and the eigenvalues of its Hessian encode the mass spectrum of the moduli fields and their fermionic superpartners. 
 Thus, the conditioning methods presented in this paper can be immediately translated into this context,  allowing to study  the statistical properties of the $4d$ effective theory when  constrained to satisfy one or various conditions  (see \cite{Denef:2004ze,Denef:2004cf}), e.g. the existence of a vacuum with a particular  supersymmetry breaking scale, or to have a mass spectrum containing a certain number of light modes.

\section{Acknowledgements}

We are grateful to Alex Vilenkin, Masaki Yamada and Jeremy M. Wachter for useful discussions, and to Jonathan Frazer  for discussions and 
collaboration at the early stages of this project.  This work was 
supported in part by the Spanish Ministry MINECO grant (FPA2015-64041-C2-1P), the MCIU/AEI/FEDER
grant (PGC2018-094626-B-C21), the Basque Government grant (IT-979-16), the University of the Basque Country grant (PIF17/74), the Basque 
Foundation for Science (IKERBASQUE) and the Czech science foundation GA\v{C}R grant (19-01850S). The numerical work necessary to carry out this research
has been possible thanks to the computing infrastructure of the ARINA cluster at the University
of the Basque Country, UPV/EHU.

\appendix

\section{Construction of Slepian models}

Throughout this Appendix, we will give a detailed description of the tools and derivations needed in order to generate conditioned Gaussian random fields, such as the ones we have been using throughout the main text. We will be mainly following \cite{Lindgren,adler2009random}.

\subsection{Introductory remarks and some properties of Gaussian random variables}

A random variable $x$ is said to follow a \emph{normal} or \emph{Gaussian distribution} if its probability distribution function (PDF) is given by
\begin{equation}
f(x) = \frac{1}{\sqrt{2\pi \sigma^2}} e^{-\frac{(x-\mu)^2}{2\sigma^2}}
\end{equation}
where $\mu = \left\langle x \right\rangle$ and $\sigma = \left\langle  x^2 \right\rangle$ are the mean and variance of the distribution, respectively. Likewise, a p-dimensional vector $\vec{x}^T = (x_1, \ldots, x_p)$ is defined as a Gaussian random vector (composed of jointly Gaussian variables) if every linear combination satisfies
\begin{equation}
\vec{a} \cdot \vec{x} = \sum_{i=1}^{p} a_i x_i \sim N(\tilde{\mu}, \tilde{\sigma}),
\label{def}
\end{equation}
that is, it follows a normal distribution. The PDF of the whole vector is
\begin{equation}
f(\vec{x}) = \frac{1}{(2\pi)^{p/2} \sqrt{\det \Sigma}} \exp \left[ -\frac{1}{2} (\vec{x}-\vec{\mu})^T \Sigma^{-1} (\vec{x}-\vec{\mu}) \right]
\label{vector_pdf_app}
\end{equation}
where $\vec\mu = \left\langle \vec{x} \right\rangle$ is the mean \emph{vector} and $\Sigma$ is the (non-degenarate) \emph{covariance matrix}, whose elements are given by 
\begin{equation}
\Sigma_{ij}= \left\langle (x_i - \mu_i)(x_j-\mu_j) \right\rangle.
\end{equation}

\subsection{Conditioned Gaussian random vectors}
\label{sec:conditioning}

Let $A$ be a $p \times p$ matrix and $\vec{x}^T = (x_1, \ldots, x_p)$ a Gaussian random vector. Then, by definition, 
\begin{equation}
\vec{y} = A \vec{x} \quad \rightarrow \quad y_j = A_{ij} x_i 
\label{y}
\end{equation}
is also a Gaussian random vector with mean $\vec{\mu}'$ and covariance matrix $\Sigma'$. Since (\ref{y}) is a linear transformation, the new mean is given by
\begin{align}
\vec{\mu}' = A \vec{\mu},
\end{align}
whereas the new covariance matrix is
\begin{align}
\Sigma'_{ij} &= \left\langle (y_i - \mu'_i)(y_j - \mu'_j) \right\rangle = \left\langle (x_a A_{ai} - \mu_b A_{bi})(x_c A_{cj} - \mu_d A_{dj}) \right\rangle \nonumber \\
&= \langle x_a x_c \rangle A_{ai} A_{cj} - \mu_d \langle x_a \rangle A_{ai} A_{dj} - \mu_b \langle x_c \rangle A_{bi} A_{cj} + \mu_b \mu_d A_{bi} A_{dj} \nonumber \\
&= \left\langle (x_a - \mu_a)(x_b - \mu_b) \right\rangle A_{ai} A_{bj} = (A^T)_{ia} \Sigma_{ab} A_{bj}
\end{align}
or, more compactly,
\begin{equation}
\Sigma' = A^T \Sigma A.
\end{equation}

In order to introduce conditional probability notions to jointly Gaussian random variables, let us discuss some interesting properties 
of grouped random variables. If we split some Gaussian vector $\vec{x}$ into two parts, namely,
\begin{equation}
\vec{x} = \left( \vec{x}_1 , \vec{x}_2 \right) = \left( (x_1, \ldots, x_d), (x_{d+1}, \ldots, x_p) \right)
\end{equation}
then the mean vector and covariance matrix will also split accordingly:
\begin{align}
\vec{\mu} &= \left( \vec{\mu}_1 , \vec{\mu}_2 \right) = \left( (\mu_1, \ldots, \mu_d), (\mu_{d+1}, \ldots, \mu_p) \right) \\[7pt]
\Sigma &= \left(
\begin{array}{cc}
\Sigma_{11} & \Sigma_{12} \\
\Sigma_{21} & \Sigma_{22}
\end{array}
\right),
\end{align}
each block in $\Sigma$ having the proper dimensions to accommodate the covariances among the vectors $\vec{x}^1$ and $\vec{x}^2$. 

With these remarks at hand, let us perform a linear transformation on $\vec{x}$, choosing 
\begin{equation}
A= \left(
\begin{array}{cc}
\mathbb{1}_d & -\Sigma_{12} \Sigma_{22}^{-1} \\
0 & \mathbb{1}_{p-d}  
\end{array}
\right). 
\label{A}
\end{equation}
After some straightforward algebra, one can show that the new Gaussian vector $\vec{y}$ is
\begin{equation}
\vec{y}^T = \left( \vec{x}_1 -\Sigma_{12} \Sigma_{22}^{-1} \vec{x}_2,\vec{x}_2 \right) = (\vec{y}_1,\vec{x}_2)
\end{equation}
whose associated mean vector and covariance matrix are
\begin{align}
\vec{\mu}'^T &= \left( \vec{\mu}_1 -\Sigma_{12} \Sigma_{22}^{-1} \vec{\mu}_2,\vec{\mu}_2 \right) \\[7pt]
\Sigma' &= \left(
\begin{array}{cc}
\Sigma_{11} - \Sigma_{12} \Sigma_{22}^{-1} \Sigma_{21}& 0 \\
0 & \Sigma_{22} 
\end{array}
\right),
\end{align}
meaning that the new $\vec{y}_1$ and $\vec{x}_2$ are uncorrelated and, therefore, independent. 

Given a bivariate joint probability distribution function $f(x_1,x_2)$, the conditional probability $f'(x_1|x_2=\tilde{x})$ is defined by \cite{riley2006mathematical}
\begin{equation}
f'(x_1|x_2=\tilde{x}) \equiv \frac{f(x_1,\tilde{x})}{\int d x_1 \ f(x_1,\tilde{x})} = \frac{\int dx_2 \ \delta (x_2-\tilde{x}) f(x_1,x_2)}{\int dx_1 \ dx_2 \ \delta (x_2-\tilde{x}) \ f(x_1,x_2)}.
\label{cond}
\end{equation}
Let $\vec{x}$ be a Gaussian random vector,  a subset of which has been set to $\vec{x}_2=\tilde{\vec{x}}$. We could, in principle, substitute the value of the variables $\vec{x}_1$ into (\ref{vector_pdf_app}) and proceed with the remaining (and normalized) expression. However, more interesting conclusions can be drawn if the above results are applied. Instead of working with $\vec{x} = (\vec{x}_1, \vec{x}_2)$, let us use the PDF associated to $\vec{y} = A \vec{x}$, where A is given by (\ref{A}):
\begin{align}
f(\vec{y}) &= \frac{1}{(2\pi)^{p/2} \sqrt{\det \Sigma_{22}} \sqrt{ \det (\Sigma_{11} - \Sigma_{12} \Sigma_{22}^{-1} \Sigma_{21}) }}  \nonumber \\
& \hspace{1cm} \exp \left[-\frac{1}{2} (\vec{y}_1 - \vec{\mu'}_1)^T (\Sigma_{11} - \Sigma_{12} \Sigma_{22}^{-1} \Sigma_{21})^{-1} (\vec{y}_1 - \vec{\mu'}_1) \right] \exp \left[-\frac{1}{2} (\vec{x}_2 - \vec{\mu}_2)^T \Sigma_{22}^{-1} (\vec{x}_2 - \vec{\mu}_2) \right] \nonumber \\
&= \tilde{f}(\vec{x}_1, \vec{x}_2)  
\end{align}

Fixing $\vec{x}_2 = \vec{\tilde{x}}$ and applying (\ref{cond}) to the resulting probability distribution function, we find
\small
\begin{align}
\tilde{f'}(\vec{x}_1|\vec{x}_2=\tilde{\vec{x}}) = \frac{1}{(2\pi)^{d/2} \sqrt{\det \tilde{\Sigma} }} \exp \left[ - \frac{1}{2} \left( \vec{x}_1 - \vec{\tilde\mu} \right)^T \tilde{\Sigma}^{-1}  \left( \vec{x}_1 - \vec{\tilde\mu} \right) \right]
\end{align}
\normalsize
where
\begin{align}
\vec{\tilde{\mu}} &= \vec{\mu}_1 + \Sigma_{12} \Sigma_{22}^{-1} (\vec{\tilde{x}}-\vec{\mu}_2) \label{cond_mean} \\[5pt]
\tilde{\Sigma} &= \Sigma_{11} - \Sigma_{12} \Sigma_{22}^{-1} \Sigma_{21} \label{cond_cov}
\end{align}

From the expression above, we can conclude that \emph{conditioned Gaussian random vectors retain their Gaussian nature} with mean and covariance matrix given by $\vec{\tilde{\mu}}$ and $\tilde\Sigma$ respectively.

\subsection{Gaussian random fields}

The idea of Gaussian random vectors can be generalized to random variables dependent on a certain set of parameters. Instead of having $p$ Gaussian variables, we will have an infinite amount of them; the mean vector and covariance matrix will thus transform into a mean and covariance \emph{functions}.

A Gaussian random field (GRF) $\lbrace V(\vec{t}) , t \in \mathbb{R}^n \rbrace$ is defined as a function satisfying
\begin{equation}
\sum_{i=1}^{r} a_i V(\vec{t}_i) \sim N(\tilde{\mu},\tilde{\sigma}) \quad \forall r \in \mathbb{N}, \quad \forall a_i \in \mathbb{R}
\end{equation}
at every point of its domain.
The mean function will be given by $\mu (\vec{t}) = \langle V(\vec{t}) \rangle$ whereas the covariance function must satisfy $C(\vec{t},\vec{s})=\langle V(\vec{t}) V(\vec{s}) \rangle$.
If $C(\vec{t},\vec{s}) = f(\vec{t}-\vec{s})$ the GRF is said to be \emph{homogeneous}; if, on the other hand, $C(\vec{t},\vec{s}) = g(\vec{t} \cdot \vec{s},|\vec{t}|,|\vec{s}|)$ the field is \emph{isotropic}. GRFs which are both homogeneous and isotropic are referred to as \emph{stationary}, and satisfy
\begin{equation}
C(\vec{t},\vec{s}) = C(|\vec{t} - \vec{s}|).
\end{equation}
In the main text, we will we working with this last type of covariance function.

Finally, note that any GRF $V(\vec{t})$ with mean $\mu (\vec{t})$ can always be decomposed as
\begin{equation}
V(\vec{t}) = \mu (\vec{t}) + W(\vec{t})
\end{equation}
where $W(\vec{t})$ is a mean-zero GRF sharing the same covariance function $V(\vec{t})$. This construction will be useful to construct GRFs numerically (see Appendix \ref{sec:numerical}).

\subsection{Useful correlations}
\label{sec:correlations}

Since linear combinations of Gaussian variables are Gaussian as well, it is straightforward to see that the derivatives of Gaussian random fields at any point of their domain are Gaussian too. Some of the most important covariance functions relating different Gaussian variables are the following \cite[sect. 5.5]{adler2009random}:
\begin{align}
\left\langle \frac{\partial^{\alpha + \beta} V(\bs{\phi})}{\partial^{\alpha} {\phi}_i \partial^{\beta} {\phi}_j} \ \frac{\partial^{\gamma + \delta} V(\bs{\phi})}{\partial^{\gamma} {\phi}_k \partial^{\delta} {\phi}_l} \right\rangle &= \left. (-1)^{\alpha + \beta} \ \frac{\partial^{\alpha + \beta + \gamma + \delta} }{\partial^{\alpha} {\phi}_i \partial^{\beta} {\phi}_j \partial^{\gamma} {\phi}_k \partial^{\delta} {\phi}_l} C(\bs{\phi}) \right|_{\bs{\phi} = \bs{0}}.
\end{align}
Let us change the notation to $\partial_{\phi_j} V(\bs{0}) = V'_{j} (\bs{0})$ and evaluate the previous expression for some useful cases:
\small
\begin{align}
\left\langle V(\bs{0}) V(\bs{0}) \right\rangle &= U_0^2 \label{cov1} \\[5pt]
\left\langle V(\bs{0})  V'_i(\bs{0}) \right\rangle &= \left\langle  V'_i(\bs{0}) V''_{jk}(\bs{0}) \right\rangle = 0 \\[5pt]
\left\langle V'_i (\bs{0}) V'_j (\bs{0}) \right\rangle &= - \left\langle V(\bs{0}) V''_{ij} (\bs{0}) \right\rangle = -\frac{\partial^2 C(\bs{0})}{\partial {\phi}_i \partial {\phi}_j} = \alpha_2 \delta_{ij} \\[5pt]
\left\langle V''_{ij} (\bs{0}) V''_{kl} (\bs{0}) \right\rangle &= \frac{\partial^4 C(\bs{0})}{\partial {\phi}_i \partial {\phi}_j \partial {\phi}_k \partial {\phi}_l} = \left\lbrace
\begin{array}{lcl}
\alpha_{22} & \text{if} & i=j\neq k=l \text{ (and perms.)} \\
\alpha_{4} & \text{if} & i=j=k=l \\
0 & \text{otherwise.} & 
\end{array} 
\right. \\[5pt]
\left\langle V (\bs{0}) V'''_{jkl} (\bs{0}) \right\rangle &= \left\langle V''_{ij} (\bs{0}) V'''_{klm} (\bs{0}) \right\rangle = 0 \\[7pt]
\left\langle V'_i (\bs{0}) V'''_{jkl} (\bs{0}) \right\rangle &= - \left\langle V''_{ij} (\bs{0}) V''_{kl} (\bs{0}) \right\rangle = \left\lbrace
\begin{array}{lcl}
-\alpha_{22} & \text{if} & i=j\neq k=l \text{ (and perms.)} \\
-\alpha_{4} & \text{if} & i=j=k=l \\
0 & \text{otherwise.} & 
\end{array} 
\right. \\[7pt]
\left\langle V'''_{ijk} (\bs{0}) V'''_{lmn} (\bs{0}) \right\rangle &= - \frac{\partial^6 C (\bs{0})}{\partial {\phi}_i \partial  {\phi}_j \partial  {\phi}_k \partial  {\phi}_l \partial  {\phi}_m \partial  {\phi}_n} = \left\lbrace
\begin{array}{lcl}
\alpha_{222} & \text{if} & i=j\neq k=l \neq m=n \text{ (and perms.)} \\
\alpha_{24} & \text{if} & i=j\neq k=l=m=n \text{ (and perms.)} \\
\alpha_6 & \text{if} & i=j=k=l=m=n \\
0 & \text{otherwise.} & 
\end{array}
\right.
\end{align} 
\normalsize
In the above expressions, $\alpha_{i}$, $\alpha_{ij}$ and $\alpha_{ijk}$ are numerical constants which depend only on the covariance function of the (unconstrained) Gaussian random field.
Note that in the two-dimensional case $\alpha_{222}$ will be absent from all derivations, since the indices appearing in the correlation function between the third derivatives can only take two different values. 

Note also that odd derivatives of the GRF are uncorrelated with even ones when they are evaluated at the same point in field space. This is due to the isotropy of the covariance function: if it is written as a power series, only even powers such as $\phi_i^2, \phi_i^2 \phi_j^2$ will be involved. Therefore, only those correlations which end up involving even derivatives of the covariance function are non-zero.

This however, does not mean the fields $V(\vec{\phi})$ and, say, $V'_i(\vec{\phi})$ are completely uncorrelated. If we evaluate them at different points in field space, it can be shown \cite[theorem 2.3]{lindgren2012stationary} that
\begin{align}
\left\langle V(\vec{\phi}) V'_i(\vec{0}) \right\rangle &= - \frac{\partial}{\partial \phi_i} C(\vec{\phi}) \\[5pt]
\left\langle V(\vec{\phi}) V''_{ij}(\vec{0}) \right\rangle &= \frac{\partial^2}{\partial \phi_i \partial \phi_j} C(\vec{\phi}) \\[5pt]
\left\langle V (\bs{ {\phi}}) V'''_{ijk} (\bs{0}) \right\rangle &= - \frac{\partial^3}{\partial  {\phi}_i \partial  {\phi}_j \partial  {\phi}_k} C(\bs{ {\phi}}) 
\label{cov2}
\end{align}
therefore, a GRF and any of its derivatives are correlated as processes.

\subsection{The Kac-Rice formula and conditioned Gaussian random fields}
\label{sec:KR}

Consider a Gaussian random \emph{vector} field with components $\vec{\mathcal{V}}(\vec{\phi})=\left\lbrace V_1 (\vec{\phi}) , \ldots , V_n(\vec{\phi}) \right\rbrace $. The multidimensional\footnote{Note that this formula is only valid for fields mapping $\mathbb{R}^n \rightarrow \mathbb{R}^n$.} Kac-Rice formula for this field gives us the expected number of times a certain event, say, $\vec{\mathcal{V}}(\vec{\phi})=\vec{u}$, happens in an interval $\vec{\phi}\in I$ of volume $\mathbb{V}$:
\begin{equation}
\mathbb{E}_{\#,I}\left[ \vec{\mathcal{V}}(\vec\phi) = \vec{u} \right] = \left\langle \int_I d\vec{\phi} \ | \det \vec{\mathcal{V'}}(\vec{\phi}) | \ \delta (\vec{\mathcal{V}}(\vec{\phi}) - \vec{u}) \right\rangle
\end{equation} 
where $\det \vec{\mathcal{V}}'(\vec{\phi})$ stands for the Jacobian determinant of the vector field\footnote{For critical points, the Jacobian is identical to the Hessian of the GRF at the critical point.}, that is,
\begin{align}
\vec{\mathcal{V}'}(\vec\phi) = 
\begin{pmatrix}
\partial_{\phi_1} V_1 (\vec\phi)& \cdots & \partial_{\phi_1} V_n (\vec\phi)\\
\vdots & & \vdots \\
\partial_{\phi_n} V_1 (\vec\phi)& \cdots & \partial_{\phi_n} V_n(\vec\phi)
\end{pmatrix} ~.
\end{align}
If the field is \emph{stationary}, that is, homogeneous and isotropic, we can simplify the expression above. Denoting $\vec{\mathcal{V}_0} = \vec{\mathcal{V}}(\vec{0})$ and $\vec{\mathcal{V}'_0} = \vec{\mathcal{V'}}(\vec{0})$, we find, assuming ergodicity, 
\begin{align}
\mathbb{E}_{\#,I}\left[ \vec{\mathcal{V}}(\vec\phi) = \vec{u} \right] = \mathbb{\mathcal{V}} \int d\vec{\mathcal{V}_0} \ d\vec{\mathcal{V}_0}' \ | \det \vec{\mathcal{V}_0}' | \ \delta (\vec{\mathcal{V}_0}-\vec{u}) \ P(\vec{\mathcal{V}_0},\vec{\mathcal{V}_0}')
\label{eq:kr_simple}
\end{align}
where the integral is performed over the whole domain of $\vec{\mathcal{V}_0}$ and $\vec{\mathcal{V}_0'}$ and $P(\vec{\mathcal{V}_0},\vec{\mathcal{V}_0}')$ is the joint PDF of $\vec{\mathcal{V}_0}$ and its derivatives. 

More than one simultaneous event can be considered in the expressions above by enlarging the vector $\vec{\mathcal{V}}$ and introducing more Dirac deltas representing each event\footnote{See, however, \cite[ch.8]{lindgren2012stationary} for a discussion on different types of conditioning events and how to deal with them. The reason why we consider the $V_0 = u$ event simply with a Dirac delta is that it is a \emph{vertical window conditioning event}}. 
\\

While the above expression can certainly be used to obtain the number of times a certain event happens in a given interval, it can also be used to obtain distribution functions. More specifically, applying ergodicity theorems, it can be shown \cite{lindgren2012stationary} that the probability of an event $A$ happening, given that $B$ has happened, that is, $P(A|B)$, can be obtained by
\begin{align}
P(A|B) = \frac{\mathbb{E}_{\#,I}\left[ A \cap B \right]}{\mathbb{E}_{\#,I}\left[ B \right]}.
\label{eq:pab}
\end{align}
If $A$ depends on continuous parameters (such as the position in field space of the GRF), then the expression above represents a probability distribution function.

\subsection{Conditioned Gaussian random field for a critical point}
\label{sec:conditioned_grf_min}

With the tools presented in the sections above, we are now ready to begin conditioning GRFs. We can begin applying (\ref{eq:pab}) and specializing it for critical points. We denote by $A$ the event describing the field $V(\vec\phi)$ taking a particular configuration, while $B$ imposes $V(\vec{0}) \equiv V_0 = u$ and $V'_i (\vec{0}) \equiv \eta_i = 0$, that is, a critical point lying in the center of  field space at height $u$. In order to proceed more easily, we shall discretize $V(\vec{\phi})$ as $\lbrace V(\vec{\phi_1}), \ldots, V(\vec{\phi_q}) \rbrace \equiv \lbrace V_1 , \ldots V_q \rbrace \equiv \vec{V}$. 

In this case, the conditioning event involves the Gaussian random vector field $\vec{\mathcal{V}}=\vec\nabla V$, whose Jacobian is the Hessian of the original field $V$ evaluated at $\vec{\phi}=\vec{0}$. Therefore, its determinant is simply the product of the eigenvalues of the Hessian evaluated at the origin, $\prod_{i=1}^{n} \lambda_i$.

Applying the Kac-Rice formula (\ref{eq:kr_simple}) into (\ref{eq:pab}) yields
\begin{align}
& P\Big( V(\vec\phi) \Big| V_0=u , \nabla V_0  = \vec{0} \Big) \equiv P_{cp} [V(\vec\phi)] = \nonumber \\[8pt]
&= \frac{\displaystyle \int \prod_{i=1}^n \Big( d\eta_i \delta(\eta_i) d\lambda_i  |\lambda_i| \Big) \ \Delta (\vec\lambda) \ \delta(V_0-u) \ \prod_{j=1}^q \left( d\tilde{V}_j \delta (\tilde{V}_j - V_j) \right) P \Big( V_0, \vec{V}, \vec{\eta}, \vec{\lambda} \Big) }{\displaystyle \int \prod_{i=1}^n \Big( d\eta_i \delta(\eta_i) d\lambda_i  |\lambda_i| \Big) \ \Delta (\vec\lambda) \ \delta(V_0-u) \ P \Big( V_0, \vec{\eta}, \vec{\lambda} \Big) } \label{eq:slepian_1} \\[8pt]
&= \mathcal{N} \  \int \prod_{i=1}^n \left( d\lambda_i  |\lambda_i| \right) \Delta(\vec{\lambda}) P \Big( V(\vec{\phi}), \lambda_1,\ldots,\lambda_n  \ \Big| \ V_0=u, \ \nabla V_0=\vec{0} \Big)
\end{align}
where the integration domain will depend on the kind of critical point we are working with. $\Delta(\vec\lambda) \propto \prod_{i<j} |\lambda_i - \lambda_j|$ is the Jacobian of the variable change from components of the Hessian matrix to its eigenvalues, the proportionality constant depending on the dimensionality of the field space. For simplicity, the denominator in (\ref{eq:slepian_1}) has been considered as a normalization factor for the distribution in the numerator. 

We can rewrite (\ref{eq:slepian_1}) in a more useful way:
\begin{align}
P_{cp} [V(\vec\phi)] = \prod_i \int d\lambda_i \ q_u (\lambda_1,\ldots,\lambda_n) \  P \Big( V(\vec{t}) \ \Big| \ V_0=u, \ \nabla V_0=\vec{0}, \ \lambda_1,\ldots,\lambda_n \Big)
\label{eq:slepian_2}
\end{align}
where 
\begin{align}
q_u (\lambda_1, \ldots, \lambda_n) =  \prod_i | \lambda_i | \ \Delta(\vec{\lambda}) \ P \Big( \lambda_1,\ldots,\lambda_n  \ \Big| \ V_0=u, \ \nabla V_0=\vec{0} \Big)
\label{eq:q_dist}
\end{align}
represents the distribution of the Hessian eigenvalues at the origin for a critical point of height $u$. However, due to the homogeneous and isotropic nature of the original GRF, the latter distribution is valid for \emph{any} critical point in the GRF, thus giving us a distribution for the parameters at critical points in the unconstrained field.

Equations (\ref{eq:slepian_2}) and (\ref{eq:q_dist}) are central results in this derivation. Note that the $\prod_i | \lambda_i | \ \Delta(\vec{\lambda})$ factor is a direct consequence of the Kac-Rice formula, and as we shall explicitly see in Appendix \ref{sec:numerical}, it carries important consequences in the distribution of the eigenvalues at critical points.

We can now see the power of this method. Assuming we have discretized our field space, we can readily compute the conditional probability distributions in (\ref{eq:slepian_2}) and (\ref{eq:q_dist}) using the results from section \ref{sec:conditioning}. This leads, together with (\ref{eq:q_dist}), to a distribution from which we can draw eigenvalues for a minimum of height $u$. These can be plugged in (\ref{eq:slepian_2}) to generate iterations of GRFs with a minimum (or any other critical point) at their origin.

In order to apply all this machinery, let us introduce the following Gaussian random vector:
\begin{align}
\lbrace  V(\bs{\phi}_1), \ldots , V(\bs{\phi}_q), V(\bs{0}) ,V'_1 (\bs{0}) , \ldots,  V'_n (\bs{0}), V''_{11} (\bs{0}) , \ldots, V''_{nn} (\bs{0}), \underbrace{ V''_{12} (\bs{0}), \ldots, V''_{(n-1)n} (\bs{0}) }_{V''_{ij} (\bs{0}) \quad i<j} \rbrace
\label{full_vector}
\end{align}
where we denote by $\vec{\phi}_q$ the position in field space of a discrete set of points whose center is located at
${\bs 0}$, $V'_i (\bs{0})$ describes the first derivative along $\phi_i$ and $V''_{ij}(\bs{0})$ is the $(i,j)$-th element of the Hessian matrix. In order to unclutter the notation, we will compactify the previous vector as
\begin{align}
\lbrace \bs{V}, V(\bs{0}) ,\bs{V'}(\bs{0}) , \bs{V}''(\bs{0}) \rbrace
\end{align}
which has dimension $q + 1 + n + n + \frac{1}{2} n (n-1)$. The mean of (\ref{full_vector}) is zero, and the covariance matrix of these quantities can be computed from the results in section \ref{sec:correlations}:
\begin{align}
\Sigma = \left(
\begin{array}{cccc}
S_{VV} & S_{V0} & S_{V1} & S_{V2} \\ 
S_{0V} & U_0^2	& \bs{0} & S_{02} \\
S_{1V} & \bs{0} & S_{11} & \bs{0} \\
S_{2V} & S_{20} & \bs{0} & S_{22}
\end{array}
\right)
\label{full_cov}
\end{align}
where
\begin{align}
S_{02} &= \left(
\begin{array}{cccccc}
-\alpha_2 & \cdots & -\alpha_2 & 0 & \cdots & 0
\end{array}
\right) = S_{20}^T \label{eq:matrix_first}\\
S_{11} &= \alpha_2 \times \mathbb{1}_n \label{S11}\\[10pt]
S_{22} &= \left(
\begin{array}{cccc|ccc}
\alpha_{4} & \alpha_{22} & \cdots & \alpha_{22} & & & \\
\alpha_{22} & \alpha_{4} & \cdots & \alpha_{22} & & 0 & \\
\vdots		 & \vdots		& \ddots & \vdots		& & & \\
\alpha_{22} & \alpha_{22} & \cdots & \alpha_{4} & & & \\ \hline
& & & & \alpha_{22} & & 0 \\
& & 0 & &  & \ddots &   \\
& & & & 0 & & \alpha_{22} 
\end{array}
\right)
\label{S22} \\
S_{VV} &= \left(
\begin{array}{cccc}
C(\bs{0}) & C(\bs{{\phi}_1}-\bs{{\phi}_2}) & \cdots & C(\bs{{\phi}_1} -\bs{{\phi}_q}) \\
C(\bs{{\phi}_2}-\bs{{\phi}_1}) & C(\bs{0}) & \cdots & C(\bs{{\phi}_2} -\bs{{\phi}_q}) \\
\vdots & \vdots & \ddots & \vdots \\
C(\bs{{\phi}_q} - \bs{{\phi}_1}) & C(\bs{{\phi}_q} - \bs{{\phi}_2}) & \cdots & C(\bs{0})
\end{array}
\right) \\[10pt]
S_{0V} &= \left(
\begin{array}{cccc}
C(\bs{{\phi}_1}) & C(\bs{{\phi}_2}) & \cdots & C(\bs{{\phi}_q})
\end{array}
\right) = S_{V0}^T \\[10pt]
S_{1V} &= \left(
\begin{array}{cccc}
-C'_1(\bs{{\phi}_1}) & -C'_1(\bs{{\phi}_2}) & \cdots & -C'_1(\bs{{\phi}_q}) \\
-C'_2(\bs{{\phi}_1}) & -C'_2(\bs{{\phi}_2}) & \cdots & -C'_2(\bs{{\phi}_q}) \\
\vdots & \vdots & \ddots & \vdots \\
-C'_n(\bs{{\phi}_1}) & -C'_n(\bs{{\phi}_2}) & \cdots & -C'_n(\bs{{\phi}_q})
\end{array}
\right) = S_{V1}^T \\[10pt]
S_{2V} &= \left(
\begin{array}{ccc}
C''_{11}(\bs{{\phi}_1}) & \cdots & C''_{11}(\bs{{\phi}_q})  \\
\vdots				& \ddots & \vdots  \\
C''_{nn}(\bs{{\phi}_1}) & \cdots & C''_{nn}(\bs{{\phi}_q})  \\
C''_{12}(\bs{{\phi}_1}) & \cdots & C''_{12}(\bs{{\phi}_q})  \\
\vdots				& \ddots & \vdots  \\
C''_{(n-1)n}(\bs{{\phi}_1}) & \cdots & C''_{(n-1)n}(\bs{{\phi}_q})  
\end{array}
\right) = S_{V2}^T \label{eq:matrix_last}
\end{align}

In order to simplify the notation, since the jointly Gaussian probability distribution in the end depends on two-point functions, we can actually write\footnote{We basically have evaluated the first row for a given $\bs{\phi_1}$ and the first column for a given $\bs{\phi_2}$, just as in \cite{Lindgren}. Doing so allows us to treat the independent variable as a continuous one, rather than a discrete one.} (\ref{full_cov}) in the following way:
\begin{align}
\Sigma = \left(
\begin{array}{cc|ccc}
U_0^2 & C(\bs{\phi_1}-\bs{\phi_2}) & ~~C(\bs{\phi_1}) & ~~S_{V1}(\bs{\phi_1}) & ~~S_{V2}(\bs{\phi_1}) \\ 
C(\bs{\phi_2}-\bs{\phi_1}) & U_0^2 & ~C(\bs{\phi_2}) & ~~S_{V1}(\bs{\phi_2}) & ~~S_{V2}(\bs{\phi_2}) \\ \hline
C(\bs{\phi_1}) & C(\bs{\phi_2}) & U_0^2	& \bs{0} & S_{02} \\
S_{1V} (\bs{\phi_1}) & S_{1V} (\bs{\phi_2}) & \bs{0} & S_{11} & \bs{0} \\
S_{2V} (\bs{\phi_1}) & S_{2V} (\bs{\phi_2}) & S_{20} & \bs{0} & S_{22}
\end{array}
\right)
\label{cov_matrix}
\end{align}
where
\begin{align}
S_{V1} (\bs{\phi}) &= \left(
\begin{array}{ccc}
- C'_1 (\bs{\phi}) & \cdots & -C'_n (\bs{\phi})
\end{array}
\right) = S_{1V}^T \\[10pt]
S_{V2} (\bs{\phi}) &= \left(
\begin{array}{cccccc}
 C''_{11} (\bs{\phi}) & \cdots & C''_{nn} (\bs{\phi}) & C''_{12} (\bs{\phi}) & \cdots & C''_{(n-1)n} (\bs{\phi})
\end{array}
\right) = S_{2V}^T 
\end{align}
With these arrangements, the Gaussian random vector corresponding to (\ref{cov_matrix}) is
\begin{align}
\left\lbrace V(\vec{\phi}_1) , V(\vec{\phi}_2) , V(\bs{0}) ,\bs{V'}(\bs{0}) , \bs{V''}(\bs{0}) \right\rbrace .
\end{align}

\thispagestyle{empty}

We have decomposed (\ref{cov_matrix}) into blocks so it can be plugged into (\ref{eq:cond_mean}) and (\ref{eq:cond_cov}) to obtain the mean function and covariance matrix of the conditioned process\footnote{Strictly speaking, we should be getting the mean and covariance of the random vector $\lbrace V(\vec\phi_1) , V(\vec{\phi}_2) \rbrace$. Due to the isotropy of the GRF, $\vec\phi_1$ and $\vec\phi_2$ can be any points in field space. Thus, in order to unclutter the notation, we will only keep track of a single component of the resulting mean vector. Likewise, we will only keep the $\langle V(\vec\phi_1) V(\vec\phi_2) \rangle$ component of the covariance matrix.}. Using the results given above, one gets that the expectation value for the
GRF around a critical point where $V_0 = u$ and $\bs{V_0'} = 0$, is given by,
\begin{align}
\tilde{\mu} (\bs{\phi}) &= \mu (\bs{\phi}) + \left(
\begin{array}{ccc}
C(\bs{\phi}) & S_{V1} (\bs{\phi}) & S_{V2} (\bs{\phi}) 
\end{array} 
\right) \left(
\begin{array}{ccc}
U_0^2 & \bs{0} & S_{02} \\
\bs{0} & S_{11} & \bs{0} \\
S_{20} & \bs{0} & S_{22}
\end{array}
\right)^{-1} \left(
\begin{array}{c}
u \\ \bs{0} \\ \bs{h}
\end{array}
\right) \nonumber \\[7pt]
&= \left(
\begin{array}{cc}
C(\bs{\phi}) & S_{V2} (\bs{\phi}) 
\end{array} 
\right) \left(
\begin{array}{ccc}
U_0^2  & S_{02} \\
S_{20} & S_{22}
\end{array}
\right)^{-1} \left(
\begin{array}{c}
u  \\ \bs{h}
\end{array}
\right) 
\label{eq:cond_mean}
\end{align}
where $\bs{h} = \left\lbrace h_{11}, \ldots, h_{nn}, h_{12}, \ldots , h_{(n-1)n} \right\rbrace$ represents a certain configuration of
the Hessian components of the field around the origin.

Furthermore, the covariance function for the conditioned GRF is now 
\begin{align}
\tilde{C}(\bs{\phi_1} , \bs{\phi_2}) &= C(\bs{\phi_1}-\bs{\phi_2}) -\left(
\begin{array}{ccc}
C(\bs{\phi_1}) & S_{V1} (\bs{\phi_1}) & S_{V2} (\bs{\phi_1}) 
\end{array} 
\right) \left(
\begin{array}{ccc}
U_0^2 & \bs{0} & S_{02} \\
\bs{0} & S_{11} & \bs{0} \\
S_{20} & \bs{0} & S_{22}
\end{array}
\right)^{-1} \left( 
\begin{array}{c}
C(\bs{\phi_2}) \\ S_{1V} (\bs{\phi_2}) \\ S_{2V} (\bs{\phi_2})
\end{array}
\right) \nonumber \\[7pt]
&=  C(\bs{\phi_1}-\bs{\phi_2}) -\left(
\begin{array}{ccc}
C(\bs{\phi_1}) & S_{V2} (\bs{\phi_1}) 
\end{array} 
\right) \left(
\begin{array}{cc}
U_0^2 & S_{02} \\
S_{20} & S_{22}
\end{array}
\right)^{-1} \left( 
\begin{array}{c}
C(\bs{s}) \\ S_{2V} (\bs{\phi_2})
\end{array}
\right) \nonumber\\ &\hspace{77pt}- S_{V1}(\bs{\phi_1}) S_{11}^{-1} S_{1V}(\bs{\phi_2}) 
\label{eq:cond_cov}
\end{align}

We can also obtain (\ref{eq:q_dist}), the distribution of eigenvalues at a critical point of a given height $u$, following the same steps as above, using as initial covariance matrix the bottom-right block of (\ref{cov_matrix}).  

\subsubsection{Analysis of a conditioned 2D Gaussian field}

Let us apply these expressions to a two-dimensional isotropic and homogeneous
 GRF with covariance function 
\beq
\label{2d-cf}
C(\bs{\phi}) = U_0^2 \exp \left( - {{ {\bs{\phi}}^2}\over {2\Lambda^2}} \right)~.
\eeq
and zero mean. For this case, we obtain the conditioned mean from (\ref{eq:cond_mean}), which gives
\begin{align}
\tilde{\mu}(\bs{\phi}) = e^{-\frac{\bs{\phi}^2}{2\Lambda^2}} \left[ u \left( 1 + \frac{\bs{\phi}^2}{2\Lambda^2} \right) + \frac{1}{2} 
\left(
\begin{array}{cc}
\phi_1 & \phi_2
\end{array}
\right) \left(
\begin{array}{cc}
h_{11} & h_{12} \\ h_{21} & h_{22}
\end{array}
\right) \left(
\begin{array}{c}
\phi_1 \\ \phi_2
\end{array}
\right) \right],
\label{eq:mu_tild_1}
\end{align}
where $h_{21}=h_{12}$, by definition. 
Since we are free to choose the basis of $\bs{\phi}$, in order to simplify the expression we will employ the eigenvector basis of the Hessian matrix, therefore transforming (\ref{eq:mu_tild_1}) to
\begin{align}
\tilde{\mu}(\bs{\phi}) = e^{-\frac{\bs{\phi}^2}{2\Lambda^2}} \left[ u \left( 1 + \frac{\bs{\phi}^2}{2\Lambda^2} \right) + \frac{1}{2} \sum_{i=1}^2  \lambda_i \phi_i^2 \right].
\end{align}
where $\lambda_i$ denote the two eigenvectors, drawn from (\ref{eq:q_dist}) specialized to this case (see below).
As for the conditioned covariance, from (\ref{eq:cond_cov}) we obtain
\begin{align}
\tilde{C}(\bs{\phi_1},\bs{\phi_2}) = U_0^2 \exp \left[ - \frac{|\bs{\phi_1}|^2 + |\bs{\phi_2}|^2}{2\Lambda^2} \right] \left( \exp \left[  \frac{\bs{\phi_1}\cdot \bs{\phi_2}}{\Lambda^2}  \right] - 1 - \frac{\bs{\phi_1}\cdot\bs{\phi_2}}{\Lambda^2}  - \frac{(\bs{\phi_1}\cdot\bs{\phi_2})^2}{2\Lambda^4}  \right)~.
\end{align}
Note that the covariance function of the conditioned process is not homogeneous anymore! This, however, makes complete sense. We have actually made the center of every realization \emph{special}, meaning that homogeneity is broken in this sense. In fact, the new covariance is isotropic with respect to $\bs{\phi}=\bs{0}$, further stating that the center of the GRF is somehow different from the rest of the points.

All the presented machinery works not only for minima, but also for maxima and saddle points as well; the only difference among these being the sign of each $\lambda_i$. 

\subsubsection{Distribution of heights and eigenvalues of the Hessian at a critical point}

In order to calculate the probability distribution of the eigenvalues of the Hessian at a certain height of the potential at critical points we should pay attention to two ingredients. The first one is the fact that the height and the second derivatives are correlated, so we need to calculate the multivariate covariance function for these quantities together. Furthermore, we also want to calculate this at critical points which can be done with the use of the generalized Kac-Rice formula. 

Assuming a critical point located at $\vec{\phi}=\vec{0}$, the probability distribution to be computed is
\begin{align}
	P \Big( V_0 , \lambda_1 , \lambda_2 \Big| \nabla V_0 = \vec{0} \Big)
\end{align}
We can easily compute the PDF by conditioning the following random vector:
\begin{align}
	\{ V_0 , h_{11}, h_{22}, h_{12} , \eta_1 , \eta_2 \}
\end{align}
of mean zero and covariance matrix
\begin{align}
	\left(
	\begin{array}{cc|c}
	U_0^2 & S_{02} & \bs{0} \\
	S_{20} & S_{22} & \bs{0} \\ \hline
	\bs{0} & \bs{0} & S_{11} 
	\end{array}
	\right)
\end{align}

Applying (\ref{cond_mean}) and (\ref{cond_cov}) to obtain the  mean and covariance of the conditioned process and plugging them into (\ref{eq:q_dist}), we get 
\begin{align}
P_{cp} (V_0,\lambda_1,\lambda_2) \ du \prod_{i=1}^2 d\lambda_i &= \mathcal{N} \ |\lambda_1| |\lambda_2| \ \Delta(\vec{\lambda}) \ P \Big( V_0 , \lambda_1 , \lambda_2 \Big| \nabla V_0 = \vec{0} \Big) \label{eq:norm_kr}\\
&=\mathcal{N} |\lambda_1 - \lambda_2| |\lambda_1| |\lambda_2| \exp \left[ -\frac{V_0^2}{2 U_0^2} \right]   \exp \left[ - \left( \frac{\Lambda^2 \lambda_i + V_0}{2U_0} \right)^2 \right] d\lambda_i \ dV_0
\label{P_u_lambda}
\end{align}
where $\mathcal{N}$ is a normalization factor and, in this two-dimensional example, $\Delta (\vec{\lambda})=|\lambda_1 - \lambda_2|\cdot \pi/2$. 

Setting $V_0$ to a constant value, say $V_0 = u$, in (\ref{P_u_lambda}) yields the distribution $q_u (\lambda_1 , \lambda_2)$, defined in (\ref{eq:q_dist}). On the other hand, integrating out either $V_0$ or the eigenvalues, gives the marginal distribution for the remaining variables in critical points (see Appendix \ref{sec:plots} for more detail).

Another interesting application of (\ref{eq:norm_kr}) is that it can be used to count the expected number of critical points in a certain region of field space. For example, to compute the expected number of minima per correlation volume $\Lambda^2$ in the example above, a direct application of (\ref{eq:kr_simple}) yields
\begin{align}
	\frac{\mathbb{E}(\#_{min})}{\Lambda^2} &= \int_{-\infty}^{+\infty} du \int_{0}^{+\infty} d\lambda_1 \int_{0}^{+\infty} d\lambda_2 \frac{\pi}{2} \ \lambda_1  \lambda_2 \ |\lambda_1-\lambda_2| P \Big( V_0 , \lambda_1 , \lambda_2 \Big| \nabla V_0 = \vec{0} \Big) \nonumber \\
	&= \frac{1}{2\sqrt{3}}.
\label{eq:min_number}
\end{align}
In this case, the eigenvalues have been assumed to be positive. Setting other integration limits can give the expected number of maxima and saddle points, for example.

\subsection{Conditioned Gaussian random field for an inflection point}
\label{sec:conditioned_grf_ip}

We shall define an inflection point on our GRF as \emph{a point where the gradient of the field points in the direction of a Hessian eigenvector whose corresponding eigenvalue is zero}. Furthermore, we will also demand that the non-zero eigenvalue of the Hessian to be positive at this point. 

In order to do this we can expand the discussion of the previous section by taking into 
account the third derivatives of the GRFs along with the lower ones. In order to simplify this description 
we will give a detail account of this construction for a $2d$ GRF only. Extending this to higher dimensions 
is straightforward. In particular we will be interested in the Gaussian random vector
\begin{align}
\left\lbrace V(\bs{\phi_1}), V(\bs{\phi_2}) ,V_0 , V'_1 (\vec{0}), V'_2 (\vec{0}), V''_{11} (\vec{0}), V''_{22} (\vec{0}), V''_{12} (\vec{0}), V'''_{111} (\vec{0}), V'''_{122} (\vec{0}), V'''_{222}(\vec{0}) , V'''_{112}(\vec{0})  \right\rbrace
\end{align}
whose components have zero mean. As for the covariance matrix, it can be expressed as
\begin{align}
\Sigma = \left(
\begin{array}{cc|cccc}
U_0^2 & C(\vec\phi_1 - \vec\phi_2) & C(\bs{\phi_1}) & S_{V1}(\bs{\phi_1}) & S_{V2}(\bs{\phi_1}) & S_{V3} (\bs{\phi_1}) \\ 
C(\bs{\phi_2}-\bs{\phi_1}) & U_0^2 & C(\vec\phi_2) & S_{V1}(\bs{\phi_2}) & S_{V2}(\bs{\phi_2}) & S_{V3} (\bs{\phi_2}) \\ \hline
C(\bs{\phi_1}) & C(\bs{\phi_2}) & U_0^2 & \bs{0} & S_{02} & \bs{0}\\
S_{1V} (\bs{\phi_1}) & S_{1V} (\bs{\phi_2}) & \bs{0} & S_{11} & \bs{0} & S_{13} \\
S_{2V} (\bs{\phi_1}) & S_{2V} (\bs{\phi_2}) & S_{20} & \bs{0} & S_{22} & \bs{0} \\
S_{3V} (\bs{\phi_1}) & S_{3V} (\bs{\phi_2}) & \bs{0} & S_{31} & \bs{0} & S_{33}
\end{array}
\right)
\label{cov_matrix2}
\end{align}
where (for the 2D case)
\begin{align}
S_{V3} (\bs{\phi}) &= \left(
\begin{array}{cccc}
-C'_{111} (\bs{\phi}) & -C'_{122} (\bs{\phi}) & -C'_{222} (\bs{\phi}) & -C'_{112} (\bs{\phi}) 
\end{array}
\right) = S_{3V}^T \\[7pt]
S_{13} &= \left(
\begin{array}{cccc}
-\alpha_4 & -\alpha_{22} & 0 & 0 \\
0 & 0 & -\alpha_4 & -\alpha_{22}  
\end{array}
\right) = S_{31}^T \\[7pt]
S_{33} &= \left(
\begin{array}{cccc}
\alpha_6 & \alpha_{24} & 0 & 0 \\
\alpha_{24} & \alpha_{24} & 0 & 0 \\
0 & 0 & \alpha_6 & \alpha_{24} \\
0 & 0 & \alpha_{24} & \alpha_{24} 
\end{array}
\right)
\end{align}
and the other matrix blocks have been defined in (\ref{eq:matrix_first} - \ref{eq:matrix_last}).

Following the same steps as in the critical point case, we can obtain (for the covariance function (\ref{2d-cf}))  the expression for
a GRF once we conditioned everything up to the third derivative. In order to do this we can first compute the
mean value of the GRF in the vicinity of our inflection point, which is given by
\begin{align}
\tilde{\mu}(\bs{\phi}) &=  0 + \left(
\begin{array}{cccc}
C(\bs{\phi}) & S_{V1}(\bs{\phi}) & S_{V2}(\bs{\phi}) & S_{V3} (\bs{\phi}) 
\end{array}
\right) \left(
\begin{array}{cccc}
U_0^2	& \bs{0} & S_{02} & \bs{0}\\
\bs{0} & S_{11} & \bs{0} & S_{13} \\
S_{20} & \bs{0} & S_{22} & \bs{0} \\
\bs{0} & S_{31} & \bs{0} & S_{33}
\end{array}
\right)^{-1} \left(
\begin{array}{c}
u \\ \bs{\eta} \\ \bs{h} \\ \bs{\rho} 
\end{array}
\right) \\[7pt]
&= \left(
\begin{array}{cc}
C(\bs{\phi}) & S_{V2}(\bs{\phi}) 
\end{array}
\right) \left(
\begin{array}{cc}
U_0^2  & S_{02} \\
S_{20} & S_{22}  
\end{array}
\right)^{-1} \left(
\begin{array}{c}
u \\ \bs{h} 
\end{array}
\right) + \left(
\begin{array}{cc}
S_{V1}(\bs{\phi}) & S_{V3} (\bs{\phi}) 
\end{array}
\right) \left(
\begin{array}{cccc}
S_{11} & S_{13} \\
S_{31} & S_{33}
\end{array}
\right)^{-1} \left(
\begin{array}{c}
\bs{\eta}  \\ \bs{\rho} 
\end{array}
\right) \nonumber \\[7pt]
&= \exp \left[ - \frac{\bs{\phi}^2}{2\Lambda^2} \right] \left( (u + \bs{\phi}\cdot\bs{\eta}) \left( 1 + \frac{\bs{\phi}^2}{2\Lambda^2} \right) + \frac{1}{2} \sum_{i=1}^2 \lambda_i \phi_i^2  + \frac{1}{6} \sum_{i,j,k=1}^{2} \phi_i \phi_j \phi_k \rho_{ijk} \right)~,
\label{slepian_mean_inf}
\end{align}
where the basis of $\bs{\phi}$ has been chosen to be the eigenbasis of the Hessian matrix (whose components are described by $\bs{h}$ and its eigenvalues by $\lambda_i$) and we have denoted by $\bs{\eta}$ and $\bs{\rho}$ the components of the first and third derivatives at the origin along the eigenbasis. 

The conditioned covariance, on the other hand, reads
\small 
\begin{align}
\tilde{C}(\bs{\phi_1},\bs{\phi_2}) &= C(\bs{\phi_1} - \bs{\phi_2}) - \left(
\begin{array}{cccc}
C(\bs{\phi_1}) & S_{V1}(\bs{\phi_1}) & S_{V2}(\bs{\phi_1}) & S_{V3} (\bs{\phi_1}) 
\end{array}
\right) \left(
\begin{array}{cccc}
U_0^2	& \bs{0} & S_{02} & \bs{0}\\
\bs{0} & S_{11} & \bs{0} & S_{13} \\
S_{20} & \bs{0} & S_{22} & \bs{0} \\
\bs{0} & S_{31} & \bs{0} & S_{33}
\end{array}
\right)^{-1} \left(
\begin{array}{c}
C(\bs{\phi_2}) \\ S_{1V}(\bs{\phi_2}) \\ S_{2V}(\bs{\phi_2}) \\ S_{3V} (\bs{\phi_2})
\end{array}
\right) \nonumber \\[7pt] 
&= U_0^2 \exp \left[- \frac{|\bs{\phi_1}|^2 + |\bs{\phi_2}|^2}{2\Lambda^2} \right] \left( \exp \left[ \frac{\bs{\phi_1}\cdot \bs{\phi_2}}{\Lambda^2} \right] - 1 - \frac{\bs{\phi_1}\cdot\bs{\phi_2}}{\Lambda^2} - \frac{(\bs{\phi_1}\cdot\bs{\phi_2})^2}{2\Lambda^4} - \frac{(\bs{\phi_1}\cdot\bs{\phi_2})^3}{6\Lambda^6} \right)
\label{slepian_covariance}
\end{align}
\normalsize
which, once again, is isotropic around the origin of the field.

\subsubsection{Probability distribution for the inflection point parameters}

We can extend the treatment for the eigenvalues of the hessian that we did for
the critical points to inflection points. The difference is that we will now impose
that one of the eigenvalues vanishes while the other one is positive. Furthermore
we will also impose that the gradient in the second eigenvalue direction also vanishes.
These conditions have to be included in the calculation of the PDF
of the parameters of the inflection points $(V_0,\eta_1,\lambda_2,{\bs \rho})$. Using a generalized version of the 
Kac-Rice procedure we arrive to,
\begin{align}
P_{\text{inf}} ~ dV_0~ d\lambda_2 ~d\eta_1 ~d{\bs \rho}= \mathcal{N} |\lambda_2|^2 |\rho_{111}| \ P \Big( V_0, \ \lambda_2 \ \left|  \ \lambda_1 = 0 \right. \Big) \ P \left(\eta_1, \rho_{ijk} \ \left| \ \eta_2 = 0 \right. \right)
\label{eq:pinf}
\end{align}
where
\begin{align}
	P \Big( V_0, \ \lambda_2 \ \left| \ \lambda_1 = 0 \right. \Big) \ dV_0 \ d\lambda_2 &= \mathcal{N} \exp \left[ - \frac{4V_0^2 - 2\Lambda^2 V_0 \lambda_2 - \Lambda^4 \lambda_2^2}{2U_0}  \right] \ dV_0 \ d\lambda_2 \\
P \left( \eta_1, \rho_{ijk} \ \left| \ \eta_2 = 0 \right. \right) \ d\eta_1 \ d\rho_{ijk} &= \nonumber \\
& \hspace{-100pt} \mathcal{N} \exp \left[ - \frac{\Lambda^2}{12 U_0^2} \left( 18 \eta_1^2 + 6 \Lambda^2 \eta_1 (\rho_{111} + \rho_{122}) + \Lambda^4 \sum_{i,j,k=1}^2 \rho_{ijk}^2 \right) \right] \ d\eta_1 \ d\rho_{ijk}
\end{align}
In (\ref{eq:pinf}), one of the $|\lambda_2|$ factors comes from the Jacobian of the variable change to the eigenbasis of the Hessian (though with $\lambda_1 = 0$); the remaining $|\lambda_2||\rho_{111}|$ factor is just the determinant appearing in Kac-Rice's expression. 

These last expressions can be used as in (\ref{eq:min_number}) to compute the expected number of inflection point per correlation volume $\Lambda^2$, which yields, for our choice of covariance function,
\begin{align}
\frac{\mathbb{E}(\#_{ip})}{\Lambda^2} &= \frac{\sqrt{5}-\sqrt{3}}{3\pi}.
\end{align}

\section{Numerical implementation and tests of the probability distributions \label{sec:plots}}
\label{sec:numerical}
\subsection{Generation of Gaussian random fields: Karhunen-Loève expansion}

In order to generate realizations of two-dimensional Gaussian random fields, we resorted to the so-called spectral or Karhunen-Loève decomposition, due to its mathematical and computational simplicity.  

Given a certain mean function $\mu (\vec{t})$, covariance function $C(\vec{t},\vec{s})$ and a discretized space $\lbrace \vec{t}_a \rbrace$ (where $a$ runs over all $n$ points in the lattice space) of a GRF, we can build the matrix $C_{ab}=C(\vec{t}_a,\vec{t}_b)$, which by construction is symmetric and positive definite; therefore, we can always decompose $C_{ab}$ as
\begin{align}
C= U \Lambda U^T
\end{align} 
where $\Lambda = \text{diag}(\lambda_1, \ldots, \lambda_n)$ is the diagonal eigenvalue matrix, consisting of non-negative entries, and $U$ is constructed by inserting all eigenvectors along its rows. Since $\Lambda>0$, we can further decompose $C$ as
\begin{align}
C= U \sqrt{\Lambda} \sqrt{\Lambda} U^T = \left( U \sqrt{\Lambda} \right) \left( U \sqrt{\Lambda} \right)^T = L \ L^T.
\end{align}
This procedure is tantamount to performing a Cholesky decomposition \cite{press2007numerical} on $C$; which is by far the most expensive step in this algorithm, in terms of computational cost. 

Once we have computed $L$, constructing the GRF on the discretized space is straightforward. We only need to construct a random vector $\vec{\xi}$ of length $n$ whose entries are \emph{independently} distributed as Gaussian variables of zero mean and unit variance, and introduce the following variables:
\begin{align}
V_a = \mu_a + L_{ab} \xi_b,
\end{align} 
where $\mu_a = \mu (\vec{t}_a)$. It can be easily shown that this gives the correct  correlations  among the values of the GRF evaluated at different points $\vec{t}_a$, 
\begin{align}
\langle (V_a - \mu_a)(V_b - \mu_b) \rangle &= \langle L_{ac} \xi_c L_{bd} \xi_d \rangle = L_{ac} L_{bd} \langle \xi_c \xi_d \rangle \nonumber \\ 
&= L_{ac} L_{bd} \delta_{cd} = L_{ac} L_{bc} = L_{ac} L^T_{cb} = (LL^T)_{ab} = C_{ab} =C(\vec{t}_a,\vec{t}_b).
\end{align}

The main advantage of using this procedure to generate GRFs is that the main computationally costly step, constructing the $L$ matrix, needs to be performed only once. The rest of the algorithm is highly trivial from this perspective and allows for further simplification, as we have seen. 

\subsection{Numerical evaluations of Critical points}

\begin{figure}
	\centering
	\subfloat[]{
		\includegraphics[width=0.45\textwidth]{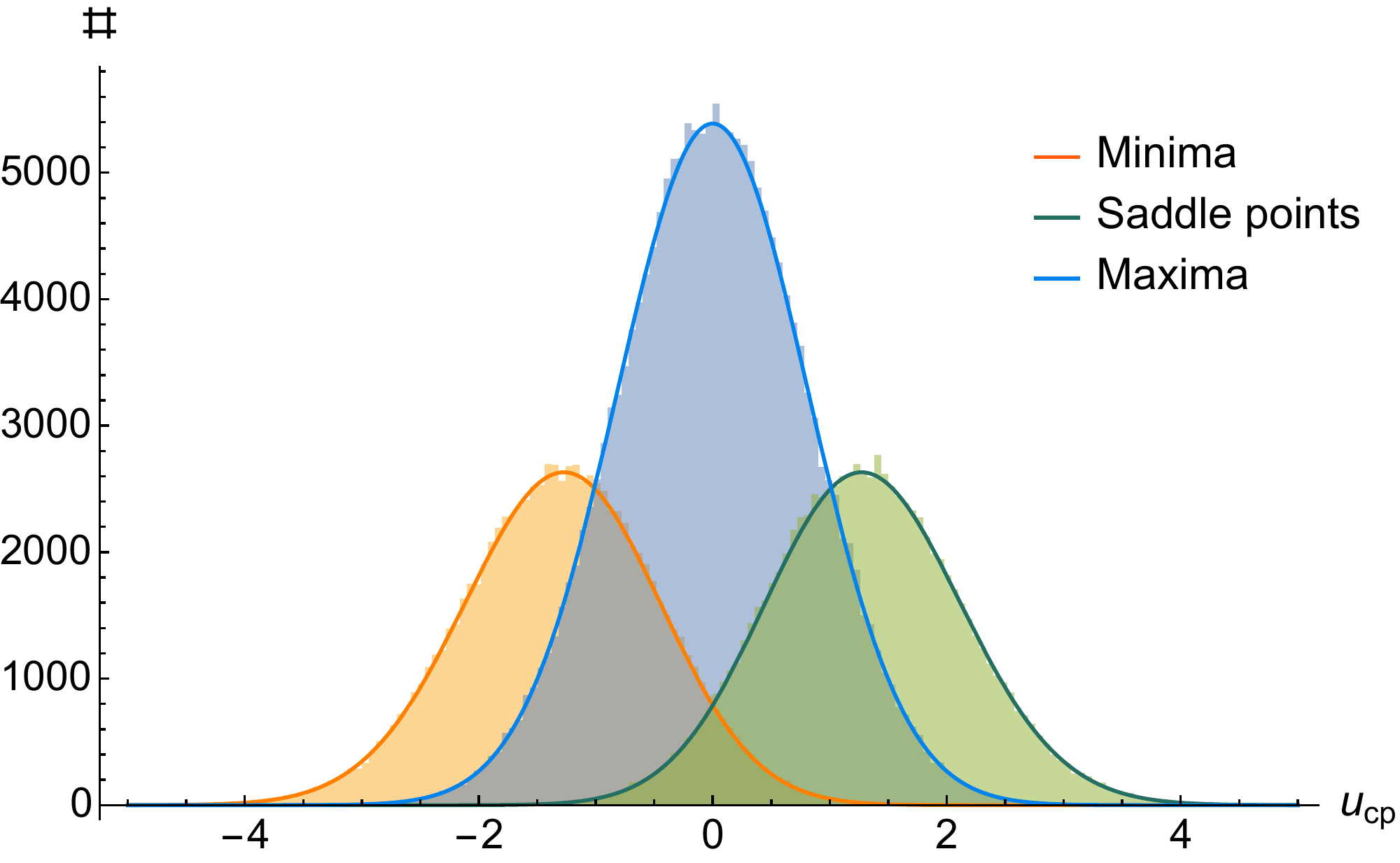}
	}
	\hfill	
	\subfloat[]{
		\includegraphics[width=0.45\textwidth]{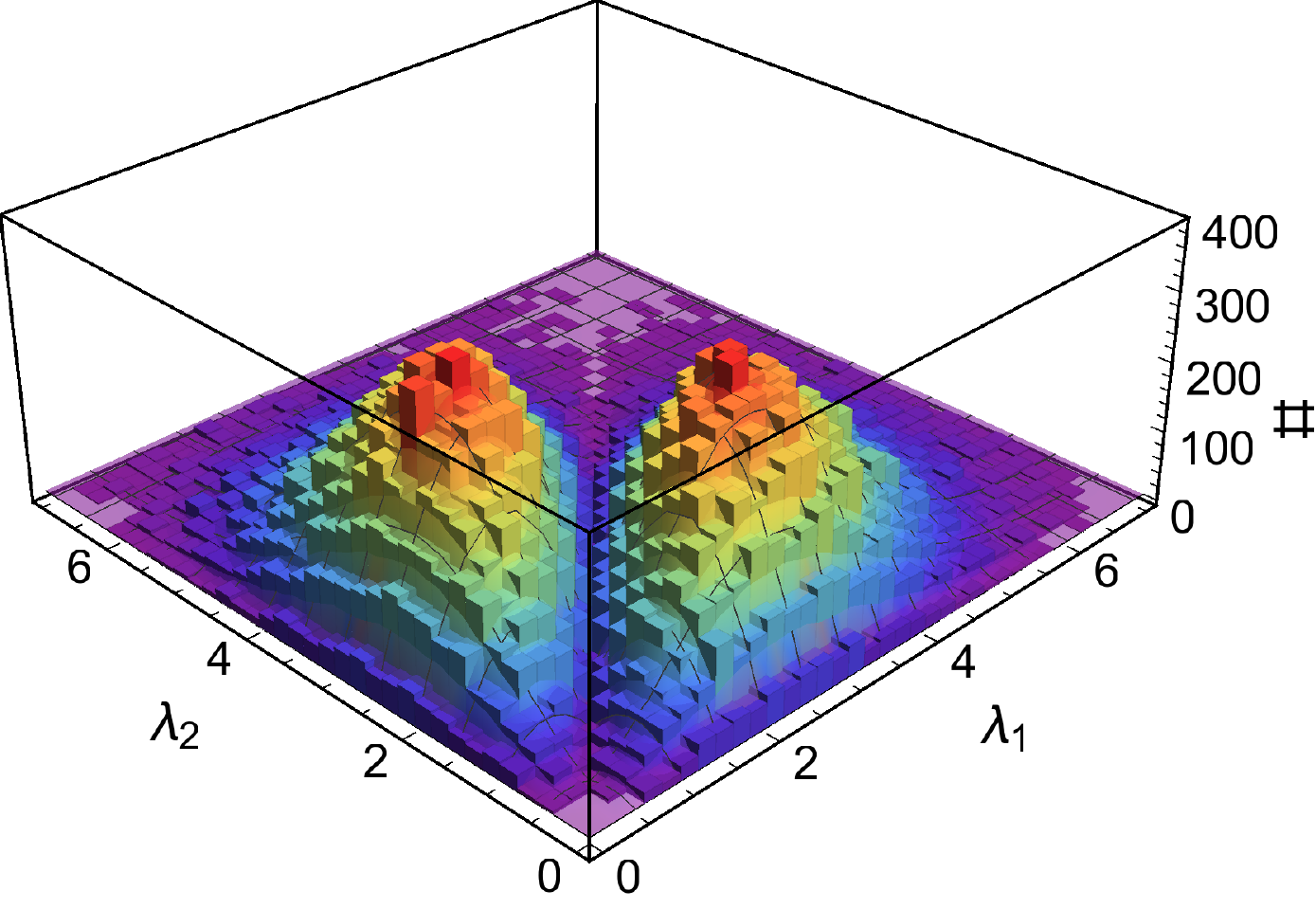}
	}
	\\[7pt]
	\subfloat[]{
		\includegraphics[width=0.45\textwidth]{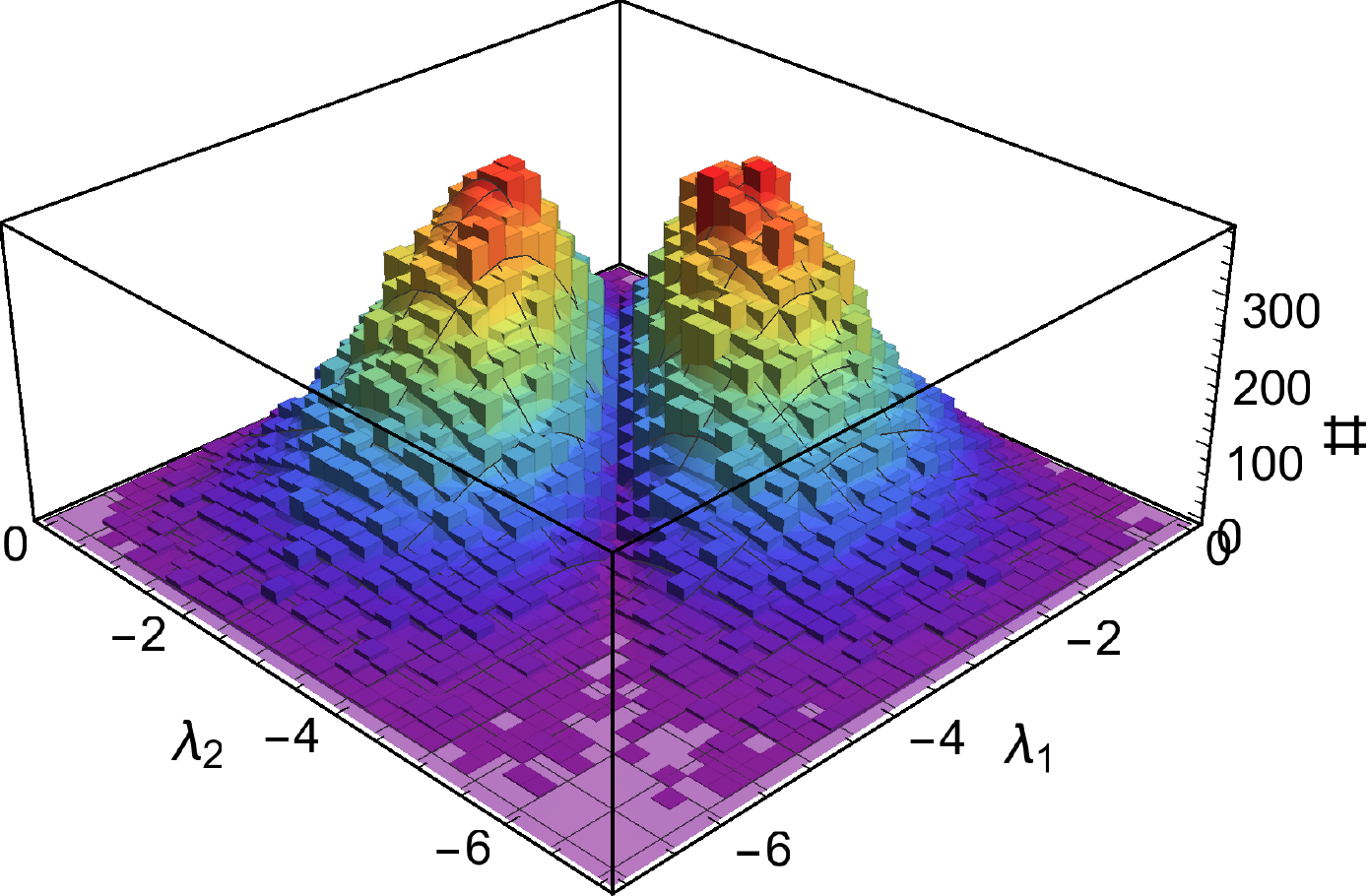}
	}
	\hfill	
	\subfloat[]{
		\includegraphics[width=0.45\textwidth]{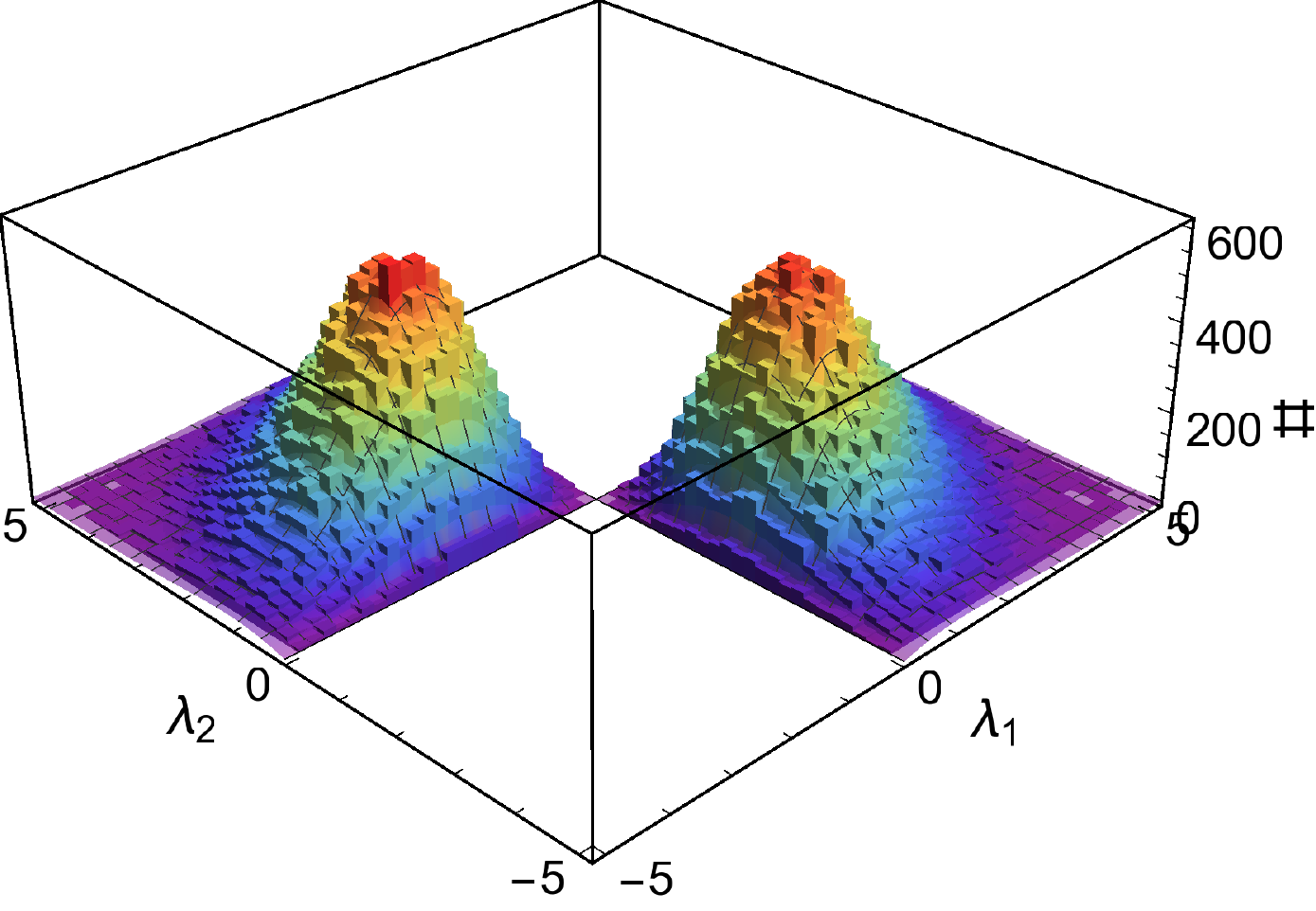}
	}
	\caption{Histograms of (a) heights and (b-d) eigenvalues for critical points, normalized to expected values, from a $10^5 \Lambda^2$ GRF. Distributions (\ref{u_sp}) and (\ref{eq:lambda_dist_cp}) are plotted along with their respective histograms, normalized with respect to (\ref{eq:min_number}).}
	\label{cp_histograms}
\end{figure}

Using the expressions above we can compute the normalized distribution of heights of 
 minima, maxima and saddle points for a $2d$ GRF, 
\small
\begin{align}
P_{u,min} \ du &= \frac{\sqrt{3}}{4\pi U_0} e^{-u^2/U_0^2} \left( -\frac{2u}{U_0} + 2\sqrt{\pi} \ e^{u^2/4U_0^2} \text{ erfc} \left[\frac{u}{2 U_0}\right] + \sqrt{2\pi} \left( \frac{u^2}{U_0^2}-1 \right) e^{u^2/2U_0^2} \text{ erfc} \left[ \frac{u}{\sqrt{2}U_0} \right] \right) \ du  \nonumber \\
P_{u,max} \ du &= \frac{\sqrt{3}}{4\pi U_0} e^{-u^2/U_0^2} \left( \frac{2u}{U_0} + 2\sqrt{\pi} \ e^{u^2/4U_0^2} \text{ erfc} \left[-\frac{u}{2 U_0}\right] + \sqrt{2\pi} \left( \frac{u^2}{U_0^2}-1 \right) e^{u^2/2U_0^2} \text{ erfc} \left[- \frac{u}{\sqrt{2}U_0} \right] \right) \ du \nonumber \\
P_{u,sp} ~\ du &= \frac{\sqrt{3}}{2\sqrt{\pi} U_0} \exp \left[ - \frac{3u^2}{4U_0^2} \right] \label{u_sp}~.
\end{align} 
\normalsize
Furthermore, we can also compute the marginal distribution for the Hessian eigenvalues at critical points regardless
of their height. This distribution is given by,
\small
\begin{align}
P_{sp,\lambda_i} \ d\lambda_1 \ d\lambda_2 &= \sqrt{\frac{3}{\pi}} \frac{\Lambda^{10}}{32 U_0^5} \ \prod_{i=1}^2 \left( |\lambda_i| \exp \left[ - \frac{\Lambda^4}{8U_0^2} \lambda_i^2 \right] \right) \ |\lambda_1-\lambda_2| \exp \left[ - \frac{\Lambda^4}{16U_0^2} (\lambda_1-\lambda_2)^2 \right] \ d\lambda_1 \ d\lambda_2 \nonumber\\
	&= \frac{1}{2} P_{min,max,\lambda_i} \ d\lambda_1 \ d\lambda_2.
\label{eq:lambda_dist_cp}
\end{align}
\normalsize
We have checked the distributions above with numerical realizations of unconstrained Gaussian random fields in Mathematica. Regarding the heights of critical points, the numerical results fit the analytical prediction perfectly, as shown in figure \ref{cp_histograms}(a). 

As for the eigenvalue distribution, Fig. \ref{cp_histograms}(b)-(d) shows that the histograms fit the analytical predictions 
perfectly once again. An important feature of these distributions is the fact that critical points with one of the eigenvalues 
close to zero or both eigenvalues close to each other are very rare; this effect (referred to as \emph{eigenvalue repulsion}) is 
a direct consequence of the presence of the Vandermonde determinant in the distributions, as well as the Jacobian of the 
gradient field in the Kac-Rice formula. 

\subsection{Numerical evaluations of Inflection points}

\begin{figure}
	\centering
	\subfloat[]{
		\includegraphics[width=0.45\textwidth]{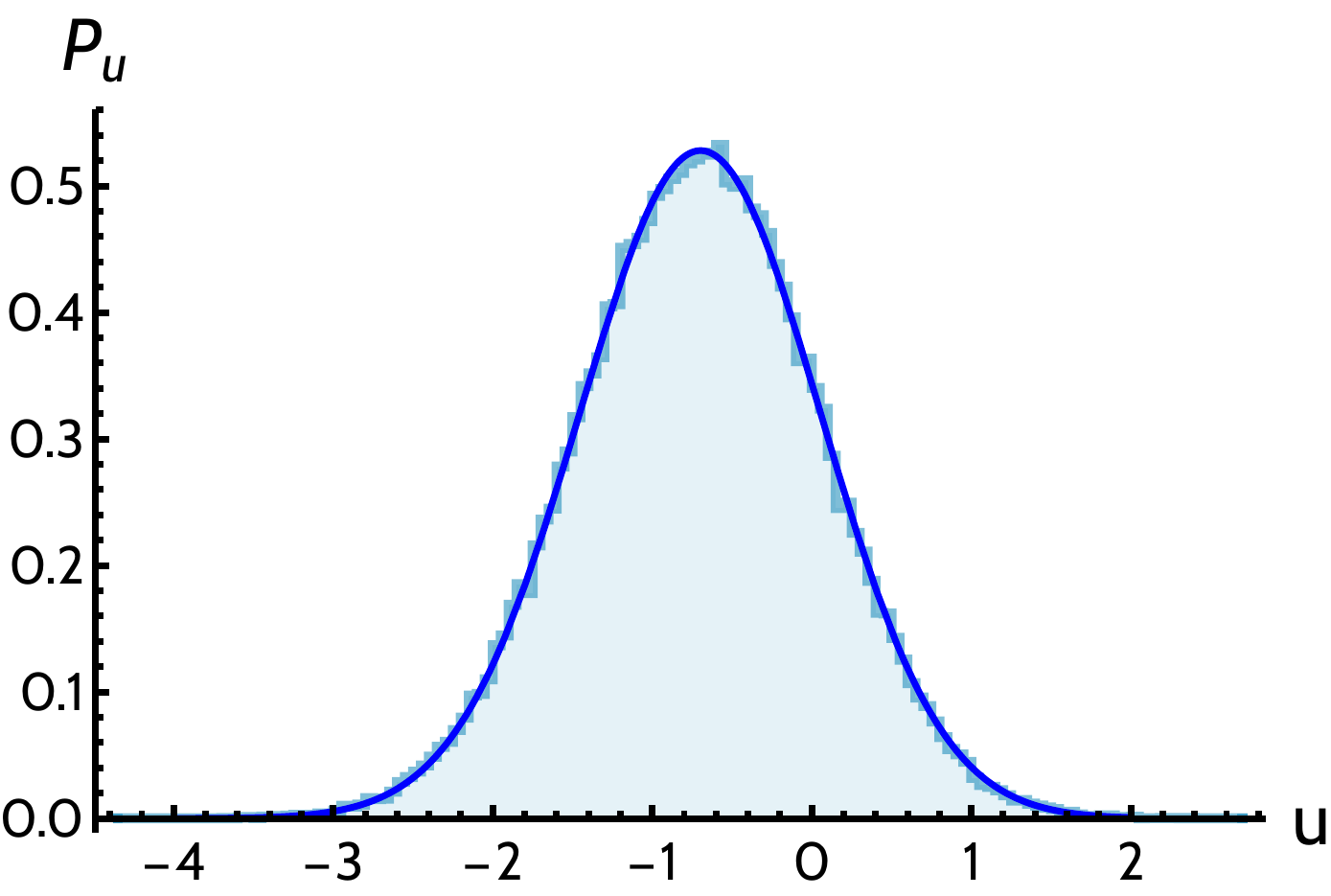}
	}
	\hfill	
	\subfloat[]{
		\includegraphics[width=0.45\textwidth]{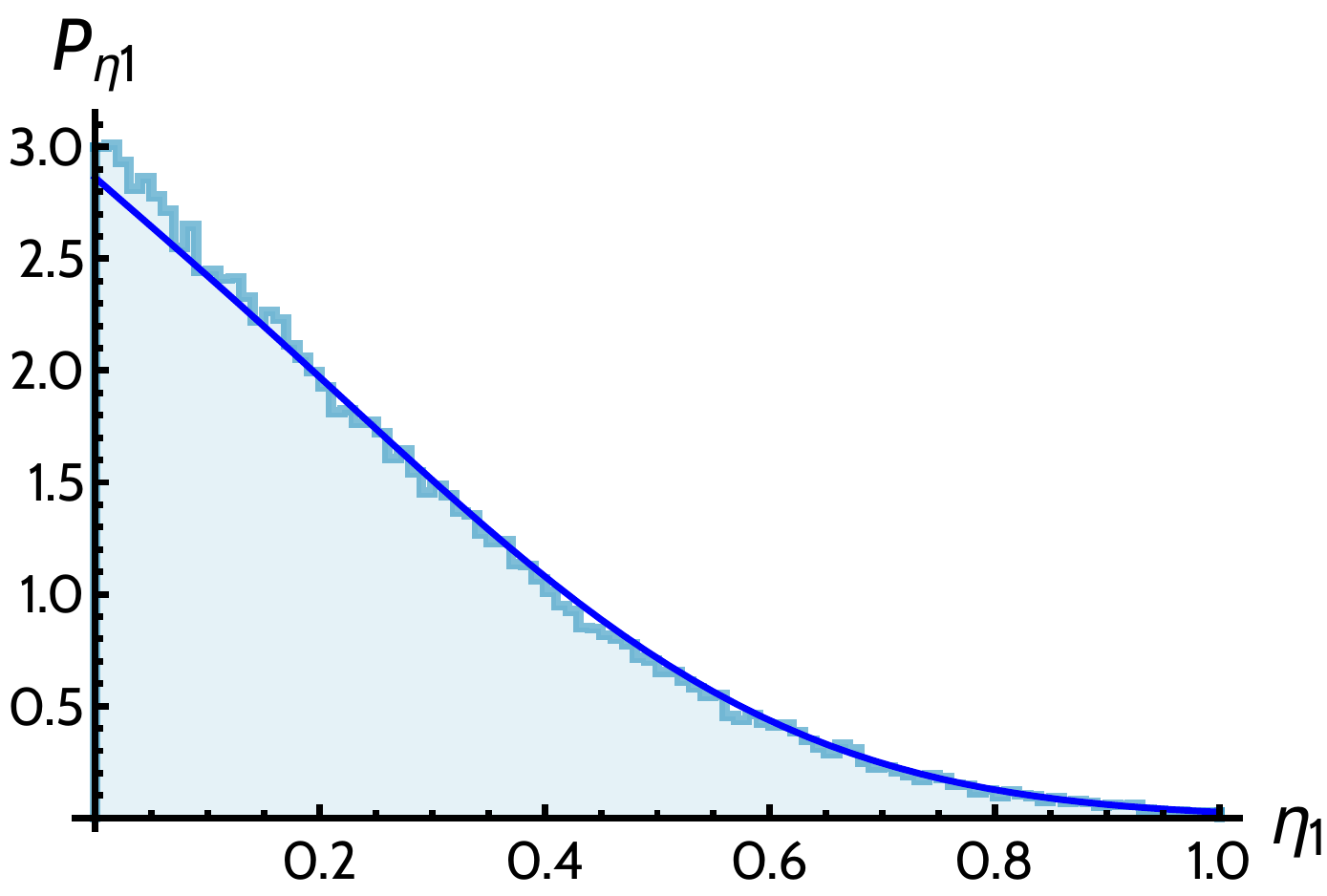}
	}
	\\[7pt]
	\subfloat[]{
		\includegraphics[width=0.45\textwidth]{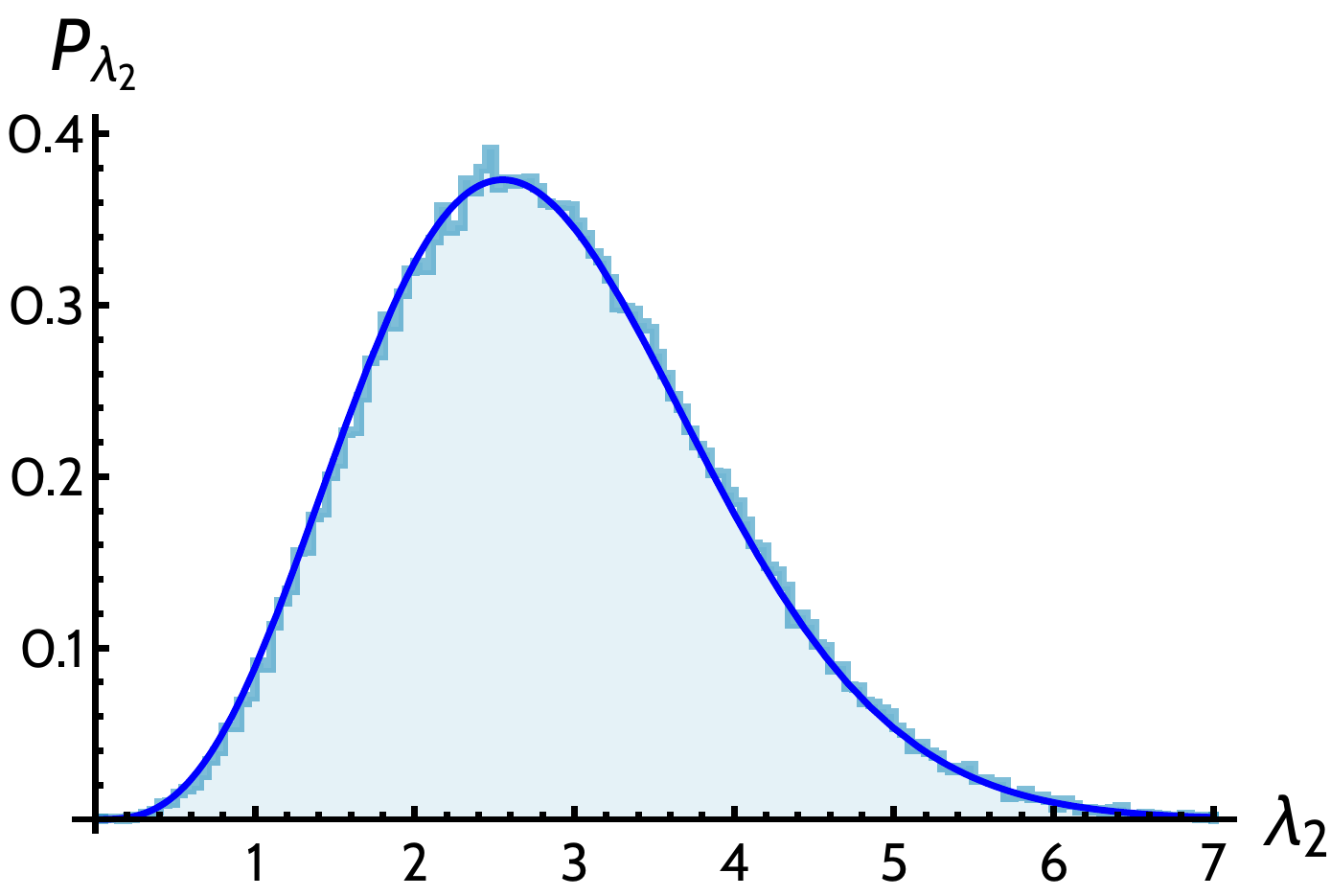}
	}
	\hfill	
	\subfloat[]{
		\includegraphics[width=0.45\textwidth]{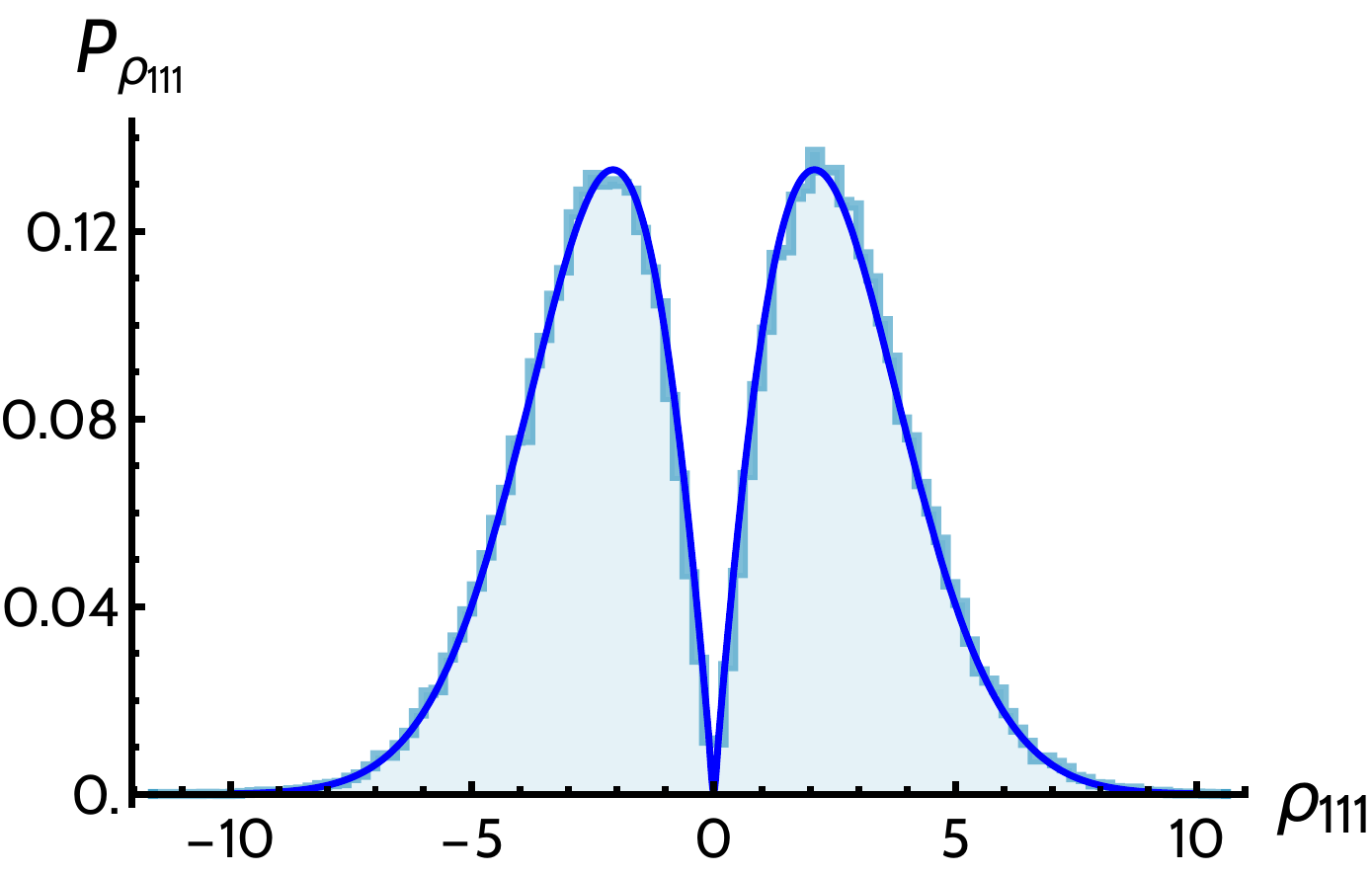}
	}
	\caption{Normalized histograms of (a) height (b) $\eta_1$ (c) $\lambda_2$ (d) $\rho_{111}$ for inflection points and their expected PDFs (\ref{eq:u_ip_dist})-(\ref{eq:r111_ip_dist}), constrained by the condition $\lambda_2/\eta_1> 4$.}
	\label{fig:ip_dist}
\end{figure}

Using the results given above, we can obtain the following distributions for the parameters of the
inflection points in a typical GRF.

\begin{align}
P_{u} \ du~~ &=  \frac{3\sqrt{3}}{16 \pi U_0^3} \ \exp \left[ -\frac{u^2}{U_0^2} \right] \ \left( -2U_0 u + \sqrt{\pi} (u^2 + 2U_0^2) \exp \left[ \frac{u^2}{4U_0^2} \right] \text{erfc}\left[ \frac{u}{2U_0} \right] \right) \ du \label{eq:u_ip_dist} \\
P_{\lambda_2} \ d\lambda_2 &= \sqrt{\frac{3}{\pi}} \frac{3}{16} \frac{\Lambda^6}{U_0^3} \ \lambda_2^2 \ \exp \left[ - \frac{3\Lambda^4 }{16 U_0^2} \lambda_2^2 \right] d\lambda_2  \\
P_{\eta_1} \ d\eta_1 ~&= \frac{(3 + \sqrt{15})\Lambda}{12 U_0^2} \ \exp \left[-\frac{5\Lambda^2}{4U_0^2}\eta_1^2 \right] \left( \sqrt{\frac{12}{\pi}} U_0 - 3\Lambda \ |\eta_1| \ \exp \left[ \frac{3\Lambda^2}{4 U_0^2} \eta_1^2 \right] \text{erfc}\left[ \frac{\sqrt{3} \Lambda}{2U_0} \ |\eta_1| \right] \right) \ d\eta_1  \\
P_{\rho_{111}} \ d\rho_{111} &= \frac{(5 + \sqrt{15}) \Lambda^6}{60 U_0^2} \ |\rho_{111}| \ \exp \left[ - \frac{\Lambda^6}{30 U_0^2} \rho_{111}^2 \right] \ \text{erfc} \left[ \frac{\Lambda^3}{2\sqrt{5} U_0} |\rho_{111}| \right] \ d\rho_{111} \label{eq:r111_ip_dist}
\end{align}
where the complementary error function is defined as
\begin{align*}
\text{erfc} (x) = \frac{2}{\sqrt{\pi}} \int_{x}^{\infty} dt \ e^{-t^2}.
\end{align*} 

Once again, we found these expressions to be fully consistent with the numerical results, as shown in figure \ref{fig:ip_dist}.

In order to find inflection points in our numerically generated potentials, we looked for roots of the system
\begin{align}
	\left\lbrace
	\begin{array}{l}
	\bs{\eta}^T \ \mathcal{H} \ \bs{\eta} \\
	\bs{\eta}^T \ \mathcal{H} \ \bs{\eta}_{\perp} 
	\end{array}
	\right.
\label{find_ip}
\end{align}
where $\mathcal{H}$ is the Hessian matrix, $\bs{\eta}^T = ( \eta_1, \eta_2 )$ represents the gradient at any point of the field and $\bs{\eta}^T_{\perp} = ( - \eta_2, \eta_1 )$. It can be easily shown that simultaneous roots of Eq. (\ref{find_ip}) are either critical or inflection points.

Finding inflection points numerically is quite tricky and the algorithm sometimes incorporates spurious points
that, upon further study, are proven to be fictitious inflection points. In order to make a proper comparison to the 
general expressions we have found analytically and avoid the inclusion of those spurious inflection points, we only 
considered those points which satisfied  $\lambda_2/\eta_1> 4$ . This cut removes around 30\% of the potential 
inflection points. Note that even though we might be removing a portion of real inflection points, the distributions 
above are still in perfect agreement with the analytic computations.

\bibliography{slepian_models}

\begin{thebibliography}{53}%
\makeatletter
\providecommand \@ifxundefined [1]{%
 \@ifx{#1\undefined}
}%
\providecommand \@ifnum [1]{%
 \ifnum #1\expandafter \@firstoftwo
 \else \expandafter \@secondoftwo
 \fi
}%
\providecommand \@ifx [1]{%
 \ifx #1\expandafter \@firstoftwo
 \else \expandafter \@secondoftwo
 \fi
}%
\providecommand \natexlab [1]{#1}%
\providecommand \enquote  [1]{``#1''}%
\providecommand \bibnamefont  [1]{#1}%
\providecommand \bibfnamefont [1]{#1}%
\providecommand \citenamefont [1]{#1}%
\providecommand \href@noop [0]{\@secondoftwo}%
\providecommand \href [0]{\begingroup \@sanitize@url \@href}%
\providecommand \@href[1]{\@@startlink{#1}\@@href}%
\providecommand \@@href[1]{\endgroup#1\@@endlink}%
\providecommand \@sanitize@url [0]{\catcode `\\12\catcode `\$12\catcode
  `\&12\catcode `\#12\catcode `\^12\catcode `\_12\catcode `\%12\relax}%
\providecommand \@@startlink[1]{}%
\providecommand \@@endlink[0]{}%
\providecommand \url  [0]{\begingroup\@sanitize@url \@url }%
\providecommand \@url [1]{\endgroup\@href {#1}{\urlprefix }}%
\providecommand \urlprefix  [0]{URL }%
\providecommand \Eprint [0]{\href }%
\providecommand \doibase [0]{http://dx.doi.org/}%
\providecommand \selectlanguage [0]{\@gobble}%
\providecommand \bibinfo  [0]{\@secondoftwo}%
\providecommand \bibfield  [0]{\@secondoftwo}%
\providecommand \translation [1]{[#1]}%
\providecommand \BibitemOpen [0]{}%
\providecommand \bibitemStop [0]{}%
\providecommand \bibitemNoStop [0]{.\EOS\space}%
\providecommand \EOS [0]{\spacefactor3000\relax}%
\providecommand \BibitemShut  [1]{\csname bibitem#1\endcsname}%
\let\auto@bib@innerbib\@empty
%</preamble>
\bibitem [{\citenamefont {Ashok}\ and\ \citenamefont
  {Douglas}(2004)}]{Ashok:2003gk}%
  \BibitemOpen
  \bibfield  {author} {\bibinfo {author} {\bibfnamefont {S.}~\bibnamefont
  {Ashok}}\ and\ \bibinfo {author} {\bibfnamefont {M.~R.}\ \bibnamefont
  {Douglas}},\ }\href {\doibase 10.1088/1126-6708/2004/01/060} {\bibfield
  {journal} {\bibinfo  {journal} {JHEP}\ }\textbf {\bibinfo {volume} {01}},\
  \bibinfo {pages} {060} (\bibinfo {year} {2004})},\ \Eprint
  {http://arxiv.org/abs/hep-th/0307049} {arXiv:hep-th/0307049 [hep-th]}
  \BibitemShut {NoStop}%
%%CITATION = HEP-TH/0307049;%%
\bibitem [{\citenamefont {Denef}\ and\ \citenamefont
  {Douglas}(2004)}]{Denef:2004ze}%
  \BibitemOpen
  \bibfield  {author} {\bibinfo {author} {\bibfnamefont {F.}~\bibnamefont
  {Denef}}\ and\ \bibinfo {author} {\bibfnamefont {M.~R.}\ \bibnamefont
  {Douglas}},\ }\href {\doibase 10.1088/1126-6708/2004/05/072} {\bibfield
  {journal} {\bibinfo  {journal} {JHEP}\ }\textbf {\bibinfo {volume} {05}},\
  \bibinfo {pages} {072} (\bibinfo {year} {2004})},\ \Eprint
  {http://arxiv.org/abs/hep-th/0404116} {arXiv:hep-th/0404116 [hep-th]}
  \BibitemShut {NoStop}%
%%CITATION = HEP-TH/0404116;%%
\bibitem [{\citenamefont {Denef}\ and\ \citenamefont
  {Douglas}(2005)}]{Denef:2004cf}%
  \BibitemOpen
  \bibfield  {author} {\bibinfo {author} {\bibfnamefont {F.}~\bibnamefont
  {Denef}}\ and\ \bibinfo {author} {\bibfnamefont {M.~R.}\ \bibnamefont
  {Douglas}},\ }\href {\doibase 10.1088/1126-6708/2005/03/061} {\bibfield
  {journal} {\bibinfo  {journal} {JHEP}\ }\textbf {\bibinfo {volume} {03}},\
  \bibinfo {pages} {061} (\bibinfo {year} {2005})},\ \Eprint
  {http://arxiv.org/abs/hep-th/0411183} {arXiv:hep-th/0411183 [hep-th]}
  \BibitemShut {NoStop}%
%%CITATION = HEP-TH/0411183;%%
\bibitem [{\citenamefont {Tegmark}(2005)}]{Tegmark:2004qd}%
  \BibitemOpen
  \bibfield  {author} {\bibinfo {author} {\bibfnamefont {M.}~\bibnamefont
  {Tegmark}},\ }\href {\doibase 10.1088/1475-7516/2005/04/001} {\bibfield
  {journal} {\bibinfo  {journal} {JCAP}\ }\textbf {\bibinfo {volume} {0504}},\
  \bibinfo {pages} {001} (\bibinfo {year} {2005})},\ \Eprint
  {http://arxiv.org/abs/astro-ph/0410281} {arXiv:astro-ph/0410281 [astro-ph]}
  \BibitemShut {NoStop}%
%%CITATION = ASTRO-PH/0410281;%%
\bibitem [{\citenamefont {Aazami}\ and\ \citenamefont
  {Easther}(2006)}]{Aazami:2005jf}%
  \BibitemOpen
  \bibfield  {author} {\bibinfo {author} {\bibfnamefont {A.}~\bibnamefont
  {Aazami}}\ and\ \bibinfo {author} {\bibfnamefont {R.}~\bibnamefont
  {Easther}},\ }\href {\doibase 10.1088/1475-7516/2006/03/013} {\bibfield
  {journal} {\bibinfo  {journal} {JCAP}\ }\textbf {\bibinfo {volume} {0603}},\
  \bibinfo {pages} {013} (\bibinfo {year} {2006})},\ \Eprint
  {http://arxiv.org/abs/hep-th/0512050} {arXiv:hep-th/0512050 [hep-th]}
  \BibitemShut {NoStop}%
%%CITATION = HEP-TH/0512050;%%
\bibitem [{\citenamefont {Easther}\ and\ \citenamefont
  {McAllister}(2006)}]{Easther:2005zr}%
  \BibitemOpen
  \bibfield  {author} {\bibinfo {author} {\bibfnamefont {R.}~\bibnamefont
  {Easther}}\ and\ \bibinfo {author} {\bibfnamefont {L.}~\bibnamefont
  {McAllister}},\ }\href {\doibase 10.1088/1475-7516/2006/05/018} {\bibfield
  {journal} {\bibinfo  {journal} {JCAP}\ }\textbf {\bibinfo {volume} {0605}},\
  \bibinfo {pages} {018} (\bibinfo {year} {2006})},\ \Eprint
  {http://arxiv.org/abs/hep-th/0512102} {arXiv:hep-th/0512102 [hep-th]}
  \BibitemShut {NoStop}%
%%CITATION = HEP-TH/0512102;%%
\bibitem [{\citenamefont {Marsh}\ \emph {et~al.}(2012)\citenamefont {Marsh},
  \citenamefont {McAllister},\ and\ \citenamefont {Wrase}}]{Marsh:2011aa}%
  \BibitemOpen
  \bibfield  {author} {\bibinfo {author} {\bibfnamefont {D.}~\bibnamefont
  {Marsh}}, \bibinfo {author} {\bibfnamefont {L.}~\bibnamefont {McAllister}}, \
  and\ \bibinfo {author} {\bibfnamefont {T.}~\bibnamefont {Wrase}},\ }\href
  {\doibase 10.1007/JHEP03(2012)102} {\bibfield  {journal} {\bibinfo  {journal}
  {JHEP}\ }\textbf {\bibinfo {volume} {03}},\ \bibinfo {pages} {102} (\bibinfo
  {year} {2012})},\ \Eprint {http://arxiv.org/abs/1112.3034} {arXiv:1112.3034
  [hep-th]} \BibitemShut {NoStop}%
%%CITATION = ARXIV:1112.3034;%%
\bibitem [{\citenamefont {Marsh}\ \emph {et~al.}(2013)\citenamefont {Marsh},
  \citenamefont {McAllister}, \citenamefont {Pajer},\ and\ \citenamefont
  {Wrase}}]{Marsh:2013qca}%
  \BibitemOpen
  \bibfield  {author} {\bibinfo {author} {\bibfnamefont {M.~C.~D.}\
  \bibnamefont {Marsh}}, \bibinfo {author} {\bibfnamefont {L.}~\bibnamefont
  {McAllister}}, \bibinfo {author} {\bibfnamefont {E.}~\bibnamefont {Pajer}}, \
  and\ \bibinfo {author} {\bibfnamefont {T.}~\bibnamefont {Wrase}},\ }\href
  {\doibase 10.1088/1475-7516/2013/11/040} {\bibfield  {journal} {\bibinfo
  {journal} {JCAP}\ }\textbf {\bibinfo {volume} {1311}},\ \bibinfo {pages}
  {040} (\bibinfo {year} {2013})},\ \Eprint {http://arxiv.org/abs/1307.3559}
  {arXiv:1307.3559 [hep-th]} \BibitemShut {NoStop}%
%%CITATION = ARXIV:1307.3559;%%
\bibitem [{\citenamefont {Bachlechner}(2014)}]{Bachlechner:2014rqa}%
  \BibitemOpen
  \bibfield  {author} {\bibinfo {author} {\bibfnamefont {T.~C.}\ \bibnamefont
  {Bachlechner}},\ }\href {\doibase 10.1007/JHEP04(2014)054} {\bibfield
  {journal} {\bibinfo  {journal} {JHEP}\ }\textbf {\bibinfo {volume} {04}},\
  \bibinfo {pages} {054} (\bibinfo {year} {2014})},\ \Eprint
  {http://arxiv.org/abs/1401.6187} {arXiv:1401.6187 [hep-th]} \BibitemShut
  {NoStop}%
%%CITATION = ARXIV:1401.6187;%%
\bibitem [{\citenamefont {Sousa}\ and\ \citenamefont
  {Ortiz}(2015)}]{Sousa:2014qza}%
  \BibitemOpen
  \bibfield  {author} {\bibinfo {author} {\bibfnamefont {K.}~\bibnamefont
  {Sousa}}\ and\ \bibinfo {author} {\bibfnamefont {P.}~\bibnamefont {Ortiz}},\
  }\href {\doibase 10.1088/1475-7516/2015/02/017} {\bibfield  {journal}
  {\bibinfo  {journal} {JCAP}\ }\textbf {\bibinfo {volume} {1502}},\ \bibinfo
  {pages} {017} (\bibinfo {year} {2015})},\ \Eprint
  {http://arxiv.org/abs/1408.6521} {arXiv:1408.6521 [hep-th]} \BibitemShut
  {NoStop}%
%%CITATION = ARXIV:1408.6521;%%
\bibitem [{\citenamefont {Pedro}\ and\ \citenamefont
  {Westphal}(2017)}]{Pedro:2016sli}%
  \BibitemOpen
  \bibfield  {author} {\bibinfo {author} {\bibfnamefont {F.~G.}\ \bibnamefont
  {Pedro}}\ and\ \bibinfo {author} {\bibfnamefont {A.}~\bibnamefont
  {Westphal}},\ }\href {\doibase 10.1007/JHEP03(2017)163} {\bibfield  {journal}
  {\bibinfo  {journal} {JHEP}\ }\textbf {\bibinfo {volume} {03}},\ \bibinfo
  {pages} {163} (\bibinfo {year} {2017})},\ \Eprint
  {http://arxiv.org/abs/1611.07059} {arXiv:1611.07059 [hep-th]} \BibitemShut
  {NoStop}%
%%CITATION = ARXIV:1611.07059;%%
\bibitem [{\citenamefont {Freivogel}\ \emph {et~al.}(2016)\citenamefont
  {Freivogel}, \citenamefont {Gobbetti}, \citenamefont {Pajer},\ and\
  \citenamefont {Yang}}]{Freivogel:2016kxc}%
  \BibitemOpen
  \bibfield  {author} {\bibinfo {author} {\bibfnamefont {B.}~\bibnamefont
  {Freivogel}}, \bibinfo {author} {\bibfnamefont {R.}~\bibnamefont {Gobbetti}},
  \bibinfo {author} {\bibfnamefont {E.}~\bibnamefont {Pajer}}, \ and\ \bibinfo
  {author} {\bibfnamefont {I.-S.}\ \bibnamefont {Yang}},\ }\href@noop {} {\
  (\bibinfo {year} {2016})},\ \Eprint {http://arxiv.org/abs/1608.00041}
  {arXiv:1608.00041 [hep-th]} \BibitemShut {NoStop}%
%%CITATION = ARXIV:1608.00041;%%
\bibitem [{\citenamefont {Masoumi}\ \emph
  {et~al.}(2017{\natexlab{a}})\citenamefont {Masoumi}, \citenamefont
  {Vilenkin},\ and\ \citenamefont {Yamada}}]{Masoumi:2016eag}%
  \BibitemOpen
  \bibfield  {author} {\bibinfo {author} {\bibfnamefont {A.}~\bibnamefont
  {Masoumi}}, \bibinfo {author} {\bibfnamefont {A.}~\bibnamefont {Vilenkin}}, \
  and\ \bibinfo {author} {\bibfnamefont {M.}~\bibnamefont {Yamada}},\ }\href
  {\doibase 10.1088/1475-7516/2017/05/053} {\bibfield  {journal} {\bibinfo
  {journal} {JCAP}\ }\textbf {\bibinfo {volume} {1705}},\ \bibinfo {pages}
  {053} (\bibinfo {year} {2017}{\natexlab{a}})},\ \Eprint
  {http://arxiv.org/abs/1612.03960} {arXiv:1612.03960 [hep-th]} \BibitemShut
  {NoStop}%
%%CITATION = ARXIV:1612.03960;%%
\bibitem [{\citenamefont {Masoumi}\ \emph
  {et~al.}(2017{\natexlab{b}})\citenamefont {Masoumi}, \citenamefont
  {Vilenkin},\ and\ \citenamefont {Yamada}}]{Masoumi:2017gmh}%
  \BibitemOpen
  \bibfield  {author} {\bibinfo {author} {\bibfnamefont {A.}~\bibnamefont
  {Masoumi}}, \bibinfo {author} {\bibfnamefont {A.}~\bibnamefont {Vilenkin}}, \
  and\ \bibinfo {author} {\bibfnamefont {M.}~\bibnamefont {Yamada}},\ }\href
  {\doibase 10.1088/1475-7516/2017/07/003} {\bibfield  {journal} {\bibinfo
  {journal} {JCAP}\ }\textbf {\bibinfo {volume} {1707}},\ \bibinfo {pages}
  {003} (\bibinfo {year} {2017}{\natexlab{b}})},\ \Eprint
  {http://arxiv.org/abs/1704.06994} {arXiv:1704.06994 [hep-th]} \BibitemShut
  {NoStop}%
%%CITATION = ARXIV:1704.06994;%%
\bibitem [{\citenamefont {Bjorkmo}\ and\ \citenamefont
  {Marsh}(2018{\natexlab{a}})}]{Bjorkmo:2018txh}%
  \BibitemOpen
  \bibfield  {author} {\bibinfo {author} {\bibfnamefont {T.}~\bibnamefont
  {Bjorkmo}}\ and\ \bibinfo {author} {\bibfnamefont {M.~C.~D.}\ \bibnamefont
  {Marsh}},\ }\href@noop {} {\  (\bibinfo {year} {2018}{\natexlab{a}})},\
  \Eprint {http://arxiv.org/abs/1805.03117} {arXiv:1805.03117 [astro-ph.CO]}
  \BibitemShut {NoStop}%
%%CITATION = ARXIV:1805.03117;%%
\bibitem [{\citenamefont {Slepian}(1962)}]{slepian1962one}%
  \BibitemOpen
  \bibfield  {author} {\bibinfo {author} {\bibfnamefont {D.}~\bibnamefont
  {Slepian}},\ }\href@noop {} {\bibfield  {journal} {\bibinfo  {journal} {Bell
  System Technical Journal}\ }\textbf {\bibinfo {volume} {41}},\ \bibinfo
  {pages} {463} (\bibinfo {year} {1962})}\BibitemShut {NoStop}%
\bibitem [{\citenamefont {Bachlechner}\ \emph {et~al.}(2013)\citenamefont
  {Bachlechner}, \citenamefont {Marsh}, \citenamefont {McAllister},\ and\
  \citenamefont {Wrase}}]{Bachlechner:2012at}%
  \BibitemOpen
  \bibfield  {author} {\bibinfo {author} {\bibfnamefont {T.~C.}\ \bibnamefont
  {Bachlechner}}, \bibinfo {author} {\bibfnamefont {D.}~\bibnamefont {Marsh}},
  \bibinfo {author} {\bibfnamefont {L.}~\bibnamefont {McAllister}}, \ and\
  \bibinfo {author} {\bibfnamefont {T.}~\bibnamefont {Wrase}},\ }\href
  {\doibase 10.1007/JHEP01(2013)136} {\bibfield  {journal} {\bibinfo  {journal}
  {JHEP}\ }\textbf {\bibinfo {volume} {01}},\ \bibinfo {pages} {136} (\bibinfo
  {year} {2013})},\ \Eprint {http://arxiv.org/abs/1207.2763} {arXiv:1207.2763
  [hep-th]} \BibitemShut {NoStop}%
%%CITATION = ARXIV:1207.2763;%%
\bibitem [{\citenamefont {Easther}\ \emph {et~al.}(2016)\citenamefont
  {Easther}, \citenamefont {Guth},\ and\ \citenamefont
  {Masoumi}}]{Easther:2016ire}%
  \BibitemOpen
  \bibfield  {author} {\bibinfo {author} {\bibfnamefont {R.}~\bibnamefont
  {Easther}}, \bibinfo {author} {\bibfnamefont {A.~H.}\ \bibnamefont {Guth}}, \
  and\ \bibinfo {author} {\bibfnamefont {A.}~\bibnamefont {Masoumi}},\
  }\href@noop {} {\  (\bibinfo {year} {2016})},\ \Eprint
  {http://arxiv.org/abs/1612.05224} {arXiv:1612.05224 [hep-th]} \BibitemShut
  {NoStop}%
%%CITATION = ARXIV:1612.05224;%%
\bibitem [{\citenamefont {Masoumi}\ \emph
  {et~al.}(2017{\natexlab{c}})\citenamefont {Masoumi}, \citenamefont
  {Vilenkin},\ and\ \citenamefont {Yamada}}]{Masoumi:2017xbe}%
  \BibitemOpen
  \bibfield  {author} {\bibinfo {author} {\bibfnamefont {A.}~\bibnamefont
  {Masoumi}}, \bibinfo {author} {\bibfnamefont {A.}~\bibnamefont {Vilenkin}}, \
  and\ \bibinfo {author} {\bibfnamefont {M.}~\bibnamefont {Yamada}},\ }\href
  {\doibase 10.1088/1475-7516/2017/12/035} {\bibfield  {journal} {\bibinfo
  {journal} {JCAP}\ }\textbf {\bibinfo {volume} {1712}},\ \bibinfo {pages}
  {035} (\bibinfo {year} {2017}{\natexlab{c}})},\ \Eprint
  {http://arxiv.org/abs/1707.03520} {arXiv:1707.03520 [hep-th]} \BibitemShut
  {NoStop}%
%%CITATION = ARXIV:1707.03520;%%
\bibitem [{\citenamefont {Blanco-Pillado}\ \emph {et~al.}(2018)\citenamefont
  {Blanco-Pillado}, \citenamefont {Vilenkin},\ and\ \citenamefont
  {Yamada}}]{Blanco-Pillado:2017nin}%
  \BibitemOpen
  \bibfield  {author} {\bibinfo {author} {\bibfnamefont {J.~J.}\ \bibnamefont
  {Blanco-Pillado}}, \bibinfo {author} {\bibfnamefont {A.}~\bibnamefont
  {Vilenkin}}, \ and\ \bibinfo {author} {\bibfnamefont {M.}~\bibnamefont
  {Yamada}},\ }\href {\doibase 10.1007/JHEP02(2018)130} {\bibfield  {journal}
  {\bibinfo  {journal} {JHEP}\ }\textbf {\bibinfo {volume} {02}},\ \bibinfo
  {pages} {130} (\bibinfo {year} {2018})},\ \Eprint
  {http://arxiv.org/abs/1711.00491} {arXiv:1711.00491 [hep-th]} \BibitemShut
  {NoStop}%
%%CITATION = ARXIV:1711.00491;%%
\bibitem [{\citenamefont {Coleman}(1977)}]{Coleman:1977py}%
  \BibitemOpen
  \bibfield  {author} {\bibinfo {author} {\bibfnamefont {S.~R.}\ \bibnamefont
  {Coleman}},\ }\href {\doibase 10.1103/PhysRevD.15.2929,
  10.1103/PhysRevD.16.1248} {\bibfield  {journal} {\bibinfo  {journal} {Phys.
  Rev.}\ }\textbf {\bibinfo {volume} {D15}},\ \bibinfo {pages} {2929} (\bibinfo
  {year} {1977})},\ \bibinfo {note} {[Erratum: Phys.
  Rev.D16,1248(1977)]}\BibitemShut {NoStop}%
%%CITATION = PHRVA,D15,2929;%%
\bibitem [{\citenamefont {Callan}\ and\ \citenamefont
  {Coleman}(1977)}]{Callan:1977pt}%
  \BibitemOpen
  \bibfield  {author} {\bibinfo {author} {\bibfnamefont {C.~G.}\ \bibnamefont
  {Callan}, \bibfnamefont {Jr.}}\ and\ \bibinfo {author} {\bibfnamefont
  {S.~R.}\ \bibnamefont {Coleman}},\ }\href {\doibase 10.1103/PhysRevD.16.1762}
  {\bibfield  {journal} {\bibinfo  {journal} {Phys. Rev.}\ }\textbf {\bibinfo
  {volume} {D16}},\ \bibinfo {pages} {1762} (\bibinfo {year}
  {1977})}\BibitemShut {NoStop}%
%%CITATION = PHRVA,D16,1762;%%
\bibitem [{\citenamefont {Greene}\ \emph {et~al.}(2013)\citenamefont {Greene},
  \citenamefont {Kagan}, \citenamefont {Masoumi}, \citenamefont {Mehta},
  \citenamefont {Weinberg},\ and\ \citenamefont {Xiao}}]{Greene:2013ida}%
  \BibitemOpen
  \bibfield  {author} {\bibinfo {author} {\bibfnamefont {B.}~\bibnamefont
  {Greene}}, \bibinfo {author} {\bibfnamefont {D.}~\bibnamefont {Kagan}},
  \bibinfo {author} {\bibfnamefont {A.}~\bibnamefont {Masoumi}}, \bibinfo
  {author} {\bibfnamefont {D.}~\bibnamefont {Mehta}}, \bibinfo {author}
  {\bibfnamefont {E.~J.}\ \bibnamefont {Weinberg}}, \ and\ \bibinfo {author}
  {\bibfnamefont {X.}~\bibnamefont {Xiao}},\ }\href {\doibase
  10.1103/PhysRevD.88.026005} {\bibfield  {journal} {\bibinfo  {journal} {Phys.
  Rev.}\ }\textbf {\bibinfo {volume} {D88}},\ \bibinfo {pages} {026005}
  (\bibinfo {year} {2013})},\ \Eprint {http://arxiv.org/abs/1303.4428}
  {arXiv:1303.4428 [hep-th]} \BibitemShut {NoStop}%
%%CITATION = ARXIV:1303.4428;%%
\bibitem [{\citenamefont {Aravind}\ \emph {et~al.}(2014)\citenamefont
  {Aravind}, \citenamefont {Lorshbough},\ and\ \citenamefont
  {Paban}}]{Aravind:2014aza}%
  \BibitemOpen
  \bibfield  {author} {\bibinfo {author} {\bibfnamefont {A.}~\bibnamefont
  {Aravind}}, \bibinfo {author} {\bibfnamefont {D.}~\bibnamefont {Lorshbough}},
  \ and\ \bibinfo {author} {\bibfnamefont {S.}~\bibnamefont {Paban}},\ }\href
  {\doibase 10.1103/PhysRevD.89.103535} {\bibfield  {journal} {\bibinfo
  {journal} {Phys. Rev.}\ }\textbf {\bibinfo {volume} {D89}},\ \bibinfo {pages}
  {103535} (\bibinfo {year} {2014})},\ \Eprint {http://arxiv.org/abs/1401.1230}
  {arXiv:1401.1230 [hep-th]} \BibitemShut {NoStop}%
%%CITATION = ARXIV:1401.1230;%%
\bibitem [{\citenamefont {Aravind}\ \emph {et~al.}(2015)\citenamefont
  {Aravind}, \citenamefont {DiNunno}, \citenamefont {Lorshbough},\ and\
  \citenamefont {Paban}}]{Aravind:2014pva}%
  \BibitemOpen
  \bibfield  {author} {\bibinfo {author} {\bibfnamefont {A.}~\bibnamefont
  {Aravind}}, \bibinfo {author} {\bibfnamefont {B.~S.}\ \bibnamefont
  {DiNunno}}, \bibinfo {author} {\bibfnamefont {D.}~\bibnamefont {Lorshbough}},
  \ and\ \bibinfo {author} {\bibfnamefont {S.}~\bibnamefont {Paban}},\ }\href
  {\doibase 10.1103/PhysRevD.91.025026} {\bibfield  {journal} {\bibinfo
  {journal} {Phys. Rev.}\ }\textbf {\bibinfo {volume} {D91}},\ \bibinfo {pages}
  {025026} (\bibinfo {year} {2015})},\ \Eprint {http://arxiv.org/abs/1412.3160}
  {arXiv:1412.3160 [hep-th]} \BibitemShut {NoStop}%
%%CITATION = ARXIV:1412.3160;%%
\bibitem [{\citenamefont {Dine}\ and\ \citenamefont
  {Paban}(2015)}]{Dine:2015ioa}%
  \BibitemOpen
  \bibfield  {author} {\bibinfo {author} {\bibfnamefont {M.}~\bibnamefont
  {Dine}}\ and\ \bibinfo {author} {\bibfnamefont {S.}~\bibnamefont {Paban}},\
  }\href {\doibase 10.1007/JHEP10(2015)088} {\bibfield  {journal} {\bibinfo
  {journal} {JHEP}\ }\textbf {\bibinfo {volume} {10}},\ \bibinfo {pages} {088}
  (\bibinfo {year} {2015})},\ \Eprint {http://arxiv.org/abs/1506.06428}
  {arXiv:1506.06428 [hep-th]} \BibitemShut {NoStop}%
%%CITATION = ARXIV:1506.06428;%%
\bibitem [{\citenamefont {Freivogel}\ \emph {et~al.}(2006)\citenamefont
  {Freivogel}, \citenamefont {Kleban}, \citenamefont {Rodriguez~Martinez},\
  and\ \citenamefont {Susskind}}]{Freivogel:2005vv}%
  \BibitemOpen
  \bibfield  {author} {\bibinfo {author} {\bibfnamefont {B.}~\bibnamefont
  {Freivogel}}, \bibinfo {author} {\bibfnamefont {M.}~\bibnamefont {Kleban}},
  \bibinfo {author} {\bibfnamefont {M.}~\bibnamefont {Rodriguez~Martinez}}, \
  and\ \bibinfo {author} {\bibfnamefont {L.}~\bibnamefont {Susskind}},\ }\href
  {\doibase 10.1088/1126-6708/2006/03/039} {\bibfield  {journal} {\bibinfo
  {journal} {JHEP}\ }\textbf {\bibinfo {volume} {03}},\ \bibinfo {pages} {039}
  (\bibinfo {year} {2006})},\ \Eprint {http://arxiv.org/abs/hep-th/0505232}
  {arXiv:hep-th/0505232 [hep-th]} \BibitemShut {NoStop}%
%%CITATION = HEP-TH/0505232;%%
\bibitem [{\citenamefont {Blanco-Pillado}\ \emph {et~al.}(2013)\citenamefont
  {Blanco-Pillado}, \citenamefont {Gomez-Reino},\ and\ \citenamefont
  {Metallinos}}]{BlancoPillado:2012cb}%
  \BibitemOpen
  \bibfield  {author} {\bibinfo {author} {\bibfnamefont {J.~J.}\ \bibnamefont
  {Blanco-Pillado}}, \bibinfo {author} {\bibfnamefont {M.}~\bibnamefont
  {Gomez-Reino}}, \ and\ \bibinfo {author} {\bibfnamefont {K.}~\bibnamefont
  {Metallinos}},\ }\href {\doibase 10.1088/1475-7516/2013/02/034} {\bibfield
  {journal} {\bibinfo  {journal} {JCAP}\ }\textbf {\bibinfo {volume} {1302}},\
  \bibinfo {pages} {034} (\bibinfo {year} {2013})},\ \Eprint
  {http://arxiv.org/abs/1209.0796} {arXiv:1209.0796 [hep-th]} \BibitemShut
  {NoStop}%
%%CITATION = ARXIV:1209.0796;%%
\bibitem [{\citenamefont {Blanco-Pillado}\ \emph {et~al.}(2015)\citenamefont
  {Blanco-Pillado}, \citenamefont {Dias}, \citenamefont {Frazer},\ and\
  \citenamefont {Sousa}}]{Blanco-Pillado:2015bha}%
  \BibitemOpen
  \bibfield  {author} {\bibinfo {author} {\bibfnamefont {J.~J.}\ \bibnamefont
  {Blanco-Pillado}}, \bibinfo {author} {\bibfnamefont {M.}~\bibnamefont
  {Dias}}, \bibinfo {author} {\bibfnamefont {J.}~\bibnamefont {Frazer}}, \ and\
  \bibinfo {author} {\bibfnamefont {K.}~\bibnamefont {Sousa}},\ }\href
  {\doibase 10.1088/1475-7516/2015/08/035} {\  (\bibinfo {year} {2015}),\
  10.1088/1475-7516/2015/08/035},\ \bibinfo {note} {[JCAP1508,035(2015)]},\
  \Eprint {http://arxiv.org/abs/1503.07579} {arXiv:1503.07579 [astro-ph.CO]}
  \BibitemShut {NoStop}%
%%CITATION = ARXIV:1503.07579;%%
\bibitem [{\citenamefont {Lindgren}(2012)}]{lindgren2012stationary}%
  \BibitemOpen
  \bibfield  {author} {\bibinfo {author} {\bibfnamefont {G.}~\bibnamefont
  {Lindgren}},\ }\href@noop {} {\emph {\bibinfo {title} {Stationary stochastic
  processes: theory and applications}}}\ (\bibinfo  {publisher} {CRC Press},\
  \bibinfo {year} {2012})\BibitemShut {NoStop}%
\bibitem [{\citenamefont {Adler}\ and\ \citenamefont
  {Taylor}(2009)}]{adler2009random}%
  \BibitemOpen
  \bibfield  {author} {\bibinfo {author} {\bibfnamefont {R.~J.}\ \bibnamefont
  {Adler}}\ and\ \bibinfo {author} {\bibfnamefont {J.~E.}\ \bibnamefont
  {Taylor}},\ }\href@noop {} {\emph {\bibinfo {title} {Random fields and
  geometry}}}\ (\bibinfo  {publisher} {Springer Science \& Business Media},\
  \bibinfo {year} {2009})\BibitemShut {NoStop}%
\bibitem [{\citenamefont {Lindgren}(1972)}]{Lindgren}%
  \BibitemOpen
  \bibfield  {author} {\bibinfo {author} {\bibfnamefont {G.}~\bibnamefont
  {Lindgren}},\ }\href {\doibase https://doi.org/10.1007/BF02384809} {\bibfield
   {journal} {\bibinfo  {journal} {G. Ark. Mat.}\ }\textbf {\bibinfo {volume}
  {10}} (\bibinfo {year} {1972}),\
  https://doi.org/10.1007/BF02384809}\BibitemShut {NoStop}%
\bibitem [{\citenamefont {Bucher}\ and\ \citenamefont
  {Louis}(2012)}]{Bucher_2012}%
  \BibitemOpen
  \bibfield  {author} {\bibinfo {author} {\bibfnamefont {M.}~\bibnamefont
  {Bucher}}\ and\ \bibinfo {author} {\bibfnamefont {T.}~\bibnamefont {Louis}},\
  }\href {\doibase 10.1111/j.1365-2966.2012.21138.x} {\bibfield  {journal}
  {\bibinfo  {journal} {Monthly Notices of the Royal Astronomical Society}\
  }\textbf {\bibinfo {volume} {424}},\ \bibinfo {pages} {1694Ð1713} (\bibinfo
  {year} {2012})}\BibitemShut {NoStop}%
\bibitem [{\citenamefont {Marcos-Caballero}\ \emph {et~al.}(2017)\citenamefont
  {Marcos-Caballero}, \citenamefont {Mart\'inez-Gonz\'alez},\ and\
  \citenamefont {Vielva}}]{Marcos_Caballero_2017}%
  \BibitemOpen
  \bibfield  {author} {\bibinfo {author} {\bibfnamefont {A.}~\bibnamefont
  {Marcos-Caballero}}, \bibinfo {author} {\bibfnamefont {E.}~\bibnamefont
  {Mart\'inez-Gonz\'alez}}, \ and\ \bibinfo {author} {\bibfnamefont
  {P.}~\bibnamefont {Vielva}},\ }\href {\doibase 10.1088/1475-7516/2017/05/023}
  {\bibfield  {journal} {\bibinfo  {journal} {Journal of Cosmology and
  Astroparticle Physics}\ }\textbf {\bibinfo {volume} {2017}},\ \bibinfo
  {pages} {023Ð023} (\bibinfo {year} {2017})}\BibitemShut {NoStop}%
\bibitem [{\citenamefont {Bardeen}\ \emph {et~al.}(1986)\citenamefont
  {Bardeen}, \citenamefont {Bond}, \citenamefont {Kaiser},\ and\ \citenamefont
  {Szalay}}]{Bardeen:1985tr}%
  \BibitemOpen
  \bibfield  {author} {\bibinfo {author} {\bibfnamefont {J.~M.}\ \bibnamefont
  {Bardeen}}, \bibinfo {author} {\bibfnamefont {J.~R.}\ \bibnamefont {Bond}},
  \bibinfo {author} {\bibfnamefont {N.}~\bibnamefont {Kaiser}}, \ and\ \bibinfo
  {author} {\bibfnamefont {A.~S.}\ \bibnamefont {Szalay}},\ }\href {\doibase
  10.1086/164143} {\bibfield  {journal} {\bibinfo  {journal} {Astrophys. J.}\
  }\textbf {\bibinfo {volume} {304}},\ \bibinfo {pages} {15} (\bibinfo {year}
  {1986})}\BibitemShut {NoStop}%
%%CITATION = ASJOA,304,15;%%
\bibitem [{\citenamefont {Bertschinger}(1987)}]{Bertschinger:1987qp}%
  \BibitemOpen
  \bibfield  {author} {\bibinfo {author} {\bibfnamefont {E.}~\bibnamefont
  {Bertschinger}},\ }\href {\doibase 10.1086/185066} {\bibfield  {journal}
  {\bibinfo  {journal} {Astrophys. J.}\ }\textbf {\bibinfo {volume} {323}},\
  \bibinfo {pages} {L103} (\bibinfo {year} {1987})}\BibitemShut {NoStop}%
%%CITATION = ASJOA,323,L103;%%
\bibitem [{\citenamefont {Ganon}\ and\ \citenamefont
  {Hoffman}(1993)}]{1993ApJ415L5G}%
  \BibitemOpen
  \bibfield  {author} {\bibinfo {author} {\bibfnamefont {G.}~\bibnamefont
  {Ganon}}\ and\ \bibinfo {author} {\bibfnamefont {Y.}~\bibnamefont
  {Hoffman}},\ }\href {\doibase 10.1086/187019} {\bibfield  {journal} {\bibinfo
   {journal} {Astrophys. J.}\ }\textbf {\bibinfo {volume} {415}},\ \bibinfo
  {pages} {L5} (\bibinfo {year} {1993})}\BibitemShut {NoStop}%
%%CITATION = ASJOA,323,L103;%%
\bibitem [{\citenamefont {Wainwright}(2012)}]{Wainwright:2011kj}%
  \BibitemOpen
  \bibfield  {author} {\bibinfo {author} {\bibfnamefont {C.~L.}\ \bibnamefont
  {Wainwright}},\ }\href {\doibase 10.1016/j.cpc.2012.04.004} {\bibfield
  {journal} {\bibinfo  {journal} {Comput. Phys. Commun.}\ }\textbf {\bibinfo
  {volume} {183}},\ \bibinfo {pages} {2006} (\bibinfo {year} {2012})},\ \Eprint
  {http://arxiv.org/abs/1109.4189} {arXiv:1109.4189 [hep-ph]} \BibitemShut
  {NoStop}%
%%CITATION = ARXIV:1109.4189;%%
\bibitem [{\citenamefont {Athron}\ \emph {et~al.}(2019)\citenamefont {Athron},
  \citenamefont {Balázs}, \citenamefont {Bardsley}, \citenamefont {Fowlie},
  \citenamefont {Harries},\ and\ \citenamefont {White}}]{Athron:2019nbd}%
  \BibitemOpen
  \bibfield  {author} {\bibinfo {author} {\bibfnamefont {P.}~\bibnamefont
  {Athron}}, \bibinfo {author} {\bibfnamefont {C.}~\bibnamefont {Balázs}},
  \bibinfo {author} {\bibfnamefont {M.}~\bibnamefont {Bardsley}}, \bibinfo
  {author} {\bibfnamefont {A.}~\bibnamefont {Fowlie}}, \bibinfo {author}
  {\bibfnamefont {D.}~\bibnamefont {Harries}}, \ and\ \bibinfo {author}
  {\bibfnamefont {G.}~\bibnamefont {White}},\ }\href@noop {} {\  (\bibinfo
  {year} {2019})},\ \Eprint {http://arxiv.org/abs/1901.03714} {arXiv:1901.03714
  [hep-ph]} \BibitemShut {NoStop}%
%%CITATION = ARXIV:1901.03714;%%
\bibitem [{\citenamefont {Espinosa}(2018)}]{Espinosa:2018hue}%
  \BibitemOpen
  \bibfield  {author} {\bibinfo {author} {\bibfnamefont {J.~R.}\ \bibnamefont
  {Espinosa}},\ }\href {\doibase 10.1088/1475-7516/2018/07/036} {\bibfield
  {journal} {\bibinfo  {journal} {JCAP}\ }\textbf {\bibinfo {volume} {1807}},\
  \bibinfo {pages} {036} (\bibinfo {year} {2018})},\ \Eprint
  {http://arxiv.org/abs/1805.03680} {arXiv:1805.03680 [hep-th]} \BibitemShut
  {NoStop}%
%%CITATION = ARXIV:1805.03680;%%
\bibitem [{\citenamefont {Espinosa}\ and\ \citenamefont
  {Konstandin}(2019)}]{Espinosa:2018szu}%
  \BibitemOpen
  \bibfield  {author} {\bibinfo {author} {\bibfnamefont {J.~R.}\ \bibnamefont
  {Espinosa}}\ and\ \bibinfo {author} {\bibfnamefont {T.}~\bibnamefont
  {Konstandin}},\ }\href {\doibase 10.1088/1475-7516/2019/01/051} {\bibfield
  {journal} {\bibinfo  {journal} {JCAP}\ }\textbf {\bibinfo {volume} {1901}},\
  \bibinfo {pages} {051} (\bibinfo {year} {2019})},\ \Eprint
  {http://arxiv.org/abs/1811.09185} {arXiv:1811.09185 [hep-th]} \BibitemShut
  {NoStop}%
%%CITATION = ARXIV:1811.09185;%%
\bibitem [{\citenamefont {Masoumi}\ \emph
  {et~al.}(2017{\natexlab{d}})\citenamefont {Masoumi}, \citenamefont {Olum},\
  and\ \citenamefont {Shlaer}}]{Masoumi:2016wot}%
  \BibitemOpen
  \bibfield  {author} {\bibinfo {author} {\bibfnamefont {A.}~\bibnamefont
  {Masoumi}}, \bibinfo {author} {\bibfnamefont {K.~D.}\ \bibnamefont {Olum}}, \
  and\ \bibinfo {author} {\bibfnamefont {B.}~\bibnamefont {Shlaer}},\ }\href
  {\doibase 10.1088/1475-7516/2017/01/051} {\bibfield  {journal} {\bibinfo
  {journal} {JCAP}\ }\textbf {\bibinfo {volume} {1701}},\ \bibinfo {pages}
  {051} (\bibinfo {year} {2017}{\natexlab{d}})},\ \Eprint
  {http://arxiv.org/abs/1610.06594} {arXiv:1610.06594 [gr-qc]} \BibitemShut
  {NoStop}%
%%CITATION = ARXIV:1610.06594;%%
\bibitem [{\citenamefont {Brown}(2018)}]{Brown:2017cca}%
  \BibitemOpen
  \bibfield  {author} {\bibinfo {author} {\bibfnamefont {A.~R.}\ \bibnamefont
  {Brown}},\ }\href {\doibase 10.1103/PhysRevD.97.105002} {\bibfield  {journal}
  {\bibinfo  {journal} {Phys. Rev.}\ }\textbf {\bibinfo {volume} {D97}},\
  \bibinfo {pages} {105002} (\bibinfo {year} {2018})},\ \Eprint
  {http://arxiv.org/abs/1711.07712} {arXiv:1711.07712 [hep-th]} \BibitemShut
  {NoStop}%
%%CITATION = ARXIV:1711.07712;%%
\bibitem [{\citenamefont {Dasgupta}(1997)}]{Dasgupta:1996qu}%
  \BibitemOpen
  \bibfield  {author} {\bibinfo {author} {\bibfnamefont {I.}~\bibnamefont
  {Dasgupta}},\ }\href {\doibase 10.1016/S0370-2693(96)01685-1} {\bibfield
  {journal} {\bibinfo  {journal} {Phys. Lett.}\ }\textbf {\bibinfo {volume}
  {B394}},\ \bibinfo {pages} {116} (\bibinfo {year} {1997})},\ \Eprint
  {http://arxiv.org/abs/hep-ph/9610403} {arXiv:hep-ph/9610403 [hep-ph]}
  \BibitemShut {NoStop}%
%%CITATION = HEP-PH/9610403;%%
\bibitem [{\citenamefont {Masoumi}\ \emph
  {et~al.}(2017{\natexlab{e}})\citenamefont {Masoumi}, \citenamefont {Olum},\
  and\ \citenamefont {Wachter}}]{Masoumi:2017trx}%
  \BibitemOpen
  \bibfield  {author} {\bibinfo {author} {\bibfnamefont {A.}~\bibnamefont
  {Masoumi}}, \bibinfo {author} {\bibfnamefont {K.~D.}\ \bibnamefont {Olum}}, \
  and\ \bibinfo {author} {\bibfnamefont {J.~M.}\ \bibnamefont {Wachter}},\
  }\href {\doibase 10.1088/1475-7516/2017/10/022} {\bibfield  {journal}
  {\bibinfo  {journal} {JCAP}\ }\textbf {\bibinfo {volume} {1710}},\ \bibinfo
  {pages} {022} (\bibinfo {year} {2017}{\natexlab{e}})},\ \Eprint
  {http://arxiv.org/abs/1702.00356} {arXiv:1702.00356 [gr-qc]} \BibitemShut
  {NoStop}%
%%CITATION = ARXIV:1702.00356;%%
\bibitem [{\citenamefont {Sarid}(1998)}]{Sarid:1998sn}%
  \BibitemOpen
  \bibfield  {author} {\bibinfo {author} {\bibfnamefont {U.}~\bibnamefont
  {Sarid}},\ }\href {\doibase 10.1103/PhysRevD.58.085017} {\bibfield  {journal}
  {\bibinfo  {journal} {Phys. Rev.}\ }\textbf {\bibinfo {volume} {D58}},\
  \bibinfo {pages} {085017} (\bibinfo {year} {1998})},\ \Eprint
  {http://arxiv.org/abs/hep-ph/9804308} {arXiv:hep-ph/9804308 [hep-ph]}
  \BibitemShut {NoStop}%
%%CITATION = HEP-PH/9804308;%%
\bibitem [{\citenamefont {Bjorkmo}\ and\ \citenamefont
  {Marsh}(2018{\natexlab{b}})}]{Bjorkmo:2017nzd}%
  \BibitemOpen
  \bibfield  {author} {\bibinfo {author} {\bibfnamefont {T.}~\bibnamefont
  {Bjorkmo}}\ and\ \bibinfo {author} {\bibfnamefont {M.~C.~D.}\ \bibnamefont
  {Marsh}},\ }\href {\doibase 10.1088/1475-7516/2018/02/037} {\bibfield
  {journal} {\bibinfo  {journal} {JCAP}\ }\textbf {\bibinfo {volume} {1802}},\
  \bibinfo {pages} {037} (\bibinfo {year} {2018}{\natexlab{b}})},\ \Eprint
  {http://arxiv.org/abs/1709.10076} {arXiv:1709.10076 [astro-ph.CO]}
  \BibitemShut {NoStop}%
%%CITATION = ARXIV:1709.10076;%%
\bibitem [{\citenamefont {Baumann}\ \emph {et~al.}(2008)\citenamefont
  {Baumann}, \citenamefont {Dymarsky}, \citenamefont {Klebanov},\ and\
  \citenamefont {McAllister}}]{Baumann:2007ah}%
  \BibitemOpen
  \bibfield  {author} {\bibinfo {author} {\bibfnamefont {D.}~\bibnamefont
  {Baumann}}, \bibinfo {author} {\bibfnamefont {A.}~\bibnamefont {Dymarsky}},
  \bibinfo {author} {\bibfnamefont {I.~R.}\ \bibnamefont {Klebanov}}, \ and\
  \bibinfo {author} {\bibfnamefont {L.}~\bibnamefont {McAllister}},\ }\href
  {\doibase 10.1088/1475-7516/2008/01/024} {\bibfield  {journal} {\bibinfo
  {journal} {JCAP}\ }\textbf {\bibinfo {volume} {0801}},\ \bibinfo {pages}
  {024} (\bibinfo {year} {2008})},\ \Eprint {http://arxiv.org/abs/0706.0360}
  {arXiv:0706.0360 [hep-th]} \BibitemShut {NoStop}%
%%CITATION = ARXIV:0706.0360;%%
\bibitem [{\citenamefont {Dias}\ \emph {et~al.}(2015)\citenamefont {Dias},
  \citenamefont {Frazer},\ and\ \citenamefont {Seery}}]{Dias:2015rca}%
  \BibitemOpen
  \bibfield  {author} {\bibinfo {author} {\bibfnamefont {M.}~\bibnamefont
  {Dias}}, \bibinfo {author} {\bibfnamefont {J.}~\bibnamefont {Frazer}}, \ and\
  \bibinfo {author} {\bibfnamefont {D.}~\bibnamefont {Seery}},\ }\href
  {\doibase 10.1088/1475-7516/2015/12/030} {\bibfield  {journal} {\bibinfo
  {journal} {JCAP}\ }\textbf {\bibinfo {volume} {1512}},\ \bibinfo {pages}
  {030} (\bibinfo {year} {2015})},\ \Eprint {http://arxiv.org/abs/1502.03125}
  {arXiv:1502.03125 [astro-ph.CO]} \BibitemShut {NoStop}%
%%CITATION = ARXIV:1502.03125;%%
\bibitem [{\citenamefont {Coleman}\ and\ \citenamefont
  {De~Luccia}(1980)}]{Coleman:1980aw}%
  \BibitemOpen
  \bibfield  {author} {\bibinfo {author} {\bibfnamefont {S.~R.}\ \bibnamefont
  {Coleman}}\ and\ \bibinfo {author} {\bibfnamefont {F.}~\bibnamefont
  {De~Luccia}},\ }\href {\doibase 10.1103/PhysRevD.21.3305} {\bibfield
  {journal} {\bibinfo  {journal} {Phys. Rev.}\ }\textbf {\bibinfo {volume}
  {D21}},\ \bibinfo {pages} {3305} (\bibinfo {year} {1980})}\BibitemShut
  {NoStop}%
%%CITATION = PHRVA,D21,3305;%%
\bibitem [{\citenamefont {Freivogel}(2011)}]{Freivogel:2011eg}%
  \BibitemOpen
  \bibfield  {author} {\bibinfo {author} {\bibfnamefont {B.}~\bibnamefont
  {Freivogel}},\ }\href {\doibase 10.1088/0264-9381/28/20/204007} {\bibfield
  {journal} {\bibinfo  {journal} {Class. Quant. Grav.}\ }\textbf {\bibinfo
  {volume} {28}},\ \bibinfo {pages} {204007} (\bibinfo {year} {2011})},\
  \Eprint {http://arxiv.org/abs/1105.0244} {arXiv:1105.0244 [hep-th]}
  \BibitemShut {NoStop}%
%%CITATION = ARXIV:1105.0244;%%
\bibitem [{\citenamefont {Riley}\ \emph {et~al.}(2006)\citenamefont {Riley},
  \citenamefont {Hobson},\ and\ \citenamefont {Bence}}]{riley2006mathematical}%
  \BibitemOpen
  \bibfield  {author} {\bibinfo {author} {\bibfnamefont {K.~F.}\ \bibnamefont
  {Riley}}, \bibinfo {author} {\bibfnamefont {M.~P.}\ \bibnamefont {Hobson}}, \
  and\ \bibinfo {author} {\bibfnamefont {S.~J.}\ \bibnamefont {Bence}},\
  }\href@noop {} {\emph {\bibinfo {title} {Mathematical methods for physics and
  engineering: a comprehensive guide}}}\ (\bibinfo  {publisher} {Cambridge
  university press},\ \bibinfo {year} {2006})\BibitemShut {NoStop}%
\bibitem [{\citenamefont {Press}\ \emph {et~al.}(2007)\citenamefont {Press},
  \citenamefont {Teukolsky}, \citenamefont {Vetterling},\ and\ \citenamefont
  {Flannery}}]{press2007numerical}%
  \BibitemOpen
  \bibfield  {author} {\bibinfo {author} {\bibfnamefont {W.~H.}\ \bibnamefont
  {Press}}, \bibinfo {author} {\bibfnamefont {S.~A.}\ \bibnamefont
  {Teukolsky}}, \bibinfo {author} {\bibfnamefont {W.~T.}\ \bibnamefont
  {Vetterling}}, \ and\ \bibinfo {author} {\bibfnamefont {B.~P.}\ \bibnamefont
  {Flannery}},\ }\href@noop {} {\emph {\bibinfo {title} {Numerical recipes 3rd
  edition: The art of scientific computing}}}\ (\bibinfo  {publisher}
  {Cambridge university press},\ \bibinfo {year} {2007})\BibitemShut {NoStop}%
\end{thebibliography}%

\end{document}